\documentclass[11pt]{article}
\usepackage{graphicx} 
\usepackage[a4paper, total={7in, 9in}]{geometry}
\usepackage{amsmath,amsfonts,amssymb}
\usepackage{dutchcal}
\usepackage[maxnames=99, maxcitenames=99]{biblatex}
\usepackage{float}
\usepackage{subcaption}
\usepackage{algorithm}
\usepackage{algpseudocode}
\usepackage{hyperref}
\usepackage{multirow}
\usepackage{booktabs}
\usepackage{caption}
\usepackage{xcolor}
\usepackage{adjustbox}
\usepackage{placeins}
\usepackage{makecell}
\usepackage{appendix}
\usepackage{tikz}
\usepackage[percent]{overpic}

\newenvironment{system}%
{\left\lbrace\begin{array}{@{}l@{}}}%
{\end{array}\right.}

\newcommand{\lp}{\left(}
\newcommand{\rp}{\right)}
\newcommand{\email}[1]{\texttt{#1}}
\title{Comparing Deterministic and Stochastic Parameter Recovery Algorithms Applied to Chaotic Systems}

\author{Ashley Wang \thanks{Stanford University, Stanford, CA 
  (\email{ashmwang@stanford.edu}).}
\and Elizabeth Carlson\thanks{Oregon State University, Corvallis, OR 
  (\email{carleliz@oregonstate.edu}).}
\and Franca Hoffman \footnotemark[3]\thanks{California Institute of Technology, Pasadena, CA  
  (\email{franca.hoffmann@caltech.edu}).}
  }

\begin{document}
\maketitle
\begin{abstract}

This paper explores the effectiveness of various novel deterministic and traditional stochastic data assimilation (DA) and parameter recovery (PR) algorithms given noisy data from chaotic systems. 
We use semi-analytic methods 
to numerically construct synthetic data from the Lorenz '63 and multiscale Lorenz '96 chaotic dynamical systems, adding white noise.  
Our findings show that, for different noise levels, deterministic PR algorithms paired with deterministic DA algorithms are shown computationally to be overall more accurate and stable than stochastic PR algorithms. Additionally, deterministic PR methods have demonstrated greater speed and efficiency, requiring less computational power than stochastic PR methods. This suggests that future work should consider exploring the full potential of deterministic PR algorithms in the presence of noise.
\end{abstract}

\section{Introduction}

Mathematical models of real-world phenomena often involve knowing important physical constants.  These physical constants are usually determined by analyzing observed data obtained either from experiments or real-world measurements.  Adaptively incorporating data into a model is known as data assimilation (DA), and allows one to dynamically correct model trajectories toward the data.  DA methods tend to fall into one of two categories: deterministic methods, which do not account for any error introduced by noisy data or systems with white noise inherent in the models, and stochastic methods, which do attempt to account for said noise.  The main deterministic DA method is nudging, first introduced in \cite{Hoke_Anthes_1976_MWR} with variants made rigorous for PDEs in \cite{Azouani_Olson_Titi_2014}.  Stochastic DA methods include 3D/4DVAR, the Kalman filter and its variants, and the particle filter and its variants.  Most stochastic DA methods are either too expensive for large scale models or they are only  optimal for linear models with Gaussian white noise, and these methods are by far the most popular in practice due to their robustness in handling noise, but they are not provably optimal for nonlinear systems, which are generally the systems we are interested in predicting.

DA algorithms can also be augmented with algorithms to recover the physical parameters of the system using the data, providing a particular methodical way to find these parameters.  Parameters in many models are either unknown or approximated, and hence choosing parameters so our models best approximate our data is a long standing problem.
DA-based stochastic methods of parameter recovery (PR) on chaotic systems have been thoroughly investigated for decades (see, e.g., \cite{Dempster_Laird_Rubin_1977, Iglesias_Law_Stuart_2013, Ljung_1979}), but this is not the only approach.  However, these methods, while indirectly dependent on the dynamics, rely heavily on the existence of noise either in the data or the system itself.
The deterministic DA method of nudging has a similar form as 3DVAR and Kalman filter and its variants, and indeed similar methods as for the  convergence of a nudging variant known as the Azouani-Olson-Titi (AOT) (or continuous data assimilation (CDA) \footnote{where ``continuous" represents the fact that we are working with a differential equation, not a difference equation}) algorithm \cite{Azouani_Olson_Titi_2014} can be used to prove the convergence of 3DVAR \cite{Bessaih_Olson_Titi_2015, Biswas_Branicki_2024} and Ensemble Kalman filter \cite{Biswas_Branicki_2024} under strict assumptions.
The AOT algorithm has been shown to be robust to a variety of different types of errors, including errors due to incorrect parameters, and the analysis has been used to develop deterministic PR algorithms, solely dependent on what is happening with the dynamics, which provably converge to the true parameter in the absence of noise \cite{Carlson_Hudson_Larios_2020, Carlson_Hudson_Larios_Martinez_Ng_Whitehead_2021, Martinez_2022_pr, Martinez_Murri_Whitehead_2025, Newey_Whitehead_Carlson_2025, Pachev_Whitehead_McQuarrie_2021concurrent}.
A comparison of these methods, specifically their accuracy and computational cost, has not been done in the presence of noisy data, and at this point it seems prudent to do so.  In this paper we provide a detailed computational comparison of deterministic and stochastic DA and PR methods given noisy data from chaotic systems.

The deterministic PR algorithms we investigate in this paper were all developed in the context of the Azouani-Olson-Titi (AOT) algorithm.  These algorithms are the Carlson-Hudson-Larios (CHL) \cite{Carlson_Hudson_Larios_2020}, Pachev-Whitehead-McQuarrie (PWM)  \cite{Pachev_Whitehead_McQuarrie_2021concurrent}, and Asymptotic Variational (AV) methods \cite{Newey_Whitehead_Carlson_2025}.  Stochastic parameter recovery algorithms are developed either in tandem with or separate from the DA method they are generally paired with.  The PR algorithm of Expectation-Maximization (EM) is independent of DA as a method but requires noise to be used, so it can only be used when there is at least noise on the  data.  We combine EM with the Ensemble Kalman Filter (EnKF) \cite{Dempster_Laird_Rubin_1977, Ghahramani_Roweis_1998, Kalnay_2003_DA_book, Pei_et_al_2019, textbook_sanz_alonso, Chen_SanzAlonso_Willett_2022} as well as a version of the particle filter (PF) known as the Gaussianized optimal particle filter \cite{Pitt_Shephard_1999} and a leading ensemble filter Local Ensemble Transform Kalman Filter \cite{soaLETKF,LETKF}; in our implementation we remove the ``local" part as for lower dimensional ordinary differential equations does not improve and indeed may in fact degrade the results (see, e.g., \cite{Nerger_2021, LETKF, AlGhattas_SanzAlonso_2024}).  We also consider the case where one can treat a parameter as a 1-dimensional extension of the state vector, for both EnKF and PF.  Finally, we experimented with a version of Ensemble Kalman Inversion (EnKI) as a PR algorithm \cite{edo_reich_stuart_2024, textbook_sanz_alonso}, where the forward model can be chosen to be either the true physical model or a corresponding DA system; here we report the results with the AOT algorithm as the results were better than using the true forward model. EnKF, PF, EnKI, and ETKF make use of a discrete time representation of a differential equation for the evolution of the state, whereas AOT is referred to as a continuous DA algorithm because it uses in its formulation the continuous time dynamical system.  Note that, regardless of continuous DA or not, we are only using discrete time observations in all our implementations. We summarize the implementations we compare in Table \ref{basicTable}.


\begin{table}[h!]
\caption{\textbf{Experimental Setups}}
\label{basicTable}
\adjustbox{margin=0pt,center}{ 
\centering
\begin{tabular}{| c | c | c |}
\hline
\textbf{Data Assimilation Algorithm} & \textbf{Parameter Recovery Algorithm} & \textbf{Forward Model} \\
\hline \hline
\multirow{4}{*}{AOT (Sec. \ref{sec:AOT})} 
 & CHL (Sec. \ref{sec:CHL}) & N/A \\  \cline{2-2}
 & PWM (Sec. \ref{sec:PWM}) & N/A  \\  \cline{2-2}
 & AV (Sec. \ref{sec:AV}) & N/A \\ \cline{2-2}
 & EM (Sec. \ref{sec:EM}, Alg \ref{algorithm:EM})  & N/A  \\ \hline
\multirow{3}{*}{EnKF (Alg \ref{algorithm:EnKF})} 
 & EM  & True \\  \cline{2-2}
 & PWM & True \\ \cline{2-2}
 & CHL & True \\ \hline
N/A & EnKI (Sec. \ref{sec:EnKI}, Alg \ref{algorithm:EnKI}) & AOT \\ \hline
\multirow{4}{*}{PF (Alg \ref{gopfAlg})} 
 & EM & True \\ \cline{2-2}
 & Ext. (Sec. \ref{sec:PF_extension})& True \\ \cline{2-2}
 & CHL & AOT \\ \cline{2-2}
 & PWM & AOT \\ \hline
 \multirow{3}{*}{ETKF (Alg \ref{alg:ETKF})}
 & EM & True\\ \cline{2-2}
 & CHL & True \\ \cline{2-2}
 & PWM & True \\ \cline{2-2}
 & AV & True \\ \hline
\end{tabular}
}
\end{table}

The results of this work show that nearly all algorithms used are robust and recover the parameter even with far-from-true initial estimates.  In particular,
\begin{itemize}
    \item The errors with the lowest mean and variance were consistently obtained by the deterministic DA methods, usually when paired with either the PWM or AV algorithms.
    \item The errors that had the most consistent mean and variance values across different levels of noise in the observations were obtained when using the EM algorithm, paired with either deterministic or stochastic DA methods.
    \item EnKI resulted in the lowest parameter error means and variances among the stochastic PR algorithms.
    \item With the exception of the CHL algorithm and sometimes the EM algorithm, the PR algorithms were robust to changes in the variance of the noise of the observations.
    \item The lowest and most consistent error mean and variance of the mixed methods were given by ETKF + AV. 
\end{itemize}   
Additional takeaways include the following.
\begin{itemize}
    \item An initial comparison of the chosen, semi-analytic numerical method and the Euler method was made in initial comparisons of some of these algorithms on the Lorenz '63 system.  It was consequently observed that wider ranges of incorrect parameters can be chosen and corrected when a better numerical discretization was used.  See Section \ref{sec:L63_results} for more details.
    \item The work was also applied to data observed sparsely in time (every 10 time steps) in the AOT + CHL subcase for the Lorenz '96 case in Section \ref{sec:wide_PR_window}.  Results were approximately the same, except the $\mu$ portion of the parameter recovery had to be modified to nudge less strongly towards the true parameter.
\end{itemize} 
As a result of these observations, future investigations will look into the sensitivity of the deterministic PR algorithms to data observation frequency and numerical discretization.

The outline of the paper is as follows: in Section \ref{sec::setup}, we present the chaotic models used as the test cases, the corresponding numerical methods, and details on the different DA and PR algorithms; in Section \ref{sec:results} we present our comprehensive findings. In the appendices, we provide additional technical details; in Appendices A and B, we define the analytical derivatives computed for part of the numerical discretization algorithm, and in Appendix C, we derive the deterministic parameter recovery algorithms for the Lorenz '96 system.

\section{Background and Setup}
\label{sec::setup}
\subsection{The Chaotic Systems}
Throughout this work, we consider the general form
\begin{equation}
\begin{system}\label{referenceSys}
\dot{u} = f (u)\\ u(0) = u_{0}
\end{system}
\end{equation}
where $u_{0}\in \mathbb{R}^n$ is the initial condition we use for generating observations, and $f$ some nonlinear function of $u$.  We denote the solution operator by $u(t) = \Psi(t,u_0)$, with the usual semigroup properties that 
\begin{itemize}
    \item $\Psi(0,u_0)=u_0$
    \item $\Psi(t,\Psi(s,u_0)) = \Psi(t+s,u_0)$.
\end{itemize}

\subsubsection{Lorenz '63 System}
Let us recall that the set of (dissipative) equations for the Lorenz '63 system is, with $u = (x_1,x_2,x_3)$,
\begin{equation}
\begin{system}
\dot{x_1} = \sigma(x_2-x_1),\\ \dot{x_2} = x_1(\rho-x_3)-x_2,\\ \dot{x_3} = x_1x_2 - \beta{x_3}. 
\end{system}
\end{equation}
The model is used to describe a two-dimensional fluid layer: $x$ is the rate of convection, $y$ is the horizontal temperature variation, and $z$ is the vertical temperature variation. The parameters are related to physical dynamics of the system, with $\sigma$ proportional to the Prandtl number, $\rho$ proportional to the Rayleigh number, and $\beta$ proportional to physical properties of the fluid layer. The Prandtl and Rayleigh numbers are dimensionless quantities defined in fluid mechanics; we use the standard values $\sigma = 10, \rho = 28, \beta = \frac{8}{3}$. Our experiments focus on recovering $\sigma$, see Section~\ref{sec:L63_results}.  
The observations we will take from this system are of the $x$ variable only. In this setting, as well as in a variety of others, it is known that this system and the parameter $\sigma$ are identifiable \cite{Carlson_Hudson_Larios_Martinez_Ng_Whitehead_2021}.

\subsubsection{Lorenz '96 System}
The two-layer (dissipative) Lorenz '96 system is given by, with $u = (u_1, ..., u_4, \mathbcal{u}_{1,j}, ..., \mathbcal{u}_{4,j})$, $j \in [1,..., J]$,
\begin{equation}
\label{Lorenz96}
\begin{system}
\dot{u_{i}} = u_{i-1}(u_{i+1} - u_{i-2}) + \sum_{j=1}^{J} \gamma_{i}u_{i}\mathbcal{u}_{i,j} - \overline{d_{i}}u_{i} + F,\\ \dot{\mathbcal{u}_{i,j}} = -d_{j}\mathbcal{u}_{i,j} - \gamma_{j}u_{i}^{2},
\end{system}
\end{equation}
and it belongs to a group of conceptual models in geophysical turbulence \cite{Chen_Majda_2018}. Here we leave the noise component out as we are interested in understanding the true dynamics of the underlying system as presented in \cite{Chen_Majda_2018}, and thus all noise is reserved for the observations. We take $i = 1, 2, 3, 4$, and $u_{-1} = u_{3}, u_{0} = u_{4}, u_{1} = u_{5}$. The variables $u_{1}, u_{2}, u_{3}, u_{4}$ trace large-scale motions. We also take $j = 1,...,J$ where $J=5$, and the variables $\{\mathbcal{u}_{i,j}\}$ are small-scale variables. Moreover, $\overline{d_{i}}$ and $\gamma_{k}$ are given as damping and coupling coefficients. In particular, $\overline{d_{i}}$ reflects large-scale instability, where negative values are responsible for instabilities \cite{Majda_Lee_2014}.
The structure of Lorenz '96 sits comfortably in complexity between that of Lorenz '63 and the 2D Navier-Stokes equations \cite{Carlson_Hudson_Larios_Martinez_Ng_Whitehead_2021}.  In both of the latter two cases, the state and linear parameters are identifiable \cite{Carlson_Hudson_Larios_Martinez_Ng_Whitehead_2021, Martinez_2022}, thus leading us to presume the identifiability of the parameter of choice for the Lorenz '96 equations.  The results in this paper suggest that such a proof is possible.

Specifically, we define the coefficients, as in \cite{Chen_Majda_2018}, to be 
\begin{equation}
\overline{d_{i}} = 1 + 0.7 \text{cos}\lp\frac{2\pi i}{J}\rp    \label{d1_value}
\end{equation}
and 
\begin{equation}
\gamma_{k} = 0.1 + 0.25 \text{cos} \lp\frac{2\pi k}{J}\rp
\end{equation}
$F$ is a forcing constant, and the authors set $F = 5$. $d_{j}$ is defined as follows:
\begin{equation}
d_{1} = 0.2, d_{2} = 0.5, d_{3} = 1, d_{4} = 2, d_{5} = 5,
\end{equation}
and this implies that the damping time is shorter in the smaller scales. Our experiments focus on recovering the parameter $\overline{d_{1}}$ and show that the algorithms we use are robust and recover the parameter even with far-from-true estimates, see Section~\ref{sec:L96_results}. 
The observations we take of this system will be $u_1,u_2,u_3$, and $u_4$.

\subsubsection{Numerical Methods}
\label{sec:numerical_methods}
To test our ideas, we implement some preliminary tests on recovering the Prandtl number in the standard Lorenz '63 system, and do a detailed test on a multi-scale version of the Lorenz '96 system \cite{Chen_Majda_2018}. Data is generated by approximating solutions to these systems with a high-order numerical scheme that will be discussed in the next paragraph \cite{Carlson_Hudson_Larios_Martinez_Ng_Whitehead_2021, Chen_Majda_2018}. We apply this high-order numerical scheme to all variables in Lorenz '63 and Lorenz '96, except for the slow variables in Lorenz '96, which are modeled using the first-order Euler method. This is acceptable because fast variables govern the nonlinear dynamics, and slow variables' accuracy have almost negligible influences on results. A direct comparison in Section~\ref{sec:numerical_comparison} shows that the high-order numerical method is worth implementing since convergence of estimations of the parameter $\sigma$ to the true parameter value in the Lorenz '63 system is more stable.

Authors of \cite{aMsDTM_paper} discussed the high-order adaptive multi-step differential transformation method (adaptive MsDTM), a semi-analytic method which was derived based on the finite Taylor series approximation for a desired solution $u$:
\begin{equation}
u(t+\Delta t)\ :=  \sum^{K}_{k=0}\frac{1}{k!}u^{(k)}(t)(\Delta t)^{k}
\end{equation}
When $K = 1$, we could numerically take $u_{j+1} = u_{j} + \Delta t\cdot u^{(1)}(t_{j})$, which is the first-order Euler method, or solve for $u_{j+1} = u_{j} + \Delta t\cdot u^{(1)}(t_{j+1})$, which is the implicit Euler method. In some of the literature, approximations of the analytical derivatives were used, such as in \cite{strom_larsson_2023}. The adaptive MsDTM incorporates adaptable time steps to lower computational cost. At the $j^{\text{th}}$ numerical time step, if the Euclidean distance between the states $u_{j-1}$ and $u_{j-2}$ is smaller than a fixed threshold $\epsilon$, one would extend the previous time step by $\Delta t$ more and approximate $u_{j}$ on the extended time interval. We took $K = 5$ for simulating the Lorenz '63 system and $K = 3$ for simulating the Lorenz '96 system.  For the Lorenz '96 system, we used MsDTM to evolve the fast variables, and forward Euler to evolve the slow variables.

We work with both deterministic and stochastic DA algorithms to generate dynamically corrected states of the chaotic systems. In the case of applying deterministic AOT, we evolve the true system with adaptive MsDTM and use a predictor-corrector scheme for the nudged system as follows.  This is summarized in Algorithm \ref{pseudocode::epsilonThresholdAlg}, where
\begin{itemize}
\item $J$ is the total number of numerical time steps, and 
\item $\tilde{\Psi}$ is the MsDTM (high order Taylor expansion) update approximating $\Psi$ which is approximated by the numerical solution of \eqref{referenceSys} (with the initial conditions chosen according to a Gaussian distribution) at each time $t_{j+1}$
, using variable values from the previous step and adaptive time step sizes. 
\end{itemize}
\begin{algorithm}
\caption{Predictor-Corrector Scheme}
\label{pseudocode::epsilonThresholdAlg}
\begin{algorithmic}
    \State \textbf{Input:} $N$ the total number of variables, initial true system variables $\{(y^\dagger)_{0}^{(n)}\}_{n=1}^{N}$, initial nudged system variables $\{x_{0}^{(n)}\}_{n=1}^{N}$, relaxation parameters $\{\mu^{(n)}\}_{n=1}^{N}$, initial time step sizes $\{\Delta t_{0}^{(n)}\}_{n=1}^{N}$, adaptive MsDTM threshold $\epsilon$, and previous step variable values $\{x_{p}^{(n)}\}_{n=1}^{N}$ initialized as
    \begin{equation}
        (y^\dagger)_{p}^{(n)} = (y^\dagger)_{0}^{(n)}, \; x_{p}^{(n)} = x_{0}^{(n)}.
    \end{equation}
    \State For $j = 0,1,...,J-1$ do the following prediction and correction steps:
    \State \textbf{Prediction:} 
    Let $(z_j, z_p) \in \{(y^\dagger_j, y^\dagger_p), (x_j, x_p)\}$.
    \State If $|z_{j}^{(n)} - z_{p}^{(n)}| < \epsilon$: 
    
    \begin{equation}
        \Delta t_{j+1}^{(n)} = \Delta t_{j}^{(n)} + \Delta t_{0}^{(n)}
    \end{equation}
    \State \text{Else: }
    \begin{equation}
        z_{p}^{(n)} = z_{j}^{(n)}, \Delta t_{j+1}^{(n)} = \Delta t_{0}^{(n)}
    \end{equation}
    \State Then update true system variables
    \begin{equation}
        \hat{z}_{j+1}^{(n)} = \tilde{\Psi} (\{\Delta t_{j+1}^{(n)}\}_{n=1}^{N},\{z_{p}^{(n)}\}_{n=1}^{N})
    \end{equation}
    \State \textbf{Correction:}
    \begin{equation}
        (y^\dagger)_{j+1}^{(n)} = \hat{(y^\dagger)}_{j+1}^{(n)}
    \end{equation}
    \begin{equation}
        x_{j+1}^{(n)} = \dfrac{\hat{x}_{j+1}^{(n)} + \mu^{(n)} \Delta t_{j+1}^{(n)}(y^\dagger)_{j+1}^{(n)}}{1 + \mu^{(n)}\Delta t_{j+1}^{(n)}}
    \end{equation}
    \State \textbf{Output:} True system variables $\{(y^\dagger)_{j+1}^{(n)}\}_{n=1}^{N}$ and nudged system variables $\{x_{j+1}^{(n)}\}_{n=1}^{N}$.
\end{algorithmic}
\end{algorithm}

\subsection{Data Assimilation Algorithms}
Throughout this work, we assume real data comes from \eqref{referenceSys}.  We observe data discretely in time, a l{\'a} \cite{Foias_Mondaini_Titi_2016, Larios_Pei_Victor_2023}.



\subsubsection{Deterministic Data Assimilation: Nudging}
\label{sec:AOT}
Nudging methods were originally proposed in 1976 in \cite{Hoke_Anthes_1976_MWR}, but first made rigorous for a general class of PDEs in 2014 in \cite{Azouani_Olson_Titi_2014}.  This method takes observations from a ``true" system, as in ~\ref{referenceSys}, and inserts them into a ``nudged" version of the true system, which we call the nudged system. 
Corresponding to the true system, the nudged system is defined as
\begin{equation}
\begin{system}\label{AOTgeneral}
\dot{v} = f(v) - \mu P_{m}(v - u)\\ v(0) = v_{0}
\end{system}
\end{equation}
where $\mu$ is a relaxation parameter (large constant), and $P_{m}$ is a projection matrix onto the $m$ observed state variables; for example, in our test cases, for Lorenz '63 we have $P_m((x,y,z)) = P_1((x,y,z))= x$, and for Lorenz '96 we have $P_m((\vec{u}, \vec{\mathbcal{u}})) = P_4((\vec{u}, \vec{\mathbcal{u}})) = \vec{u}$.  Under certain conditions on which and how many variables are observed and on the size of $\mu$, predictions from the model \eqref{AOTgeneral} will converge exponentially fast in time to the solution of \eqref{referenceSys} from which the observations were taken \cite{Azouani_Olson_Titi_2014}.
In general, this is the formulation for observations without noise.  When noise is added to the observations, \eqref{AOTgeneral} becomes the SDE (see, e.g., \cite{Bessaih_Olson_Titi_2015, Biswas_Branicki_2024})
\begin{equation}
\begin{system}\label{AOTgeneral_plusnoise}
dv = f(v)dt - \mu (P_m(v)- P_m(u))  dt +\mu \gamma dW
\\
v(0) = v_{0},
\end{system}
\end{equation}
with $\gamma >0$, and $dW$ is in the range of $P_m$ \cite{Bessaih_Olson_Titi_2015}.

\subsubsection{Stochastic Data Assimilation: Ensemble Kalman Filter} \label{subsubsec::EnKF}
The most common variation of the Kalman Filter for nonlinear problems is the Ensemble Kalman Filter, first proposed in \cite{Evensen_1994}.  We present the EnKF algorithm, as presented in \cite{textbook_sanz_alonso}, in Algorithm~\ref{algorithm:EnKF}, where
\begin{itemize}
    \item $v_{j+1} = \Psi(v_j)$ represents the nonlinear solution map of \eqref{referenceSys} at time $t_{j+1}$ given the solution input at time $t_j$, 
    \item $y_{j} = y^\dagger_j + \epsilon_j$ are the observations of the true system at each step (with $\epsilon_j$ white noise representing  the error in the data), and
    \item $H$ is the observation matrix (a projection matrix selecting only the observed variables from the state vector), such that $y^\dagger_j = Hu_j$.
\end{itemize}
Note that we used the adaptive MsDTM to generate synthetic data and simulate the nonlinear dynamics given by $\Psi$. 
Moreover, note that we use an ensemble size of $N = 50$ (see Table \ref{table:96_setup}) for Lorenz '63 as it is well-established that for many more particles than this the error is not improved (see, e.g., \cite{PuHacker_2009, MoonBalik_2021}), and similarly for Lorenz '96 (see, e.g., \cite{Miyoshi_2005}).  This ensemble size is also a very standard choice for large scale weather and climate models.  

\begin{algorithm}[H]
\caption{Ensemble Kalman Filter (EnKF)}
\label{algorithm:EnKF}
\begin{algorithmic}[1]
\State \textbf{Input:} Ensemble size $N$. Initial ensemble $\{v_{0}^{(n)}\}^{N}_{n=1}$. Observations $\{y_j\}_{j=1}^J$.
\State For {$j = 0$, $1$,..., $J-1$} do the following prediction and analysis steps:
\State \textbf{Prediction:}
\begin{equation}
    \xi_{j}^{(n)} \sim N(0,\Sigma),\,\, i.i.d.,\,\,n = 1,...,N,
\end{equation}
\begin{equation}
    \hat{v}_{j+1}^{(n)} = \Psi(v_{j}^{(n)}) + \xi_{j}^{(n)},\,\,n = 1,...,N,
\end{equation}
\begin{equation}
    \hat{m}_{j+1} = \frac{1}{N} \sum_{n=1}^{N} \hat{v}_{j+1}^{(n)},
\end{equation}
\begin{equation}
    \hat{C}_{j+1} = \frac{1}{N}\sum_{n=1}^{N}(\hat{v}_{j+1}^{(n)} - \hat{m}_{j+1}) \otimes (\hat{v}_{j+1}^{(n)} - \hat{m}_{j+1}).
\end{equation}
\State \textbf{Analysis:}
\begin{equation}
    {\eta}_{j+1}^{(n)} \sim N(0,\Gamma),\,\,n = 1,...,N,
\end{equation}
\begin{equation}
    y_{j+1}^{(n)} = y_{j+1} + \eta_{j+1}^{(n)},\,\,n = 1,...,N,
\end{equation}
\begin{equation}
    v_{j+1}^{(n)} = (I - K_{j+1}H)\hat{v}_{j+1}^{(n)} + K_{j+1}y_{j+1}^{(n)},\,\,n = 1,...,N.
\end{equation}
\begin{equation}
    K_{j+1} = \hat{C}_{j+1}H^{T}\lp H\hat{C}_{j+1}H^{T} + \Gamma\rp^{-1}
\label{Kalman_Gain}
\end{equation}

\State \textbf{Output:} Ensembles $\{v_{j}^{(n)}\}_{n=1}^{N}$,\,\,$j = 0,1,...,J$.
\end{algorithmic}
\end{algorithm}


\subsubsection{Stochastic Data Assimilation: Gaussianized Optimal Particle Filter}
\label{sec:original_PF}
Another stochastic DA method we used for comparison is the Gaussianized optimal particle filter (PF), invented in \cite{Pitt_Shephard_1999}.  Particle filters follow the general idea that the ``particles" are a distribution of initial conditions $v_0^{(n)}$ evolved according to the dynamics; the new distribution is then pared down based on the closeness of the evolved particles to the observations.   
We restate the algorithm for the Gaussianized optimal particle filter,  Algorithm 12.4 in \cite{textbook_sanz_alonso}, in Algorithm \ref{gopfAlg}. Again, $j$ represents the time step indexing; $w^{(n)}$ stands for the weight assigned to each particle $v^{(n)}$. Moreover, $S^{N}$ is an operator acting on a probability density function (pdf) $\pi_{0}$ by producing an $N$-samples Dirac approximation of $\pi_{0}$:
\begin{equation}
(S^{N}\pi_{0})(v) = \sum_{n=1}^{N}w^{(n)}\delta(v - v_{0}^{(n)})
\end{equation}
where $v_{0}^{(n)}$ are i.i.d. samples from $\pi_{0}$ that are weighted uniformly (i.e. $w^{(n)} = \frac{1}{N}$).
The indicator function $\mathbf{1}$ in Step 3 of subsequent sampling in Algorithm \ref{gopfAlg} has $I_{j}^{(m)}$ as supports with widths given by weights appearing in $\pi_{j}^{N}$,
\begin{equation}
\begin{aligned}
I_{j+1}^{(m)} &= [\alpha_{j+1}^{(m-1)},\alpha_{j+1}^{(m)}), \\
\alpha_{j+1}^{(m+1)} &= \alpha_{j+1}^{(m)} + w_{j+1}^{(m)}, \\
\alpha_{j+1}^{(0)} &= 0
\end{aligned}
\end{equation}
where by construction, $\alpha_{j}^{(N)} = 1$ for all $j$. $K$ is again the Kalman gain matrix as defined in \eqref{Kalman_Gain}, with $\hat{C}_{j+1}$ replaced with $\Sigma$, and $C = (I-KH)\Sigma$. \\
\begin{algorithm}
\caption{Gaussianized Optimal Particle Filter (PF)}
\label{gopfAlg}
\begin{algorithmic}[1]
\State \textbf{Input:} Initial distribution $\mathbb{P}(v_0) = \pi_0$, number of particles $N$, observations $\{y_j\}_{j=1}^J$.
\State \textbf{Initial Sampling:} Draw $N$ particles $v_0^{(n)} \sim \pi_0$ so that $\pi_0^N = S^{N} \pi_0$.
\State \textbf{Subsequent Sampling:} For $j = 0, 1, \dots, J-1$, perform:
\begin{enumerate}
    \item Set $\overline{w}_{j+1}^{(n)} = \exp\left(-\frac{1}{2}|y_{j+1} - H \Psi(v_j^{(n)})|^2_{S}\right)$.
    \item Set $w_{j+1}^{(n)} = \frac{\overline{w}_{j+1}^{(n)}}{\sum_{n=1}^N \overline{w}_{j+1}^{(n)}}$.
    \item Set $\hat{v}_j^{(n)} = \sum_{m=1}^N \mathbf{1}_{I_{j+1}^{(m)}}(r_{j+1}^{(n)}) v_j^{(m)}$ with $r_{j+1}^{(n)} \sim \text{ Uniform } (0,1) \text{ i.i.d. }$
    \item Set $v_{j+1}^{(n)} = (I - K_{j+1} H) \Psi(\hat{v}_j^{(n)}) + K_{j+1} y_{j+1} + \zeta_{j+1}^{(n)}$ with $\zeta_{j+1}^{(n)} \sim N(0, C)$.
    \item Set $\pi_{j+1}^N(v_{j+1}) = \frac{1}{N} \sum_{n=1}^N \delta_{0}\left(v_{j+1} - v_{j+1}^{(n)}\right)$.
\end{enumerate}
\State \textbf{Output:} Particle approximations $\pi_j^N \approx \pi_j$ for $j = 1, \dots, J$.
\end{algorithmic}
\end{algorithm}

\subsubsection{Stochastic Data Assimilation: Ensemble Transform Kalman Filter}
The localized ETKF is a state of the art stochastic DA algorithm \cite{soaLETKF,LETKF}.  As noted in the introduction, for low-dimensional systems the ``local" part is unnecessary and can in fact worsen results.  As we only have 24 variables at most in our highest dimensional system (the multi-scale Lorenz '96 system), we only implement ETKF.  We explicitly point out that 
\begin{itemize}
\item $\Psi$ remains the nonlinear dynamics, $H$ remains the observation matrix, and $y_{j}$ remains the observation at each time step, as in Algorithm~\ref{algorithm:EnKF},
\item $y_{j} = y^\dagger_j + \epsilon_j$ are the observations of the true system at each step (with $\epsilon_j$ white noise representing  the error in the data),
\item $R$ is the observation error covariance matrix,
\item $\rho$ is an empirical inflation factor larger than $1$ used to magnify the underestimated forecast errors \cite{matlab,amsArticle}. 
\end{itemize}

\begin{algorithm}[H]
\caption{Ensemble Transform Kalman Filter (ETKF)}
\label{algorithm:ETKF}
\begin{algorithmic}
\State \textbf{Input:} Ensemble size $N$, initial ensemble $\{v_{0}^{(n)}\}^{N}_{n=1}$, observations $\{y_j\}_{j=1}^J$.

\For{$j = 0, 1, ..., J-1$}
\begin{equation}
    \xi_{j}^{(n)} \sim N(0,\Sigma),\,\, i.i.d.,\,\,n = 1,...,N,
\end{equation}
\begin{equation}
    \hat{v}_{j+1}^{(n)} = \Psi(v_{j}^{(n)}) + \xi_{j}^{(n)},\,\,n = 1,...,N,
\end{equation}
\begin{equation}
    \hat{m}_{j+1} = \frac{1}{N} \sum_{n=1}^{N} \hat{v}_{j+1}^{(n)},
\end{equation}
\begin{equation}
    {d}_{j+1}^{(n)} =  \hat{v}_{j+1}^{(n)} - \hat m_{j+1}.
\end{equation}
\State Let $D$ be the matrix with ${d}_{j+1}^{(n)}$ as its $N$ columns.
\begin{equation}
\bar{y_{b}} = H\hat m_{j+1},
\end{equation}
\begin{equation}
Y_b = HD,
\end{equation}
\begin{equation}
C = Y_{b}^{T}R^{-1},
\end{equation}
\begin{equation}
\bar{P_{a}} = \left(\frac{N-1}{\rho}I + CY_{b}\right)^{-1},
\end{equation}
\begin{equation}
W_{a} = \sqrt{N-1} \bar P_{a}^{\frac{1}{2}}.
\end{equation}
\State We represent each column of $W_{a}$ (a weight matrix) with $W_{a}^{(n)}$.
\begin{equation}
\bar w_{a} = \bar P_a C(y_{j+1} - \bar{y_b}),
\end{equation}
\begin{equation}
w_{a}^{(n)} = \bar w_{a} + W_{a}^{(n)},
\end{equation}
\begin{equation}
v_{j+1}^{(n)} = \hat m_{j+1} + Dw_{a}^{(n)}.
\end{equation}

\EndFor
\State \textbf{Output:} Ensembles $\{v_{j}^{(n)}\}_{n=1}^{N}$,\,\,$j = 0,1,...,J$.
\end{algorithmic}
\label{alg:ETKF}
\end{algorithm}



\subsection{Deterministic Parameter Recovery Methods} 
The deterministic PR algorithms are inspired by the AOT DA algorithm. The nudged Lorenz '63 system can be represented in the following way:
\begin{equation}
\begin{system}
\dot{\tilde{x_1}} = \tilde{\sigma}(\tilde{x_2} - \tilde{x_1}) - \mu_{1}(\tilde{x_1} - x_1) \\ 
\dot{\tilde{x_2}} = \tilde{x_1}(\rho - \tilde{x_3}) - \tilde{x_2} - \mu_{2}(\tilde{x_2} - x_2) \\  
\dot{\tilde{x_3}} = \tilde{x_1}\tilde{x_2} - \beta\tilde{x_3} - \mu_{3}(\tilde{x_3} - x_3)
\end{system}
\end{equation}
where $\mu_{1}$ is a large constant for PR on $\sigma$. Similarly, the nudged equations for fast variables in Lorenz '96 are as follows: 
\begin{equation}
\label{blah}
\begin{system}
\dot{\tilde{u_{1}}} = \tilde{u_{4}}(\tilde{u_{2}} - \tilde{u_{3}}) + \sum_{j=1}^{J} \gamma_{1}\tilde{u_{1}}\tilde{\mathbcal{u}}_{1,j} - \tilde{\overline{d_{1}}}\tilde{u_{1}} + F - \mu_{1}(\tilde{u_{1}} - u_{1}) 
\\
\dot{\tilde{u_{2}}} = \tilde{u_{1}}(\tilde{u_{3}} - \tilde{u_{4}}) + \sum_{j=1}^{J} \gamma_{2}\tilde{u_{2}}\tilde{\mathbcal{u}}_{2,j} - \overline{d_{2}}\tilde{u_{2}} + F - \mu_{2}(\tilde{u_{2}} - u_{2}) 
\\
\dot{\tilde{u_{3}}} = \tilde{u_{2}}(\tilde{u_{4}} - \tilde{u_{1}}) + \sum_{j=1}^{J} \gamma_{3}\tilde{u_{3}}\tilde{\mathbcal{u}}_{3,j} - \overline{d_{3}}\tilde{u_{3}} + F - \mu_{3}(\tilde{u_{3}} - u_{3}) 
\\
\dot{\tilde{u_{4}}} = \tilde{u_{3}}(\tilde{u_{1}} - \tilde{u_{2}}) + \sum_{j=1}^{J} \gamma_{4}\tilde{u_{4}}\tilde{\mathbcal{u}}_{4,j} - \overline{d_{4}}\tilde{u_{4}} + F - \mu_{4}(\tilde{u_{4}} - u_{4})

\end{system}
\end{equation}
We do not add nudged terms to the slow variables $\tilde{\mathbcal{u}}_{i,j}$ in the nudged Lorenz '96 system, and consequently we exclude the evolution equations for the slow variables in \eqref{blah} for brevity. We choose to recover the value of $\overline{d_{1}}$, but PR results for $\overline{d_{i}}$ in general should behave the same.
These deterministic PR algorithms were developed in the setting of continuous time DA, but have been shown in practice to work when data is assimilated discretely in time.  As it is more common in practice to receive data discretely in time, we investigate PR in this context.

\subsubsection{The Carlson-Hudson-Larios (CHL) Method}
\label{sec:CHL}
The CHL algorithm was proposed in \cite{Carlson_Hudson_Larios_2020} and convergence was proven for the Lorenz '63 system in \cite{Carlson_Hudson_Larios_Martinez_Ng_Whitehead_2021} and for the 2D incompressible Navier-Stokes equations in \cite{Martinez_2022_pr}. The general form of the algorithm was designed for a particular equation form with linear dependence on the parameters we are interested in recovering; we reproduce it here for an ODE system with partial observations in Algorithm \ref{algorithm:CHL}.
For the algorithm, we use the following general setup, with $L$ linear, $N$ nonlinear, and $w: = v-u$:
\begin{align}
\dot{u} + \vec{\theta^\dagger}\cdot  Lu 
+N(u) = 0, \dot{v} + \vec{\theta} \cdot  Lv 
+N(v) = -\mu P_m(v-u) \\
\implies \dot{w} + (\vec{\theta^
\dagger}-\vec{\theta})\cdot Lv +\vec{\theta^\dagger}\cdot Lw + (N(v) - N(w)) = -\mu P_m w
\\
    \frac12 \frac{d}{dt}|w|^2 + \langle(\vec{\theta^
\dagger}-\vec{\theta}) Lv, w\rangle +\theta^\dagger Lw\cdot w + \langle N(v) - N(u), w\rangle = -\mu P_m w^2.
\end{align}
Here, $\theta^\dagger$ is the true parameter, and $\theta$ is our guess at $\theta^\dagger$.

\begin{algorithm}[H]
\caption{CHL Method}
\label{algorithm:CHL}
\begin{algorithmic}[1]
\State \textbf{Input:} Initialization $\vec{\theta_0}$, $P_m(u)$, equation $\dot{v} + \vec{\theta_0} \cdot  Lv 
+N(v) = -\mu P_m(v-u) $
\For{$\ell = 0, 1, \ldots, L-1$} 
\State Solve $\dot{v} + \vec{\theta_\ell} \cdot  Lv 
+N(v) = -\mu P_m(v-u)$
\If{$\frac{d}{dt} (P_m(w))^2 \approx 0$ at $t_\ell >>0$ \; \textbf{and} \; $i \in \{1, ..., m\}$}
\State $\theta_{\ell+1}^i = \theta_\ell^i - \mu \frac{w_i^2}{Lv_i\cdot w_i}$
\EndIf
\EndFor
\State \textbf{Output:} Parameter $\theta_L$.
\end{algorithmic}
\end{algorithm}

As in Section 2.1 of \cite{Carlson_Hudson_Larios_Martinez_Ng_Whitehead_2021}, the parameter update equations for Lorenz '63 are 
\begin{equation}
\begin{system}
\sigma_{n+1} = \sigma_{n} - \mu_{1}\cfrac{\tilde{x_1} - x_1}{\tilde{x_2}-\tilde{x_1}}, \\ \rho_{n+1} = \rho_{n} - \mu_{2}\cfrac{\tilde{x_2} - x_2}{\tilde{x_1}}, \\ \beta_{n+1} = \beta_{n} + \mu_{3}\cfrac{\tilde{x_3} - x_3}{\tilde{x_3}},
\label{CHL_formula}
\end{system}
\end{equation}
and we update the parameter value at every time step.  The derivation of these PR algorithms follow from a formal magnitude analysis on the evolution equations for the error in kinetic energy, and taking this same approach we derive a parameter update formula for $\tilde{\overline{d_{1}}}$ in the Lorenz '96 system (see Appendix \ref{appendix:CHL} for derivation):
\begin{equation}
\tilde{\overline{d_{1}}}^{(n+1)} = \tilde{\overline{d_{1}}}^{(n)} - \frac{\sum_{j=1}^{J}\gamma_{1}\tilde{u_{1}}(\tilde{v_{1,j}} - v_{1,j}) - \mu_{1}(\tilde{u_{1}} - u_{1})}{\tilde{u_{1}}}
\label{CHL_96_formula}
\end{equation}
and update the parameter at every time step. Generally one needs the solution to relax long enough so this approximated update holds asymptotically, but is known to work in the Lorenz '63 case updating every time step \cite{Carlson_Hudson_Larios_Martinez_Ng_Whitehead_2021} and we implement the same for Lorenz'96.  In Algorithm \ref{algorithm:CHL}, we present the general implementation for this algorithm in the case where the observed data are state variables (e.g., in the ODE case, a simple selection of states, or in the PDE case, Galerkin modes).  For two different detailed derivations of \eqref{CHL_96_formula}, see \cite{Martinez_Murri_Whitehead_2025, Newey_Whitehead_Carlson_2025}.

\subsubsection{The Pachev-Whitehead-McQuarrie (PWM) Method}
\label{sec:PWM}
The PWM method is also based on the AOT DA algorithm, and was also developed in the context of equations which depend linearly on their parameters (though it may be possible to generalize the method to systems with nonlinear dependence on their parameters). For the detailed derivation and convergence statement, see Section 2.2 in \cite{Pachev_Whitehead_McQuarrie_2021concurrent}, in particular equation (2.5); we reproduce a general form in Algorithm \ref{algorithm:PWM}.  For the algorithm, we use the following general setup, with $F,G_k$ sufficiently regular functions, $k \in \{1,..., d\}$, and again $w:=v-u$:
\begin{align}
    \dot{u} +  F(u) 
+\vec{\theta^\dagger}\cdot \vec{G}(u) = 0, \; \dot{v} + F(v) + \vec{\theta} \cdot  \vec{G}v  = -\mu P_m(v-u) 
\\
\implies P_m(\dot{w}) = - P_m\dot{u} - P_m F(v) - \vec{\theta} \cdot  P_m\vec{G}(v)  -\mu P_mw
\end{align}
Again, $\theta^\dagger$ is the true parameter, and $\theta$ is our guess at $\theta^\dagger$.
\begin{algorithm}[H]
\caption{PWM Method}
\label{algorithm:PWM}
\begin{algorithmic}[1]
\State \textbf{Input:} Initialization $\vec{\theta_0}$, equation $\dot{v} + F(v) + \vec{\theta_0} \cdot  \vec{G}v  = -\mu P_m(v-u)$, 
\For{$\ell = 0, 1, \ldots, L-1$} 
\State Solve $\dot{v} + F(v) + \vec{\theta_\ell} \cdot  \vec{G}v  = -\mu P_m(v-u)$ for one time step
\State Choose $\vec{\theta_\ell}$ such that $\vec{\eta} := \langle - P_m\dot{u} - P_mF(v) - \vec{\theta_l} \cdot  P_m\vec{G}(v)), P_m w\rangle = 0$ by solving
$A\theta = b$ where $A_{i,k} = \langle e_i, G_k(v)\rangle$, $b_i = \langle e_i, P_m(-\dot{u} - F(v))\rangle$
\EndFor
\State \textbf{Output:} Parameter $\theta_L$.
\end{algorithmic}
\end{algorithm}

We derive the PWM algorithm for Lorenz '96 and the parameter update formula is as follows (see Appendix \ref{appendix:PWM} for derivation):
\begin{equation}
\tilde{\overline{d_{1}}}^{(n+1)} = \frac{u_{4}(u_{3} - u_{2}) - \sum_{j=1}^{J}\gamma_{1}u_{1}v_{1,j} + \overline{d_{1}}u_{1} + \tilde{u_{4}}(\tilde{u_{2}} - \tilde{u_{3}}) + \sum_{j=1}^{J}\gamma_{1}\tilde{u_{1}}\tilde{v_{1,j}}}{\tilde{u_{1}}}
\label{PWM_formula}
\end{equation}
where $\tilde{\overline{d_{1}}}$ is the variable for parameter estimation, and $\overline{d_{1}}$ is the true parameter value. We allow the existence of the true parameter value in our update formula because it is essentially a part of the information we gain from data. In our experiments, we are using the true parameter value to generate synthetic data. To be more explicit about this, we can also write the update formula as 
\begin{equation}
\tilde{\overline{d_{1}}}^{(n+1)} = \frac{\tilde{u_{4}}(\tilde{u_{2}} - \tilde{u_{3}}) + \sum_{j=1}^{J}\gamma_{1}\tilde{u_{1}}\tilde{v_{1,j}} + F - \dot{u_{1}}}{\tilde{u_{1}}}
\label{eqtn:PWM96approx}
\end{equation}
where $\dot{u_{1}}$ can be approximated from real data observations if we have them; otherwise, we use \eqref{PWM_formula}. 

The deterministic PR methods can tolerate relatively large perturbations in initial conditions when we are using the adaptive MsDTM for generating synthetic data and the predictor-corrector scheme for evolving the nudged system. 

\subsubsection{The Asymptotic Variational (AV) Method}
\label{sec:AV}
We also tested a third deterministic PR method proposed in \cite{Newey_Whitehead_Carlson_2025}, which we call the Asymptotic Variational (AV) method; there is no direct dependence on the system being linearly dependent on the parameters in this method. 
Here, we solve an optimization problem with the following objective function (cost function) \cite{levenberg_duke_2024}: 
\begin{equation}
\chi^{2}(\vec{a}) =  \sum_{i=1}^{J}\left[ \dfrac{y(t_{i}) - \hat{y}(t_{i}; \vec{a})}{\sigma_{y_{i}}}\right]^{2}
\label{cost_function}
\end{equation}
where 
\begin{itemize}
    \item $\vec{a}$ is the set of parameters for a model,
    \item $y(t_{i})$ is the observation value at time $t_{i}$,
    \item $\hat{y}(t_{i}; \vec{a})$ is solution to \eqref{Lorenz96} at time $t_i$ given the parameters $\vec{a}$, 
    \item $\sigma_{y_{i}}$ is the noise standard deviation for observations at time $t_{i}$, and
    \item $J$ is the total number of time steps.
\end{itemize}

Note that $y$ is the solution to \eqref{Lorenz96} at the true parameter value.  While the CHL algorithm can be derived as the Gradient Descent (GD) algorithm applied to the objective function in \eqref{cost_function} (AV-GD), we implement the AV-Levenberg-Marquardt (AV-LM) parameter update formula. It can be derived by applying the Levenberg-Marquardt algorithm to minimize the objective \eqref{cost_function} \cite{Newey_Whitehead_Carlson_2025} (see Appendix \ref{appendix:AV-LM} for derivation):
\begin{equation}
\tilde{\overline{d_{1}}}^{(n+1)} = \tilde{\overline{d_{1}}}^{(n)} - ( S_{lm}^{2}+ \lambda)^{-1}S_{lm}(\tilde{u_{1}} - u_{1})
\label{AV_formula}
\end{equation}
where
\begin{equation}
S_{lm} = \dfrac{1}{\mu_{1}}\sum_{j=1}^{5}\tilde{v_{1,j}}\tilde{u_{1}}.
\label{eqtn:AVFormulaAffiliate}
\end{equation}
We show in Section \ref{sec:results} that using the Levenberg-Marquardt algorithm improves PR performance in terms of both accuracy and stability when we add noise to observations.  The general form of AV-LM is given in Algorithm \ref{algorithm:AV-LM}; note that we have modified our original objective in the case of Lorenz '96 to incorporate variance in the data, but this does not change the general approach.  For the algorithm we use the following general setup, again with $F$ sufficiently regular:
\begin{align}
\begin{array}{cc}
\text{ }\text{minimize } & \hspace{-1in}|P_m(v(\theta) - u)|^2   \\
\quad \text{subject to } 
&  \hspace{-0.8in} \dot{u} + F(u;\vec{\theta}^\dagger) = 0, \\
&  \dot{v} + F(v;\vec{\theta})= -\mu P_m(v-u)
\end{array}
\end{align}
Here, again, $\theta^\dagger$ is the true parameter, and $\theta$ is a guess at $\theta^\dagger$.

\begin{algorithm}[H]
\caption{AV-LM Method}
\label{algorithm:AV-LM}
\begin{algorithmic}[1]
\State \textbf{Input:} Initialization $\vec{\theta_0}$, $P_m(u)$
\For{$\ell = 0, 1, \ldots, L-1$} 
\State Solve $\dot{v} + F(v;\vec{\theta_\ell} )= -\mu P_m(v-u) $
\If{$\frac{d}{dt} (P_m(v-u)) ^2 \approx 0$ at $t_\ell >>0$}
\State $\theta_{\ell+1}^i = \theta_\ell^i - \sum\limits_j (\frac{1}{\mu}\langle P_m v_{\theta^i_\ell}, P_m v_{\theta^j_\ell}\rangle + \lambda \delta_{ij})^{-1}\frac{1}{\mu}\langle P_m(v-u), P_m v_{\theta^j_\ell}\rangle$
\EndIf
\EndFor
\State \textbf{Output:} Parameter $\theta_L$.
\end{algorithmic}
\end{algorithm}

\subsection{Stochastic Parameter Recovery Methods}
\subsubsection{Expectation-Maximization Algorithm}
\label{sec:EM}
We also implemented the EM algorithm, referring to \cite{Ghahramani_Roweis_1998, textbook_sanz_alonso}. Here we use the algorithm from \cite{textbook_sanz_alonso}:

\begin{algorithm}[H]
\caption{Expectation Maximization (EM)}
\label{algorithm:EM}
\begin{algorithmic}[1]
\State \textbf{Input:} Initialization $\theta_0$, observations $Y$.
\For{$\ell = 0, 1, \ldots, L-1$} 
    \State \textbf{E-Step:} Compute
    \[
    \mathbb{E}_{V \sim P(V | Y, \theta_\ell)}
    \left[
    \log P(V, Y | \theta)
    \right]
    =
    \int
    \log P(V, Y | \theta) P(V | Y, \theta_\ell) \, dV.
    \]
    \State \textbf{M-Step:} Compute
    \[
    \theta_{\ell+1} = \arg \max_{\theta}
    \mathbb{E}_{V \sim P(V | Y, \theta_\ell)}
    \left[
    \log P(V, Y | \theta)
    \right].
    \]
\EndFor
\State \textbf{Output:} Parameter $\theta_L$.
\end{algorithmic}
\end{algorithm}
In this algorithm, $\theta$ represents the parameter value, $V$ is a model estimate, and $Y$ are the observations. Following \cite{textbook_sanz_alonso}, we use the formulas
\begin{equation}
\theta_{l+1} = \underset{\theta}{{\operatorname{argmax}}}  \; \frac{1}{N}\sum_{n=1}^{N}\text{log} \;\mathbb{P}(V^{n}, Y|\theta)
\label{EM_1}
\end{equation}
and 
\begin{equation}
\text{log} \; \mathbb{P}(V^{n}, Y|\theta) = -\frac{1}{2\gamma^{2}}\sum^{J-1}_{j=0}|y_{j+1} - v_{j+1}^{(n)}|^{2} - \frac{1}{2c_{0}^{2}}|v_{0}^{(n)} - m_{0}|^{2} - \frac{1}{2\sigma^{2}}\sum^{J-1}_{j=0}|v_{j+1}^{(n)} - \Psi_\theta (v_{j}^{(n)})|^{2} + c
\label{EM_2}
\end{equation}
where every variable in equation \eqref{EM_2} is given in the Ensemble Kalman Filter algorithm (Algorithm~\ref{algorithm:EnKF}), which allows us to combine EM with EnKF to perform PR. As can be seen in Algorithm~\ref{algorithm:EM}, the ideal case is to sum over a continuum of particles in the E-Step, but that is impractical when numerically implementing the algorithm. We instead approximate the integral by summing the log densities of every discrete particle. This summing process is expensive, because we are solving the forward model $N$ times at every numerical step while performing Expectation-Maximization.

\subsubsection{Ensemble Kalman Inversion (Transport to Finite Time)}
\label{sec:EnKI}
The next stochastic PR algorithm we investigated is the Ensemble Kalman Inversion (Transport to Finite Time) \cite{edo_reich_stuart_2024}. This method was used to recover the forcing parameter of the Lorenz '96 system in \cite{edo_reich_stuart_2024}, but we utilize the algorithm to help us recover a different physical constant, namely $\overline{d_{1}}$.


\begin{algorithm}[H]
\caption{Ensemble Kalman Inversion (Transport to Finite Time)}
\begin{algorithmic}[1]
\Statex \textbf{Input:} Data $y$, $J$, and $\Delta t$ such that $J\Delta t = 1$, ensemble size $N$, initial ensemble $\{u_{0}^{(n)}\}_{n=1}^{N}$.
\noindent \Statex \textbf{For} {$j = 0$ to $J-1$} \textbf{do}
\Statex\hspace{\algorithmicindent} \textbf{Prediction:} for $n = 1, ..., N$ do 
\Statex\hspace{7cm}$\eta_{j}^{(n)} \sim N(0, \frac{\Gamma}{\Delta t})$
\Statex\hspace{6.9cm} $\widehat{u}_{j+1}^{(n)} = u_{j}^{(n)}$
\Statex\hspace{6.9cm} $\widehat{y}_{j+1}^{(n)} = G(\widehat{u}_{j+1}^{(n)}) + \eta_{j}^{(n)}$.
\Statex\hspace{\algorithmicindent}\hspace{0.2cm}Compute
\Statex\hspace{5cm}$\widehat{m}_{j+1} = \frac{1}{N}\sum_{n=1}^{N}\widehat{u}_{j+1}^{(n)}$, $\widehat{o}_{j+1} = \frac{1}{N}\sum_{n=1}^{N} G(\widehat{u}_{j+1}^{(n)})$,
\Statex\hspace{5cm}$\widehat{C}_{j+1}^{uG} = \frac{1}{N}\sum_{n=1}^{N}(\widehat{u}_{j+1}^{(n)} - \widehat{m}_{j+1}) \otimes (G(\widehat{u}_{j+1}^{(n)}) - \widehat{o}_{j+1})$,
\Statex\hspace{5cm}$\widehat{C}_{j+1}^{GG} = \frac{1}{J} \sum_{j=1}^{J}(G(\widehat{u}_{j+1}^{(n)}) - \widehat{o}_{j+1}) \otimes (G(\widehat{u}_{j+1}^{(n)}) - \widehat{o}_{j+1})$.
\Statex\hspace{\algorithmicindent}\hspace{0.2cm}\textbf{Analysis:} for $n = 1, ..., N$ do
\Statex\hspace{\algorithmicindent}\hspace{5cm}$u_{j+1}^{(n)} = \widehat{u}_{j+1}^{(n)} + \Delta t \widehat{C}_{j+1}^{uG}(\Delta t\widehat{C}_{j+1}^{GG} + \Gamma)^{-1}(y - \widehat{y}_{j+1}^{(n)})$.
\Statex\textbf{end for}
\Statex \textbf{Output:} Ensembles $\{u_{j}^{(n)}\}_{n=1}^{N}$ for $j = 0, ..., J$.
\end{algorithmic}
\label{algorithm:EnKI}
\end{algorithm}

Here, the notation is as follows:
\begin{itemize}
    \item $y$ is for observations of states,
    \item $u_{0}^{(n)}$ are initial parameter estimate ensemble particles, 
    \item $\Gamma$ is noise covariance matrix which is scaled by $\Delta t$.
\end{itemize}   
EnKI does not have a corresponding state estimation process, and the forward evolution $G$ is the observation operator applied to the solution operator of the nudged system over a short time. In particular, $G(u) = H \Psi(T,\,\cdot\,; u)$, where $u$ is the parameter, $\Psi$ is the solution operator, $T$ represents the time when we stop the simulation, and $\cdot$ stands as a placeholder for an initial condition pulled from the initial ensemble.  The choice of using the nudged system instead of the true model is informed by experiments and the idea that we could use the predictor-corrector scheme for a more refined numerical simulation. 
\subsubsection{Gaussianized Optimal Particle Filter Extension}
\label{sec:PF_extension}

We can use the PF to not just update state estimations, but also parameter estimations at every step. For a review of the PF algorithm we are using, see Section \ref{sec:original_PF}. 
The PF Extension is done by extending the matrix that contains state estimations by one dimension and applying Algorithm \ref{gopfAlg}.  Specifically, instead of considering $v_{n+1}^{(j)}$ at each step, we instead consider a new ``extended" state $[v_{n+1}^{(j)}, \theta]$ and viewing that extra dimension as part of a larger ``state."  We initialize the parameter estimate with no variance (though this is not the only possible choice).  Since there is no evolution equation for $\theta$, the evolution is exclusively carried out by the effect of the Kalman gain matrix in Step 4.
\section{Results}
\label{sec:results}
Before we proceed into details of results, let us first clarify some use of terms.
\begin{itemize}
\item In the rest of the paper, the word ``accuracy'' refers to the mean of the parameter estimation error and ``stability'' refers to the variance of the parameter estimation error. Intuitively, high accuracy means low PR error mean, and high stability means low PR error variance.
\item ``NC'' in all tables mean non-convergent.
\end{itemize}
Furthermore, we detail the specifications of the computer we used. The CPU model was Apple M1 with 16GB of memory.

In particular, we implemented the experiments using Python 3.9.6 on Pycharm. The packages we needed for deterministic methods include \textcolor{blue}{\texttt{numpy}}, \textcolor{blue}{\texttt{matplotlib.pyplot}}, and \textcolor{blue}{\texttt{math}}. In particular, \textcolor{blue}{\texttt{np.random.normal}} and \textcolor{blue}{\texttt{np.random.multivariate\_normal}} were used to add noise. For stochastic methods, we also needed \textcolor{blue}{\texttt{minimize}} in \textcolor{blue}{\texttt{scipy.optimize}} for solving minimization problems in these algorithms. Additionally, \textcolor{blue}{\texttt{time}} and \textcolor{blue}{\texttt{Parallel}} and \textcolor{blue}{\texttt{delayed}} from \textcolor{blue}{\texttt{joblib}} were used for monitoring the speed of the code scripts.

We provide a link to our Github repository for this project \href{https://github.com/AshleyyyMWang/Parameter_Recovery_Algorithms_Applied_to_Chaotic_Systems}{here}.

\subsection{Lorenz '63}
\label{sec:L63_results}
\subsubsection{Effectiveness of CHL Parameter Recovery Algorithm as Corresponds to Choice of Numerical Method}
\label{sec:numerical_comparison}
\begin{figure}[h]
    \centering
    \begin{subfigure}[b]{0.4\textwidth}
        \centering
        \includegraphics[width=1\textwidth]{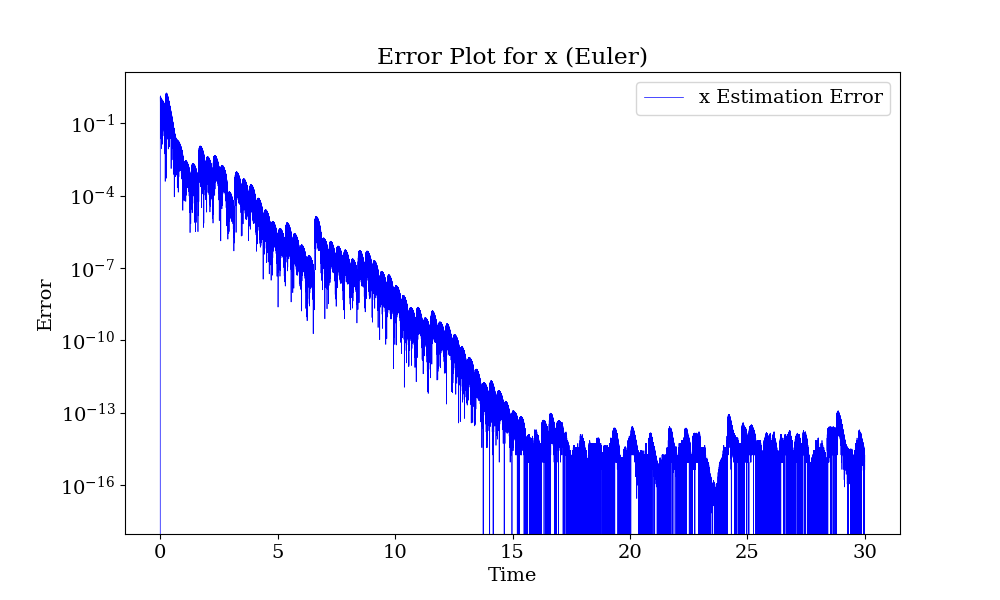}
        \caption{Euler Method}
        \label{fig:first}
    \end{subfigure}
    \hspace{0.1in}
    \begin{subfigure}[b]{0.4\textwidth}
        \centering
        \includegraphics[width=1\textwidth]{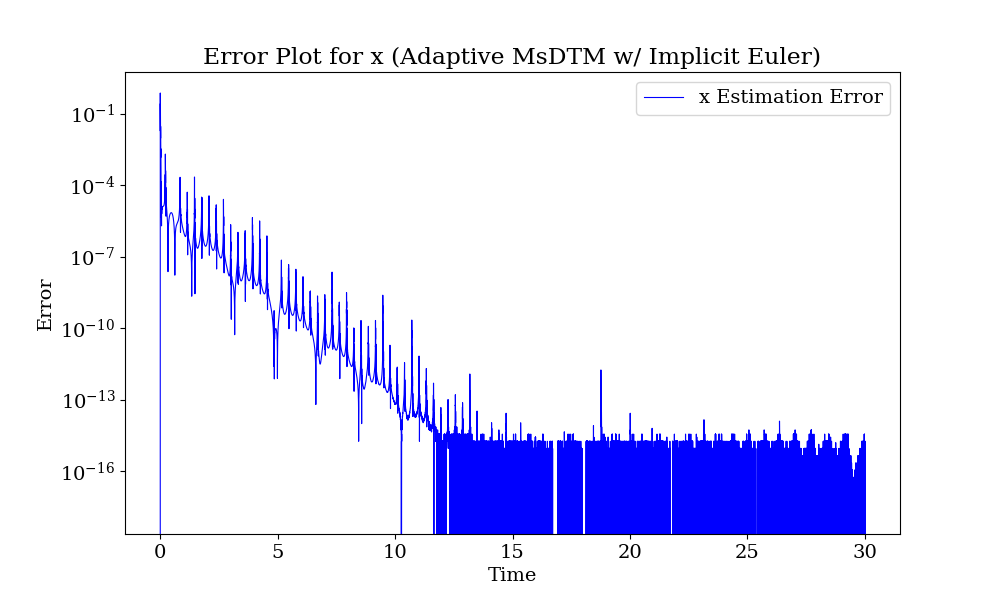}
        \caption{Adaptive MsDTM + Implicit Euler}
        \label{fig:second}
    \end{subfigure}
    \caption{Comparison for estimating $x_1$}
    \label{fig:Lorenz63x_small}
\end{figure}

\begin{figure}[H]
    \centering
    \begin{subfigure}[b]{0.4\textwidth}
        \centering
        \includegraphics[width=1\textwidth]{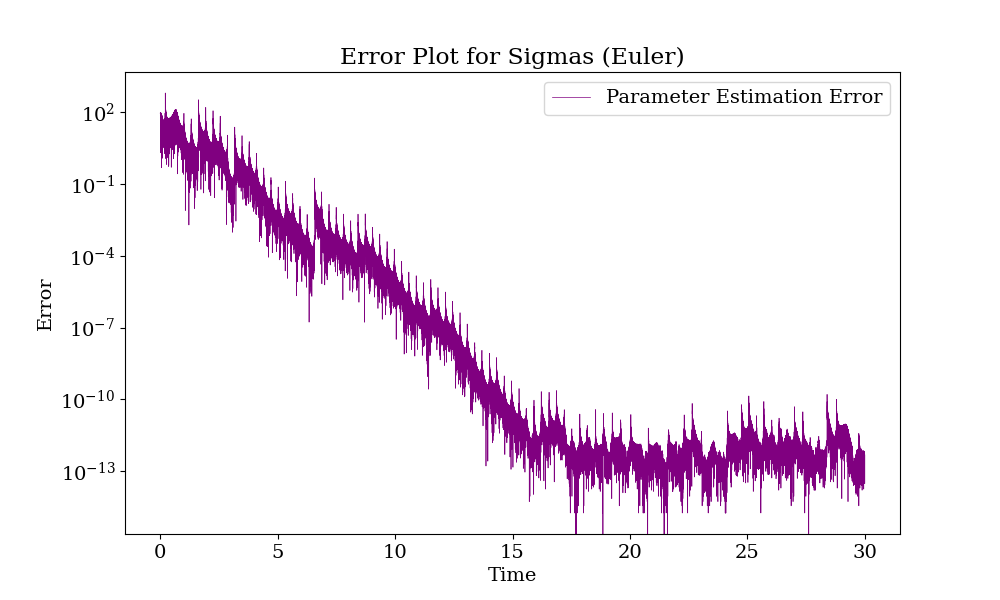}
        \caption{Euler Method}
        \label{fig:first}
    \end{subfigure}
    \hspace{0.1in}
    \begin{subfigure}[b]{0.4\textwidth}
        \centering
        \includegraphics[width=1\textwidth]{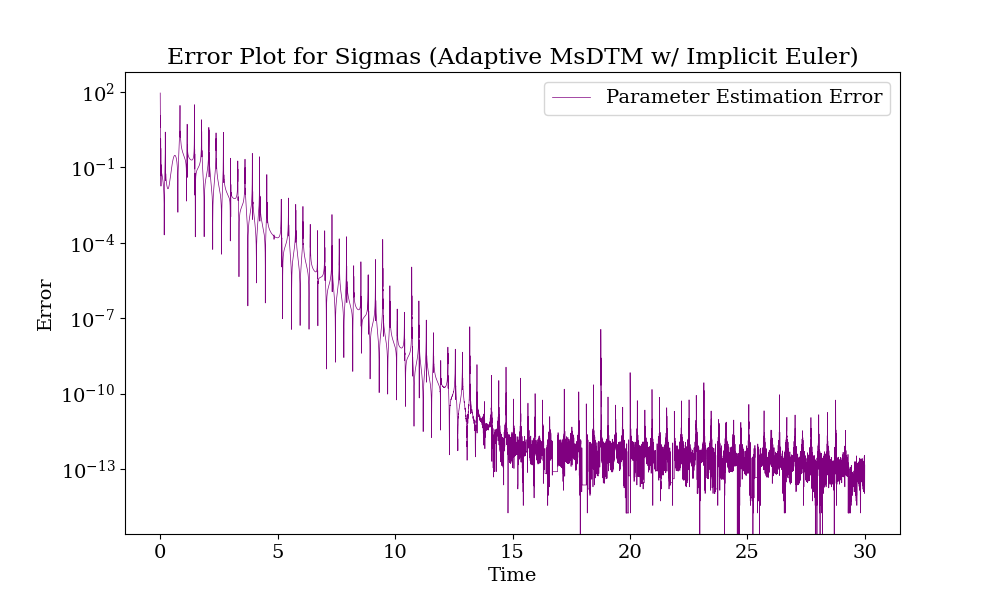}
        \caption{Adaptive MsDTM + Implicit Euler}
        \label{fig:second}
    \end{subfigure}
    \caption{Comparison for parameter estimation}
    \label{fig:Lorenz63param_small}
\end{figure}

We compare state and parameter estimation errors during deterministic PR on Lorenz '63 applying the forward Euler method and adaptive MsDTM described in Section \ref{sec:numerical_methods} (see Figures \ref{fig:Lorenz63x_small} - \ref{fig:Lorenz63param_small}), setting $\epsilon = 10^{-4}$ and initial $\sigma = 100$, and using noise-free observations. The assumption of a negligible time derivative of the error is necessary to ensure CHL convergence, so we set relaxation parameter values $\mu_{1} = 500, \mu_{2} = \mu_{3} = 0$. Section \ref{sec:wide_PR_window} discusses how a wider PR window allows lowering $\mu_{1}$, as expected (see also numerical results in \cite{Carlson_Hudson_Larios_Martinez_Ng_Whitehead_2021}). We discovered that adaptive MsDTM enables CHL to better withstand non-trivial perturbations to initial conditions, leading to more consistent convergence of parameter and state estimation errors to zero (on average), which is predicted by the theory in the noiseless setting \cite{Carlson_Hudson_Larios_Martinez_Ng_Whitehead_2021}. 
To ensure convergence with the Euler method, initial conditions are identical in all systems: (0,10,0).
\begin{figure}[H]
    \centering
    \begin{subfigure}[b]{0.4\textwidth}
        \centering
        \includegraphics[width=1\textwidth]{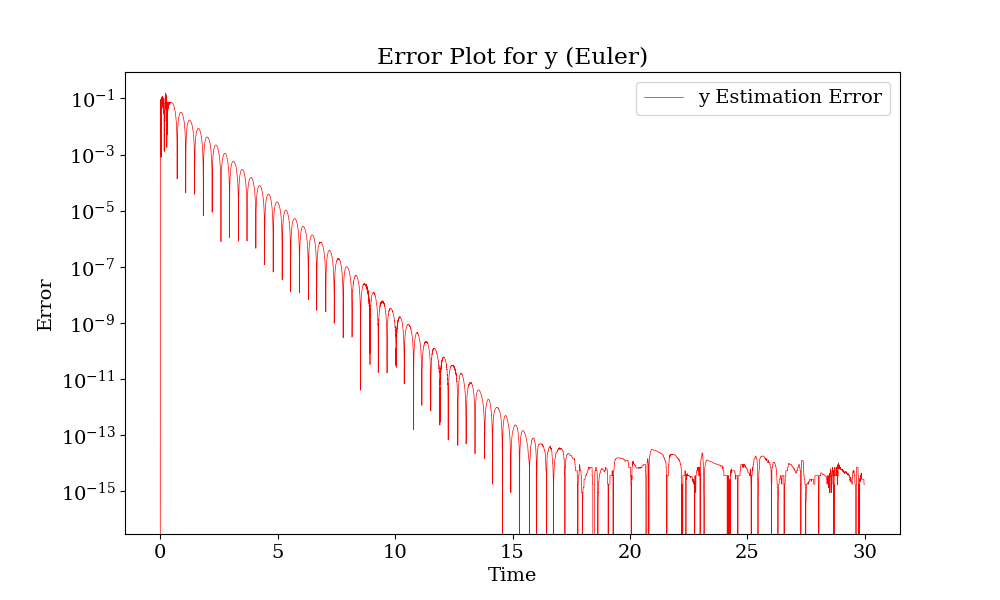}
        \caption{Euler Method}
        \label{fig:first}
    \end{subfigure}
    \hspace{0.1in}
    \begin{subfigure}[b]{0.4\textwidth}
        \centering
        \includegraphics[width=1\textwidth]{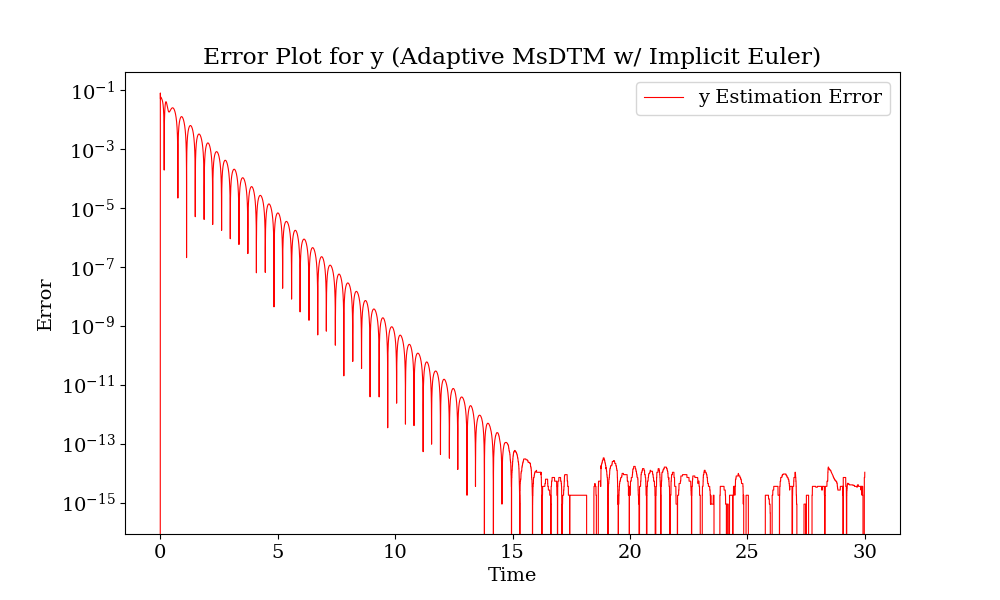}
        \caption{Adaptive MsDTM + Implicit Euler}
        \label{fig:second}
    \end{subfigure}
    \caption{Comparison for estimating $x_2$}
    \label{fig:comparison}
\end{figure}
\begin{figure}[H]
    \centering
    \begin{subfigure}[b]{0.4\textwidth}
        \centering
        \includegraphics[width=1\textwidth]{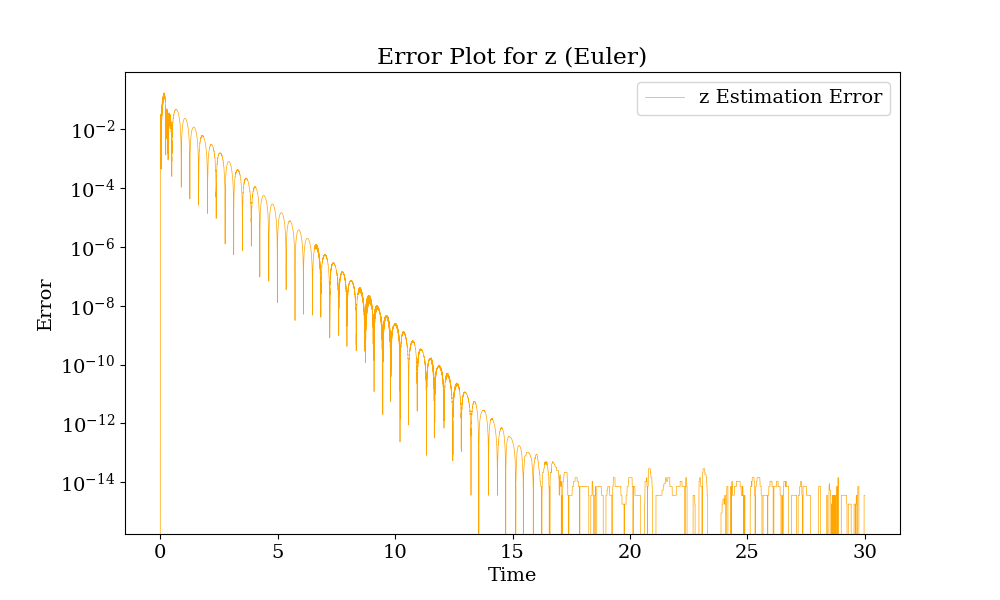}
        \caption{Euler Method}
        \label{fig:first}
    \end{subfigure}
    \hspace{0.1in}
    \begin{subfigure}[b]{0.4\textwidth}
        \centering
        \includegraphics[width=1\textwidth]{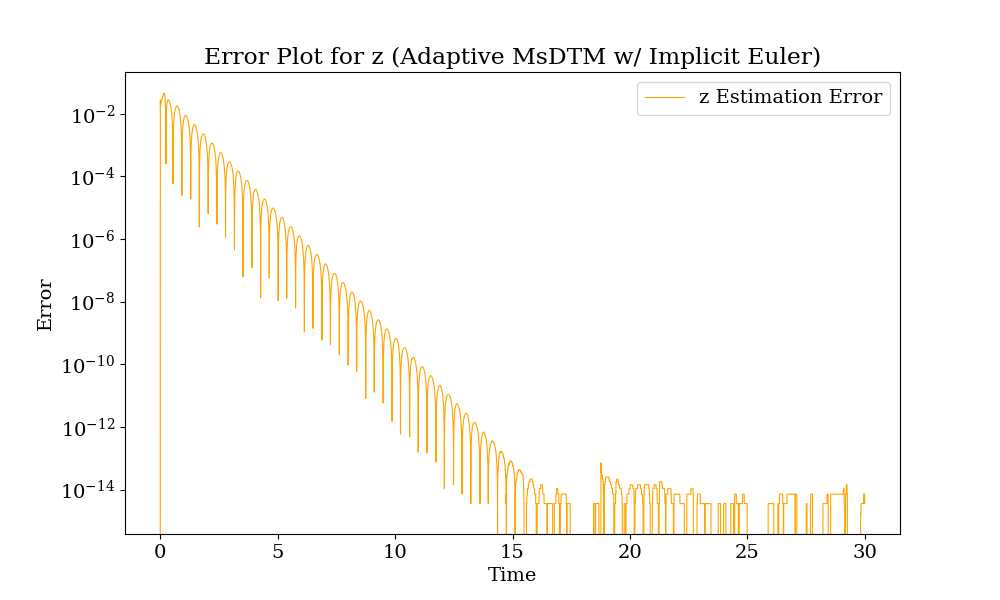}
        \caption{Adaptive MsDTM + Implicit Euler}
        \label{fig:second}
    \end{subfigure}
    \caption{Comparison for estimating $x_3$}
    \label{fig:comparison}
\end{figure}

\begin{figure}[H]
    \centering
    \begin{subfigure}[b]{0.4\textwidth}
        \centering
        \includegraphics[width=1\textwidth]{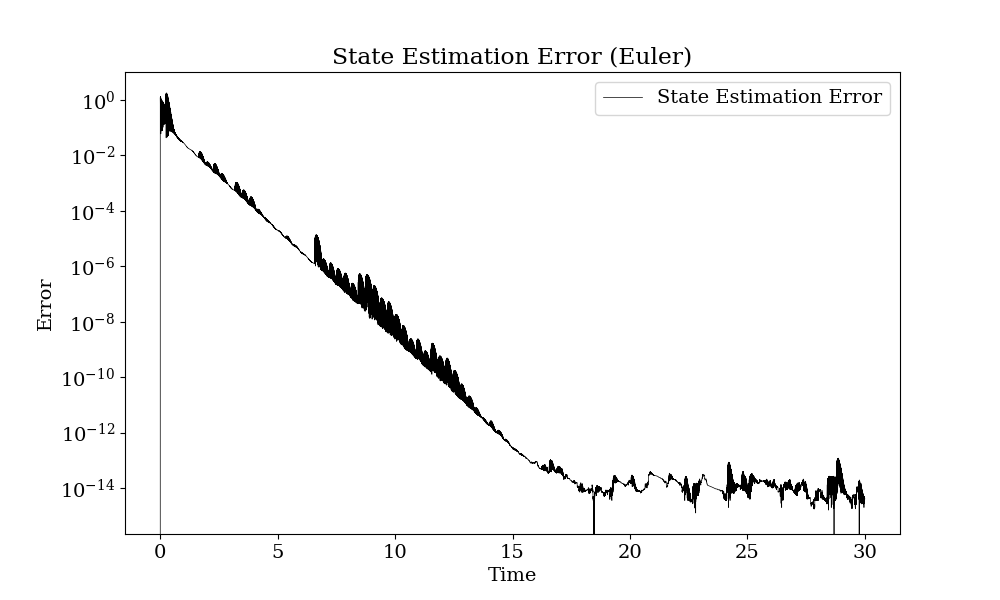}
        \caption{Euler Method}
        \label{fig:first}
    \end{subfigure}
    \hspace{0.1in}
    \begin{subfigure}[b]{0.4\textwidth}
        \centering
        \includegraphics[width=1\textwidth]{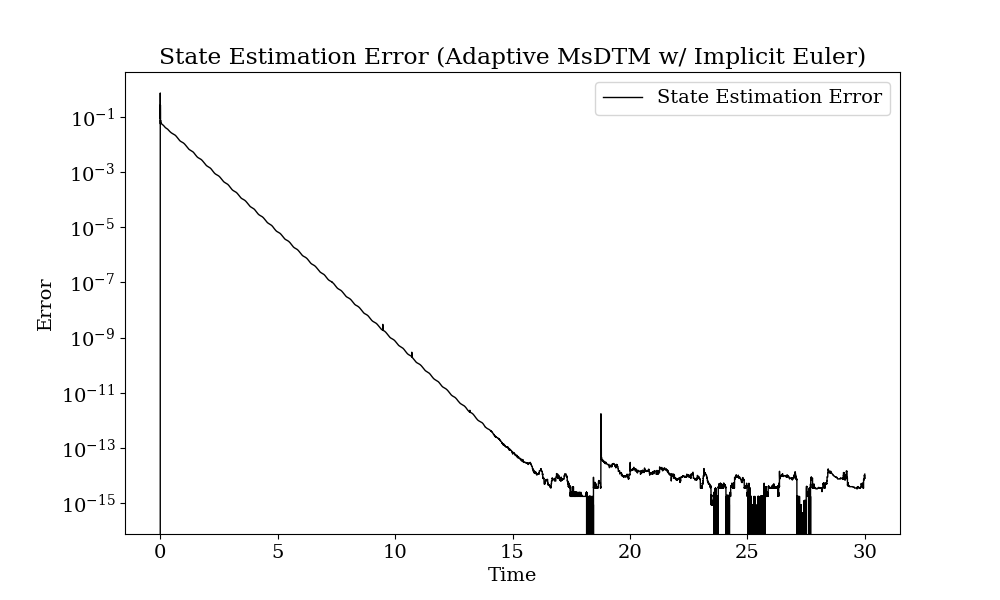}
        \caption{Adaptive MsDTM + Implicit Euler}
        \label{fig:second}
    \end{subfigure}
    \caption{Comparison for state estimation}
    \label{fig:comparison}
\end{figure}

\begin{figure}[H]
    \centering
    \begin{subfigure}[b]{0.32\textwidth}
        \centering
        \includegraphics[width=\textwidth]{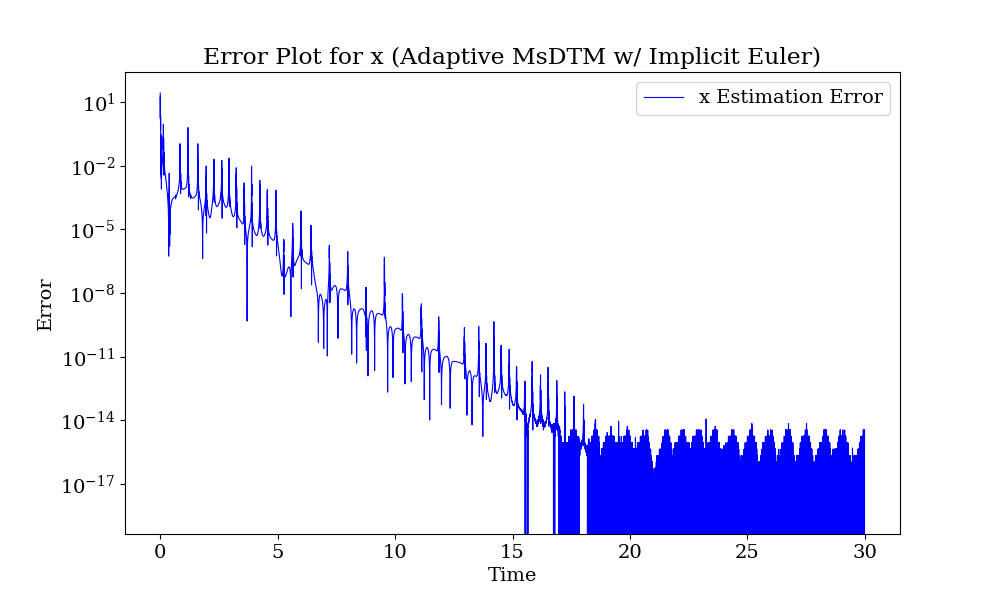}
        \caption{$x_1$ Estimation}
        \label{fig:first}
    \end{subfigure}
    \hfill
    \begin{subfigure}[b]{0.32\textwidth}
        \centering
        \includegraphics[width=\textwidth]{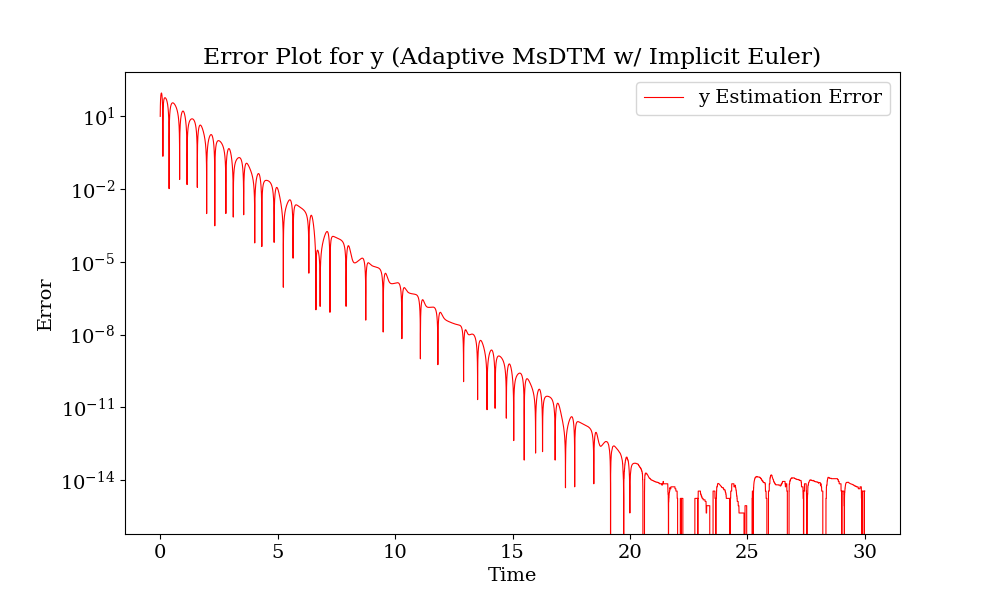}
        \caption{$x_2$ Estimation}
        \label{fig:second}
    \end{subfigure}
    \hfill
    \begin{subfigure}[b]{0.32\textwidth}
        \centering
        \includegraphics[width=\textwidth]{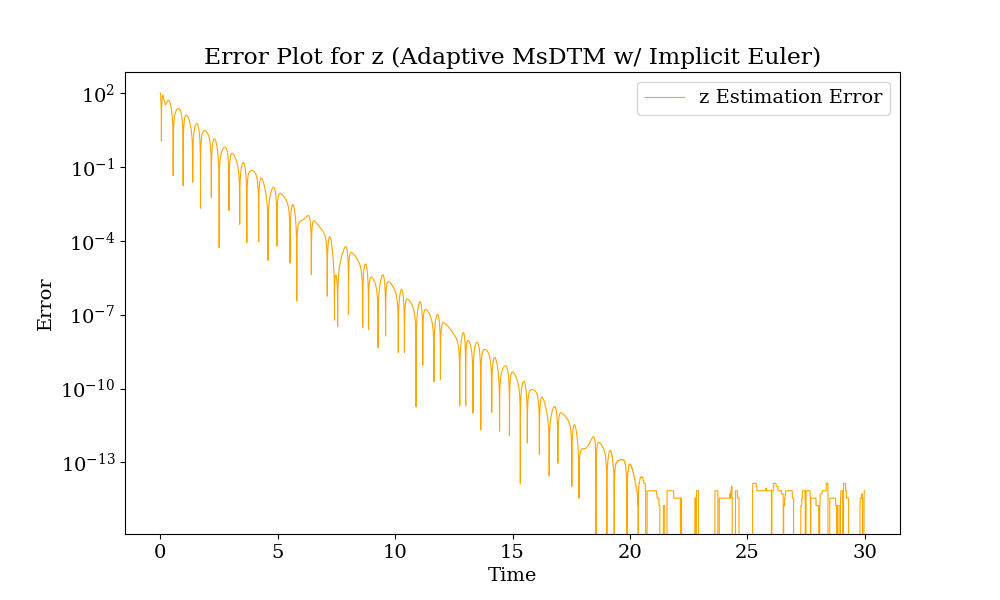}
        \caption{$x_3$ Estimation}
        \label{fig:third}
    \end{subfigure}
    \caption{Adaptive MsDTM + Implicit Euler, initial $\sigma$ = 800}
    \label{fig:msdtm_state_ests_only}
\end{figure}

\begin{figure}[H]
    \centering
    \begin{subfigure}[b]{0.32\textwidth}
        \centering
        \includegraphics[width=\textwidth]{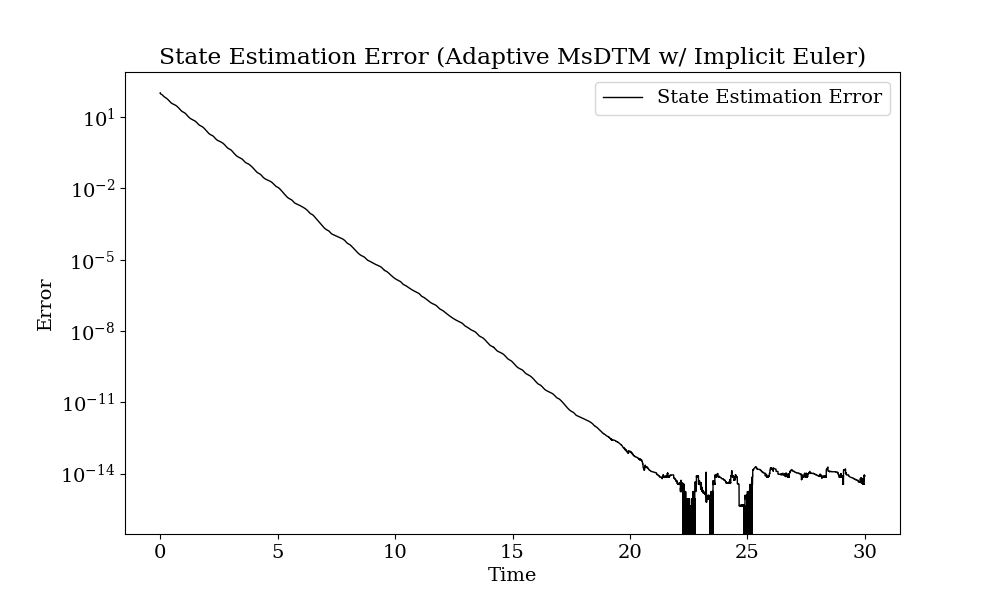}
        \caption{State Estimation}
        \label{fig:first}
    \end{subfigure}
    \hspace{0.1in}
    \begin{subfigure}[b]{0.32\textwidth}
        \centering
        \includegraphics[width=\textwidth]{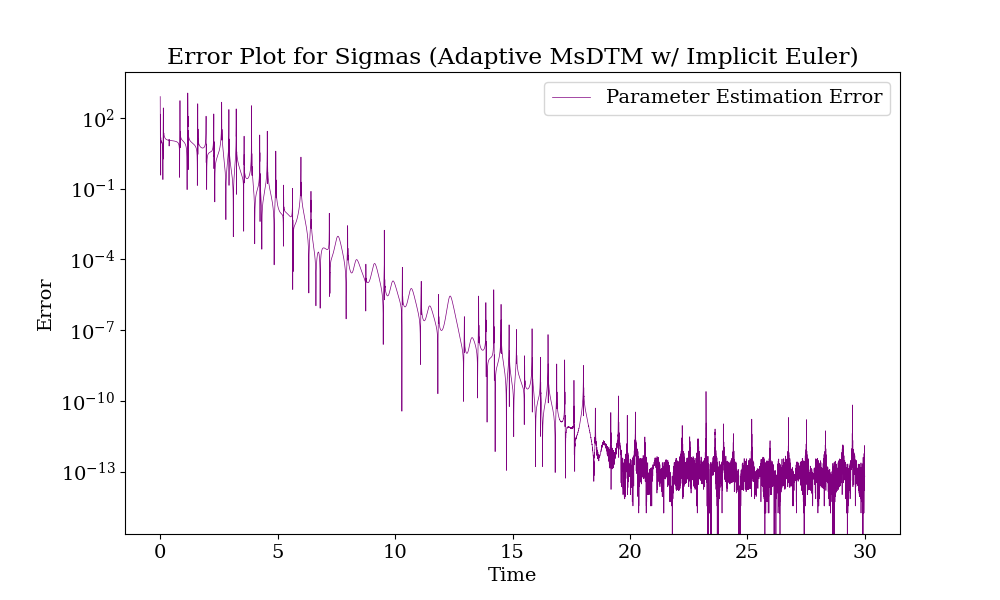}
        \caption{Parameter Estimation}
        \label{fig:second}
    \end{subfigure}
    \caption{Adaptive MsDTM + Implicit Euler, initial $\sigma$ = 800}
    \label{fig:msdtm_state_and_param_est}
\end{figure}

To demonstrate the tolerance of CHL endowed by the adaptive MsDTM to large initial errors in both parameter and state estimations, we plot the same set of errors with initial $\sigma = 800$, and initial conditions of (30,10,0) and (3,0,100) for the true and nudged systems, respectively. We purposely choose these initial conditions to demonstrate that large differences between initial conditions as such can still yield convergence of the methods for MsDTM but not for forward Euler. Convergence is not guaranteed for all initial condition pairs with the same Euclidean distance as (30,10,0) and (3,0,100). However, for pairs with smaller Euclidean distances, initial conditions can be randomly selected. The results are presented in Figures \ref{fig:msdtm_state_ests_only} and \ref{fig:msdtm_state_and_param_est}.

\subsubsection{Comparing CHL with AOT to EM with EnKF}
\label{sec:63_results}
In Section \ref{sec:numerical_methods}, we discussed how to evolve the true and nudged systems when implementing the AOT DA algorithm. To perform PR based on the AOT DA, we add a parameter update step at each numerical time step following the state update using \eqref{CHL_formula}.  For the combined algorithm of EnKF and EM applied to Lorenz '63, we use the adaptive MsDTM to numerically construct both synthetic observations and the model evolution. We use correct parameter values to generate synthetic data, but for the model we use the estimated $\sigma$ from the previous numerical time step, through \eqref{EM_1} and \eqref{EM_2}, that was obtained after the EnKF data assimilation process that updates states of all ensembles.  \\
\begin{table}[hbt!]
\caption*{\textbf{Table of Experimental Setups}}
\centering
\begin{tabular}{| l | c | c |}

\hline
DA + PR  & AOT + CHL & EnKF + EM\\
\hline \hline
Param. Init. Est. & 100 & 100 \\  \hline
Time & (0,20) & (0,20)\\  \hline
Time Step Size & 0.01 & 0.01\\  \hline
$\epsilon$ & $10^{-4}$ & $10^{-4}$\\  \hline
Obs. IC & (10,10,10) & (10,10,10)\\  \hline
Obs. Noise SD & $10^{-4}$ & $10^{-4}$ \\  \hline
Nudged System IC & (11,11,11) & N/A\\ \hline
$\mu_{1}$ & 500& N/A\\  \hline
$\mu_{2}$ & 0& N/A\\  \hline
$\mu_{3}$ & 0& N/A\\  \hline
Ensemble Size & N/A & 50\\ \hline
\end{tabular}
\caption{Lorenz '63 Experiment Conditions}
\label{table:63_experiment}
\end{table}

Notation in Table~\ref{table:63_experiment}:
\begin{itemize}
\item Param.~Init.~Est.: parameter initial estimation value.
\item Time: the time frame in which we observe the error evolutions.
\item $\epsilon$: adaptive MsDTM $\epsilon$ threshold.
\item Obs.~IC: observation initial conditions.
\item Obs.~Noise SD: observation noise standard deviation.
\item Nudged System IC: nudged system initial conditions.
\item $\mu$: relaxation parameters for algorithms that use nudging.
\item Ensemble Size: in this case, the EnKF data assimilation state ensemble size.
\end{itemize}

\begin{figure}[H]
    \centering
    \begin{subfigure}[b]{0.35\textwidth}
        \centering
        \includegraphics[width=\textwidth]{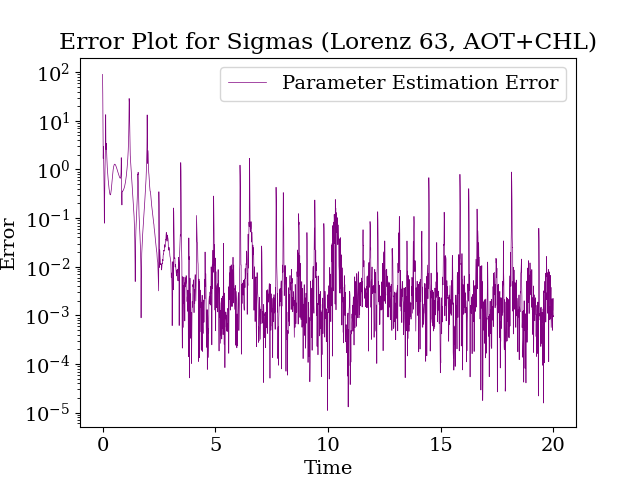}
        \caption{AOT + CHL PR Error (63)}
        \label{fig:first}
    \end{subfigure}
    \hspace{.2in}
    \begin{subfigure}[b]{0.35\textwidth}
        \centering
        \includegraphics[width=\textwidth]{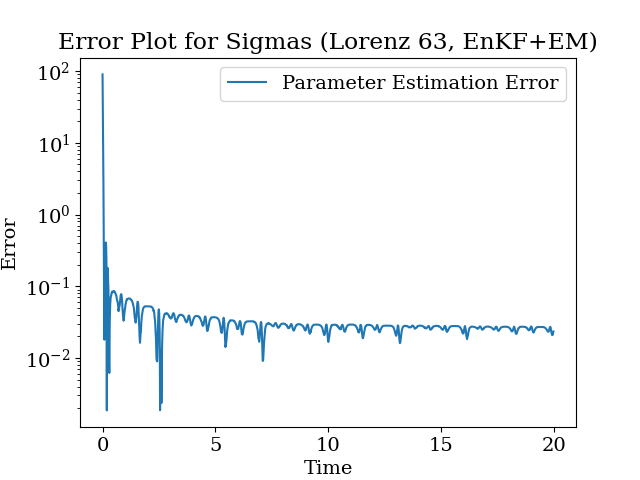}
        \caption{EnKF + EM PR Error (63)}
        \label{fig:second}
    \end{subfigure}
    \caption{Deterministic and stochastic PR comparison}
    \label{fig:CHL_EM_L63}
\end{figure}

\begin{figure}[H]
    \centering
    \begin{subfigure}[b]{0.35\textwidth}
        \centering
        \includegraphics[width=\textwidth]{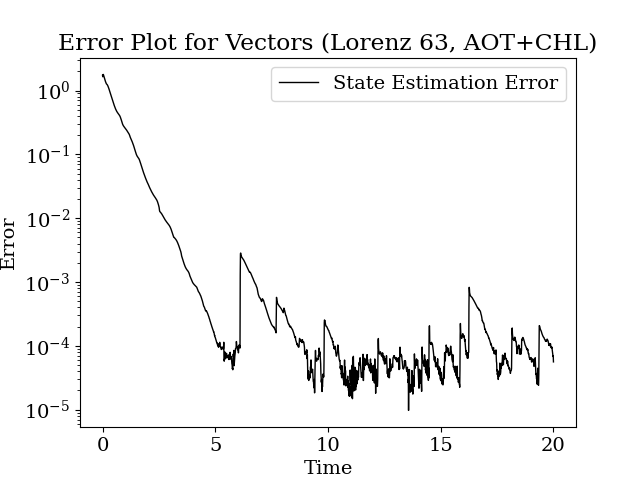}
        \caption{AOT + CHL DA Error (63)}
        \label{fig:first}
    \end{subfigure}
    \hspace{.2in}
    \begin{subfigure}[b]{0.35\textwidth}
        \centering
        \begin{overpic}[width=\textwidth,tics=10]{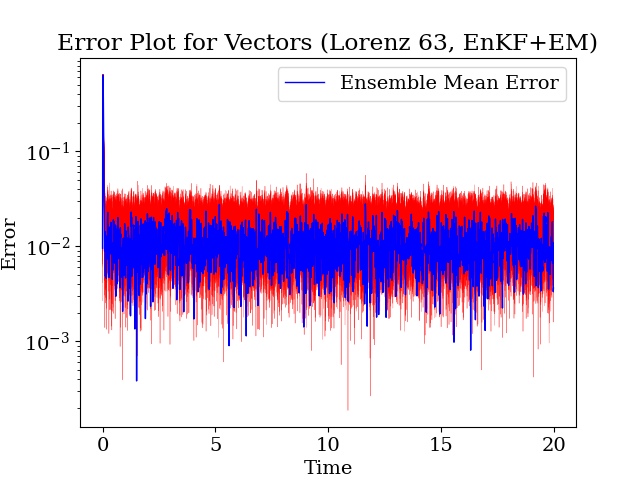}
        \put(36,55){\fontsize{5pt}{7pt}\selectfont\scalebox{1}[1.2]{\textcolor{red}{---} Ensemble Individual Errors}}
         \put(33,55){%
    \color{gray!30}%
    \fbox{\phantom{\fontsize{5pt}{7pt}\selectfont\scalebox{1}[1.2]{\textcolor{red}{---} Ensemble Individual Errors}}}}
  
        \end{overpic}
        \caption{EnKF + EM DA Error (63)}
        \label{fig:second}
    \end{subfigure}
    \caption{Deterministic and stochastic state estimation (while performing PR) comparison}
    \label{fig:L63_stateErrors_DandS}
\end{figure}

Means and variances of the errors below are calculated over the time interval (10,20), starting from the midpoint of our time frame after the errors level out. 
The accuracy of deterministic AOT + CHL appears to be higher than that of stochastic EnKF + EM at lower noise levels, and the deterministic PR process is very computationally efficient compared to the stochastic PR process. However, EnKF + EM has demonstrated greater stability in the PR process.  Results are reported in Tables \ref{table:CHL_EM_L63}, \ref{table:63_PR_mean}, and \ref{table:63_PR_variance}, and Figure \ref{fig:CHL_EM_L63}. 
The deterministic AOT + CHL has almost always shown higher or similar accuracy and stability compared to the stochastic EnKF + EM for state estimation (see Tables \ref{table:63_state_mean} and \ref{table:63_state_variance}, and Figure \ref{fig:L63_stateErrors_DandS}).


The standard deviation of the error $SD \in \{10^{-4}, 10^{-3}, 10^{-2}, 10^{-1}\}$ is comparable with the noise levels used in the standard literature; for example, in \cite{edo_reich_stuart_2024} the authors used 0.5 for noise variance for EnKI applied to Lorenz '96. As can be observed in Tables \ref{table:63_PR_mean}, \ref{table:63_PR_variance}, \ref{table:63_state_mean}, and \ref{table:63_state_variance}, the CHL starts to become unstable as noise level increases.
However, we note that this issue can possibly be addressed by adjusting the parameter update window (i.e. updating every few numerical time steps rather than at every one), which will be investigated more carefully in a separate paper. A brief investigation demonstrating this possibility is given in Section \ref{sec:wide_PR_window}.\\

\begin{table}[hbt!]
\centering
\begin{tabular}{| l | c | }

\hline
Method  & Runtime\\
\hline \hline
AOT + CHL & $\sim$ a few seconds\\  \hline
EnKF + EM & $\sim$ 6 hours\\  \hline

\end{tabular}
\caption{This table displays runtime of the AOT + CHL method and the EnKF + EM method.}
\label{table:CHL_EM_L63}
\end{table}
\begin{table}[H]
\centering
\begin{tabular}{| l | c | c | c | c |}

\hline
Parameter Estimate Error Mean  & $SD = 10^{-4}$ & $SD = 10^{-3}$ & $SD = 10^{-2}$ & $SD = 10^{-1}$ \\
\hline \hline
AOT + CHL & 0.01258 & 0.07138 & 0.83947 & NC \\  \hline
EnKF + EM & 0.02626 & 0.31245 & 0.19845 &  15.49873\\  \hline

\end{tabular}
\caption{This table displays PR error means when using AOT + CHL and EnKF + EM on Lorenz '63.}
\label{table:63_PR_mean}
\end{table}
\begin{table}[h!]
\centering
\begin{tabular}{| l | c | c | c | c |}

\hline
Parameter Estimate Error Variance & $SD = 10^{-4}$ & $SD = 10^{-3}$ & $SD = 10^{-2}$ & $SD = 10^{-1}$ \\
\hline \hline
AOT + CHL & 0.00309 & 0.05320 & 6.12530 & NC \\  \hline
EnKF + EM & $4.90 \times 10^{-6}$ & 0.00241 & 0.00110 & 0.64099\\  \hline

\end{tabular}
\caption{This table displays PR error variances when using AOT + CHL and EnKF + EM on Lorenz '63.}
\label{table:63_PR_variance}
\end{table}

\begin{table}[h!]
\centering
\begin{tabular}{| l | c | c | c | c | }

\hline
State Estimate Error Mean  & $SD = 10^{-4}$ & $SD = 10^{-3}$ & $SD = 10^{-2}$ & $SD = 10^{-1}$ \\
\hline \hline
AOT + CHL & 0.00010 & 0.00130 & 0.01157 & NC \\  \hline
EnKF + EM & 0.01054 & 0.03627 & 0.02843 &  2.73515\\  \hline

\end{tabular}
\caption{This table displays DA error means when using AOT + CHL and EnKF + EM on Lorenz '63.}
\label{table:63_state_mean}
\end{table}
\begin{table}[hbt!]
\centering
\begin{tabular}{| l | c | c | c | c | }

\hline
State Estimate Error Variance & $SD = 10^{-4}$ & $SD = 10^{-3}$ & $SD = 10^{-2}$ & $SD = 10^{-1}$ \\
\hline \hline
AOT + CHL & $1.14 \times 10^{-8}$ & 2.13 $\times 10^{-6}$ & 0.00021 & NC \\  \hline
EnKF + EM & 0.00002 & 0.00034 & 0.00019 & 2.12338\\  \hline

\end{tabular}
\caption{This table displays DA error variances when using AOT + CHL and EnKF + EM on Lorenz '63.}
\label{table:63_state_variance}
\end{table}

\subsection{Lorenz '96 System}
\label{sec:L96_results}
\subsubsection{Implementation Details for Comparison}
The idea of implementing AOT DA with deterministic PR algorithms and the stochastic EnKF + EM is the same as in Section~\ref{sec:63_results}. The deterministic PR algorithms are given in \eqref{CHL_96_formula}, \eqref{PWM_formula}, \eqref{AV_formula}. Moreover, we applied stochastic EnKI PR to Lorenz '96. All fast variables are constructed with adaptive MsDTM, but slow variables are constructed with the Euler method (see Section~\ref{sec:numerical_methods}), whose initial values are always drawn uniformly from $[0,1)$.

Apart from grouping stochastic DA methods with stochastic PR methods, and likewise for deterministic methods, we can also mix-and-match. We combine EnKF, PF, and ETKF with deterministic PR, where at each step we take the mean of ensemble states as our updated model state.
We also combine AOT with EM. This is done by evolving the model for each ensemble using AOT, and then applying EM to the ensembles as usual. 

We applied several variations of the PF to perform PR. The most straightforward way is to combine the stochastic PF with EM (similar to EnKF + EM), but we can also adapt PF to deterministic PR (see Section \ref{sec:original_PF}) by incorporating AOT DA into modeling dynamics. Additionally, one could extend the PF to perform PR, as discussed in Section \ref{sec:PF_extension}. 

\begin{table}[h!]
\caption*{\textbf{Table of Experimental Setups}}
\adjustbox{margin=0pt,center}{ 
\centering
\begin{tabular}{| l | c | c | c | c | c | c | c | c | c | c | c |}
\hline
\textbf{DA} & \textbf{PR} & $\overline{d_{1}}$ IE & Time & $\Delta t$ & $\epsilon$ & Model~IC & $\mu_{1}$ & $\mu_{2}$ & $\mu_{3}$ & $\mu_{4}$ & Size \\
\hline \hline
\multirow{4}{*}{AOT} 
 & CHL & 50 & (0,2) & $10^{-3}$ & $10^{-4}$ & (15,10,10,10) & 500 & 100 & 100 & 0 & N/A\\  \cline{2-12}
 & PWM & 50 & (0,2) & $10^{-3}$ & $10^{-4}$ & (15,10,10,10) & 500 & 100 & 100 & 0 & N/A\\  \cline{2-12}
 & AV & 50 & (0,2) & $10^{-3}$ & $10^{-4}$ & (15,10,10,10) & 500 & 100 & 100 & 0 & N/A\\ \cline{2-12}
 & EM & 50 & (0,2) & $10^{-3}$ & $10^{-4}$ & $N(0,1)$ & 500 & 100 & 100 & 0 & 50\\ \hline
\multirow{3}{*}{EnKF} 
 & EM & 50 & (0,2) & $10^{-3}$ & $10^{-4}$ & $N(0, 1)$ & N/A & N/A & N/A & N/A & 50\\  \cline{2-12}
 & CHL & 50 & (0,2) & $10^{-3}$ & $10^{-4}$ & $N(0,1)$ & 500 & N/A& N/A & N/A & 50\\ \cline{2-12}
 & PWM & 50 & (0,2) & $10^{-3}$ & $10^{-4}$ & $N(0,1)$ & N/A & N/A& N/A & N/A & 50\\ \hline
N/A & EnKI & $N(50,20^{2})$ & (0,1) & $10^{-3}$ & 0 & N/A & 20 & 20 & 20 & 20 & 100\\ \hline
\multirow{4}{*}{PF} 
 & EM & 50 & (0,2) & $10^{-3}$ & $10^{-4}$ & $N(10,1)$ & N/A & N/A & N/A & N/A & 50 \\ \cline{2-12}
 & Ext. & $N(50,0)$ & (0,2) & $10^{-3}$ & $10^{-4}$ & $N(10, 100^{2})$ & N/A & N/A & N/A & N/A & 50\\ \cline{2-12}
 & CHL & 50 & (0,2) & $10^{-3}$ & $10^{-4}$ & $N(10,100^{2})$ & 500 & 100 & 100 & 0 & 50\\ \cline{2-12}
 & PWM & 50 & (0,2) & $10^{-3}$ & $10^{-4}$ & $N(10,100^{2})$ & 500 & 100 & 100 & 0 & 50\\ \hline
 \multirow{3}{*}{ETKF}
 & EM & 50 & (0,2) & $10^{-3}$ & $10^{-4}$ & $N(0,1)$ & N/A & N/A & N/A & N/A & 50 \\ \cline{2-12}
 & CHL & 50 & (0,2) & $10^{-3}$ & $10^{-4}$ & $N(0,1)$ & 500 & N/A & N/A & N/A & 50 \\ \cline{2-12}
 & PWM & 50 & (0,2) & $10^{-3}$ & $10^{-4}$ & $N(0,1)$ & N/A & N/A & N/A & N/A & 50 \\ \cline{2-12}
 & AV & 50 & (0,2) & $10^{-3}$ & $10^{-4}$ & $N(0,1)$ & 500 & N/A & N/A & N/A & 50 \\ \hline
\end{tabular}
}
\caption{Lorenz '96 Experiment Conditions}
\label{table:96_setup}
\end{table}

Notation in Table~\ref{table:96_setup}:
\begin{itemize}
\item IE: initial estimation.
\item Time: the time frame in which we observe the error evolutions.
\item $\Delta t$: observation numerical time step size.  
\item $\epsilon$: adaptive time step threshold when applying adaptive MsDTM to fast variables.
\item IC: initial condition.
\item $\mu$: relaxation parameters from nudging. 
\item Size: state ensemble size (exception: for EnKI, it is parameter ensemble size).
\end{itemize} 

More Implementation Details:
\begin{itemize}
\item EnKI: The iteration time step size ($\Delta t$ in Algorithm~\ref{algorithm:EnKI}) was taken to be 0.01.  
\item The initial condition for the system from which we drew the observations was given by $(10,10,10,10)$ for the fast variables; for the slow variables, we generate a random set of initial values with twenty entries drawn uniformly from $[0,1)$.  This is \textit{not} a stochastic initial condition, as we only perform this sampling once.
\item For all EM implementations, we take $\gamma = c_{0} = \sigma = 1$ (see \eqref{EM_2}). The exception is PF + EM, where we tuned $\gamma = 10^{-6}, c_{0} = 100, \sigma = 10^{-6}$.
\item Model noise matrix $\Gamma$ has magnitude of $10^{-6}$ whenever it was required to exist.
\end{itemize}

\subsubsection{Comparison of Parameter and State Estimation Error Means and Variances}
\label{sec:96_results}
Same as in Section~\ref{sec:63_results}, error means and variances are calculated over the latter half of our error evolution time periods. In the rest of the article, ``purely deterministic'' methods are methods that pair deterministic DA with deterministic PR, and likewise for ``purely stochastic'' methods. Mix-and-matching generally compromises both parameter and state estimation accuracy and stability with two exceptions: AOT + EM at $SD = 10^{-1}$ and EnKF + PWM at $SD = 10^{-6}$. Nonetheless, there is always at least one purely deterministic method that outperforms mix-and-matched methods in all four Tables \ref{table:96_PR_mean}, \ref{table:96_PR_variance}, \ref{table:96_state_mean}, and \ref{table:96_state_variance}. 

\begin{figure}[h!]
    \centering
    \begin{subfigure}{0.48\textwidth}
        \centering
        \includegraphics[width=0.9\textwidth]{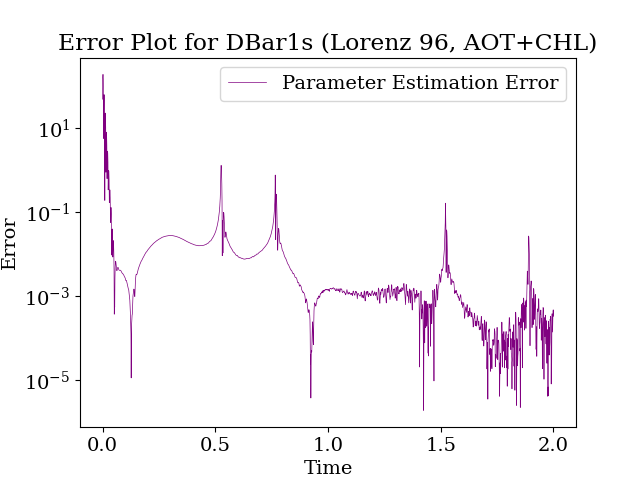} 
        \caption{AOT + CHL PR Error (96)}
        \label{fig:subfig1}
    \end{subfigure}
    \begin{subfigure}{0.48\textwidth}
        \centering
        \includegraphics[width=0.9\textwidth]{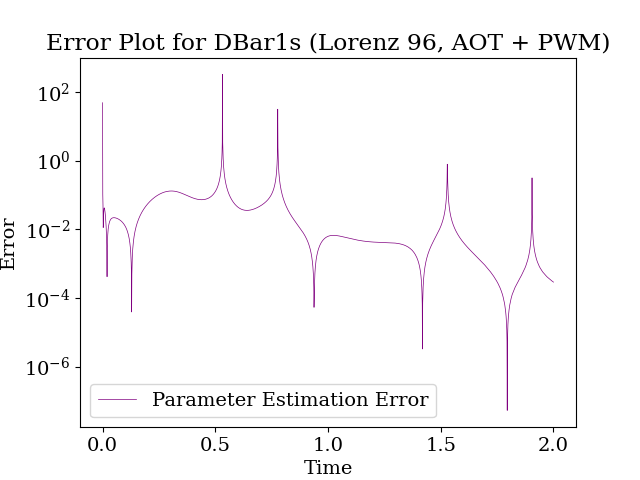} 
        \caption{AOT + PWM PR Error (96)}
        \label{fig:subfig2}
    \end{subfigure}
    \vskip\baselineskip
    \begin{subfigure}{0.48\textwidth}
        \centering
        \includegraphics[width=0.9\textwidth]{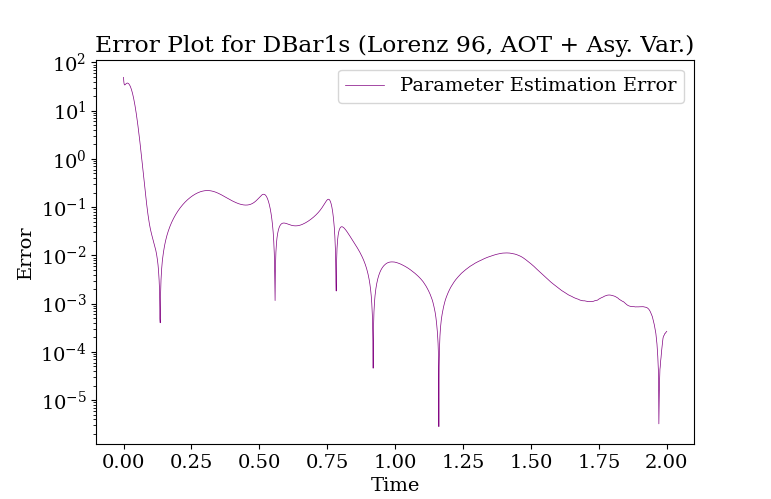} 
        \caption{AOT + AV PR Error (96)}
        \label{fig:subfig3}
    \end{subfigure}
    \begin{subfigure}{0.48\textwidth}
        \centering
        \includegraphics[width=0.95\textwidth]{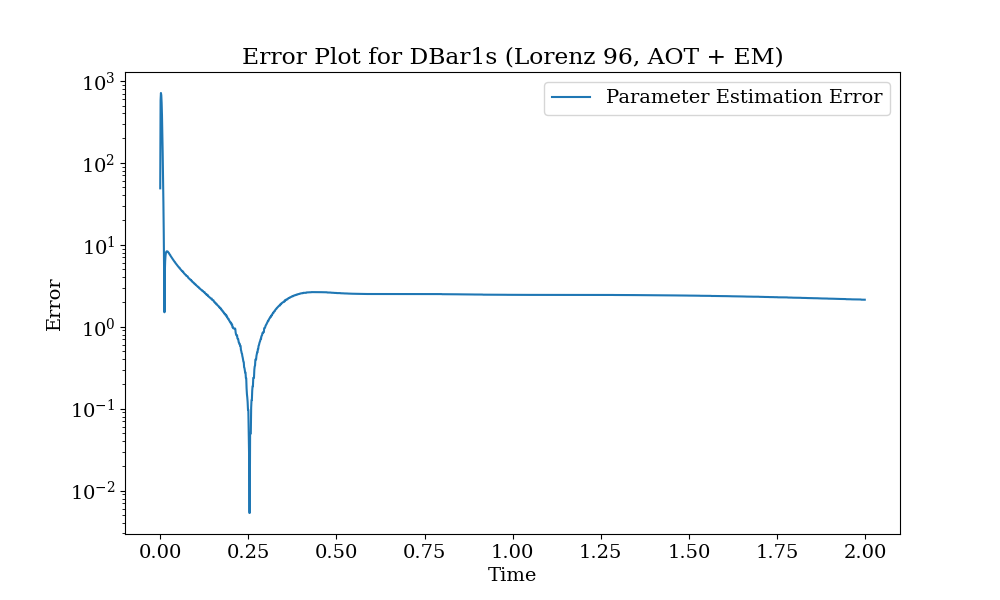} 
        \caption{AOT + EM PR Error (96)}
        \label{fig:subfig4}
    \end{subfigure}
    \caption{Different PR methods paired with deterministic AOT DA method at $SD = 10^{-6}$.}
    \label{fig:purely_deterministic_PR}
\end{figure}

\begin{figure}[h!]
    \centering
    \begin{subfigure}{0.48\textwidth}
        \centering
        \includegraphics[width=1\textwidth]{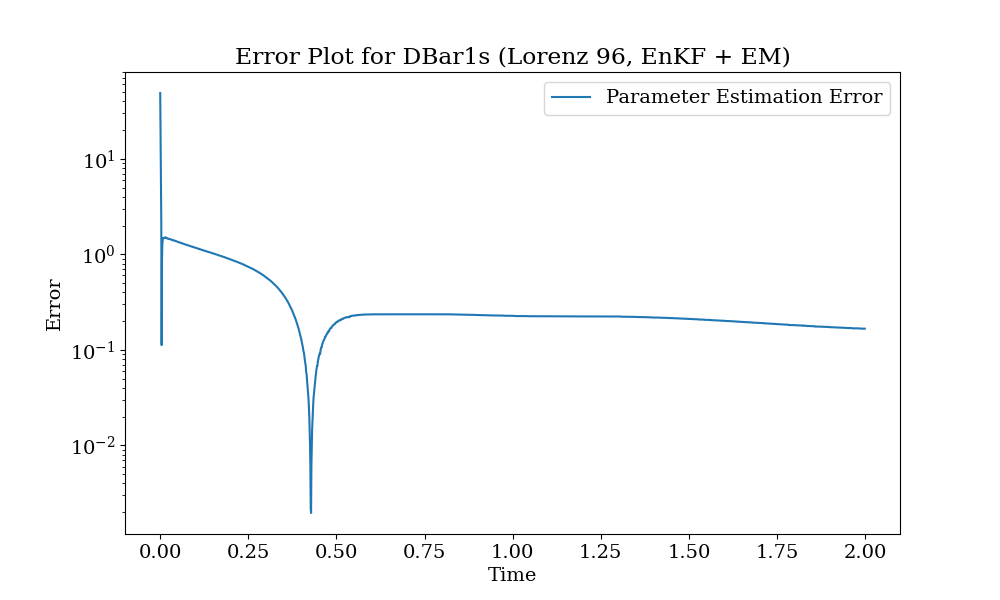} 
        \caption{EnKF + EM PR Error (96)}
        \label{fig:subfig1}
    \end{subfigure}
    \begin{subfigure}{0.48\textwidth}
        \centering
        \includegraphics[width=1\textwidth]{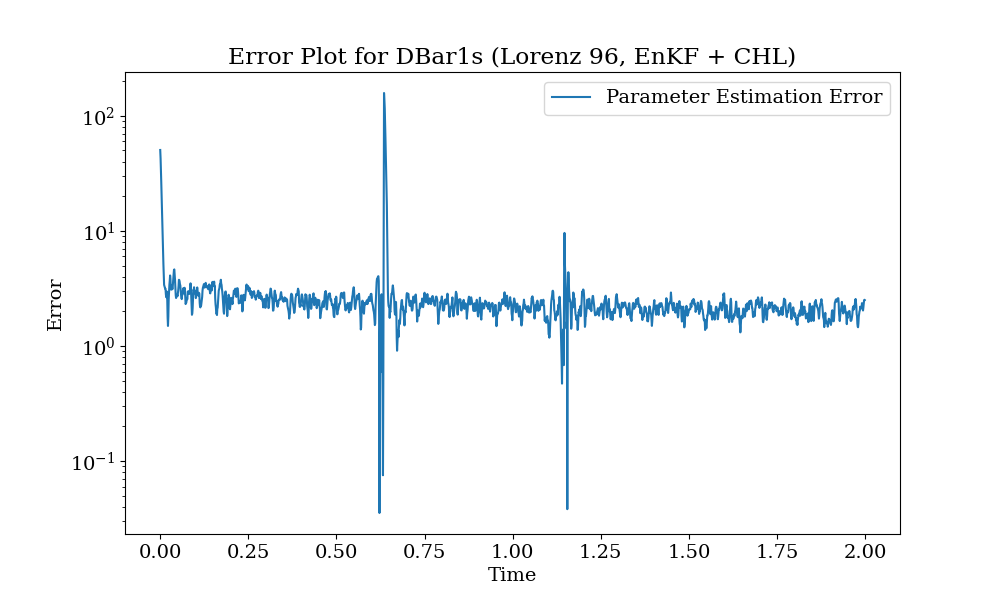} 
        \caption{EnKF + CHL PR Error (96)}
        \label{fig:subfig2}
    \end{subfigure}
        \vskip\baselineskip
    \begin{subfigure}{0.5\textwidth}
        \centering
        \includegraphics[width=\textwidth]{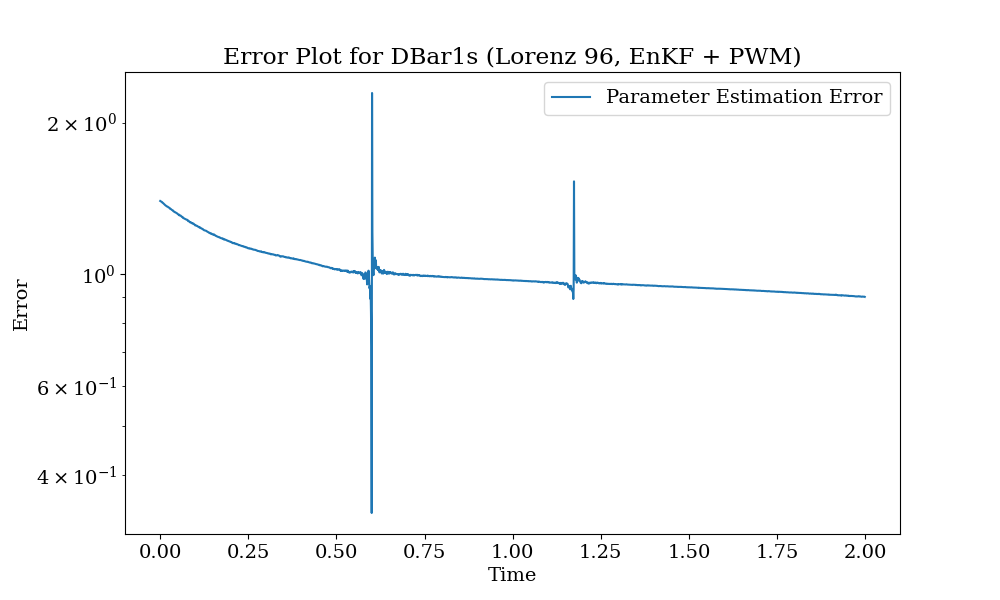} 
        \caption{EnKF + PWM PR Error (96)}
        \label{fig:subfig3}
    \end{subfigure}
    \caption{Different PR methods paired with stochastic EnKF DA method at $SD = 10^{-6}$.}
    \label{fig:EnKF_w_PR}
\end{figure}

\begin{figure}[h!]
\centering
\begin{overpic}[width=0.55\textwidth,tics=10]{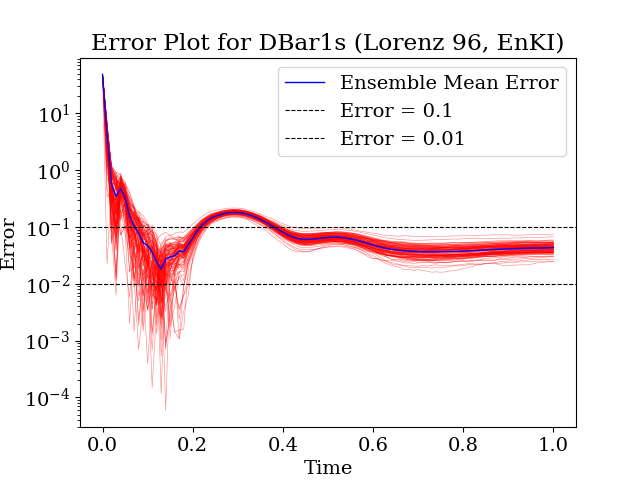}
        \put(36,45){\fontsize{5pt}{7pt}\selectfont\scalebox{1}[1.2]{\textcolor{red}{---} Ensemble Individual Errors}}
         \put(33,45){%
    \color{gray!30}%
    \fbox{\phantom{\fontsize{5pt}{7pt}\selectfont\scalebox{1}[1.2]{\textcolor{red}{---} Ensemble Individual Errors}}}}
  
        \end{overpic}
\caption{EnKI PR Error (96) at $SD = 10^{-6}$}
\label{fig:EnKI_PR_96}
\end{figure}

\begin{figure}[h!]
    \centering
    \begin{subfigure}{0.48\textwidth}
        \centering
        \includegraphics[width=1.1\textwidth]{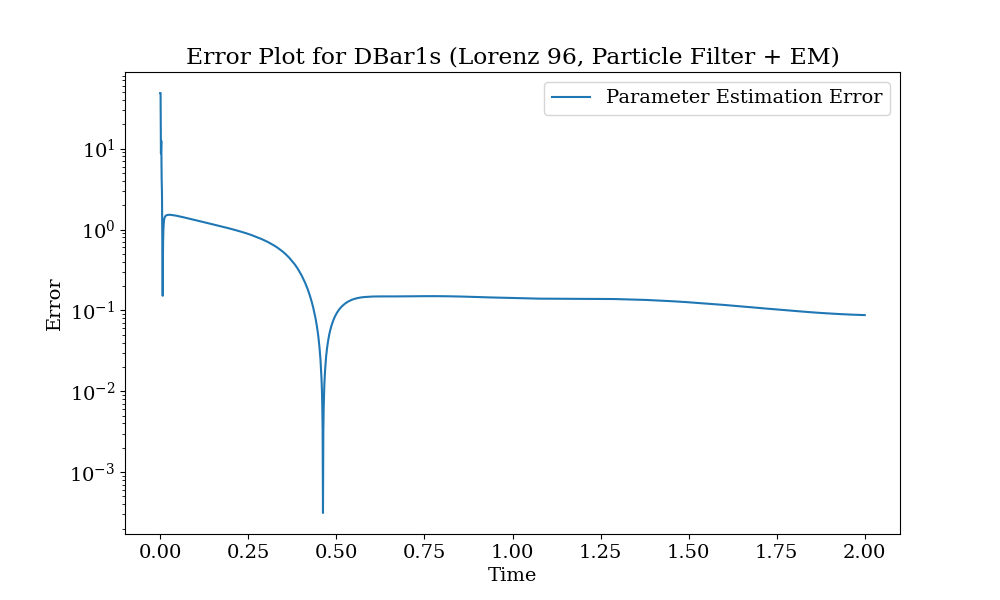} 
        \caption{PF + EM PR Error (96)}
        \label{fig:subfig1}
    \end{subfigure}
    \begin{subfigure}{0.48\textwidth}
        \centering
        \includegraphics[width=1.1\textwidth]{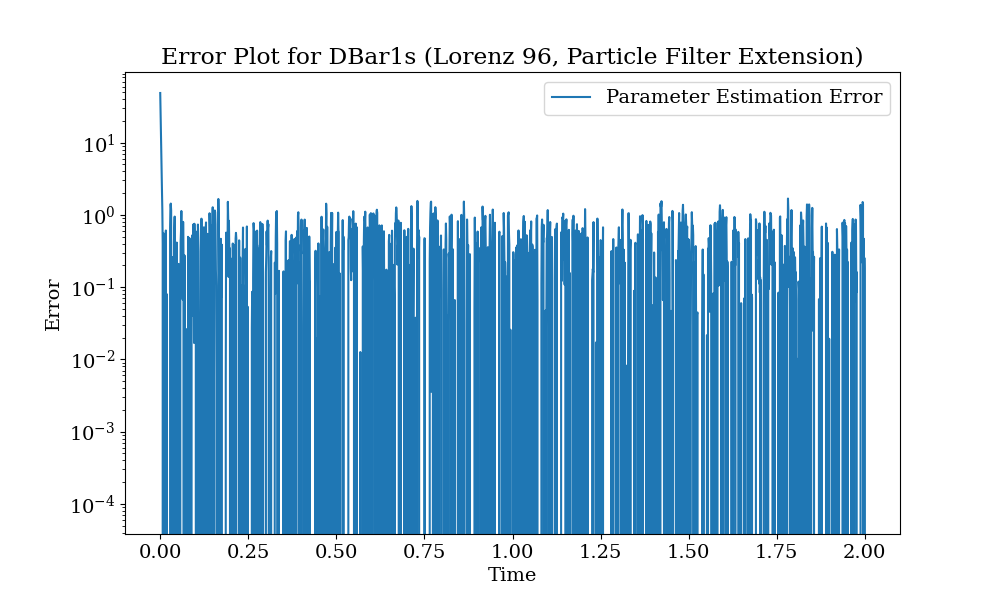} 
        \caption{PF Extension PR Error (96)}
        \label{fig:subfig2}
    \end{subfigure}
    \vskip\baselineskip
    \begin{subfigure}{0.48\textwidth}
        \centering
        \includegraphics[width=1.1\textwidth]{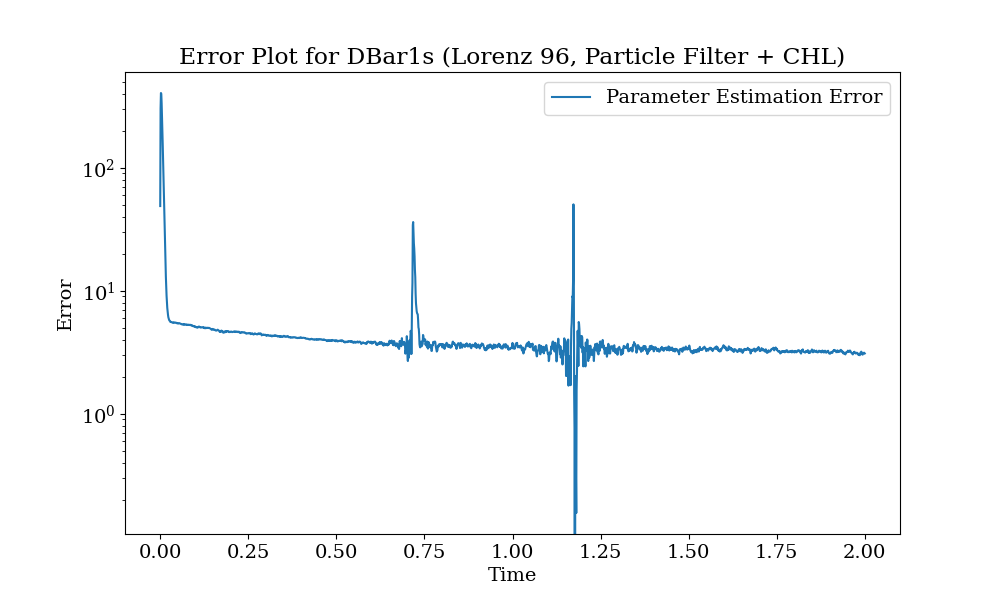} 
        \caption{PF + CHL PR Error (96)}
        \label{fig:subfig3}
    \end{subfigure}
    \begin{subfigure}{0.48\textwidth}
        \centering
        \includegraphics[width=1.1\textwidth]{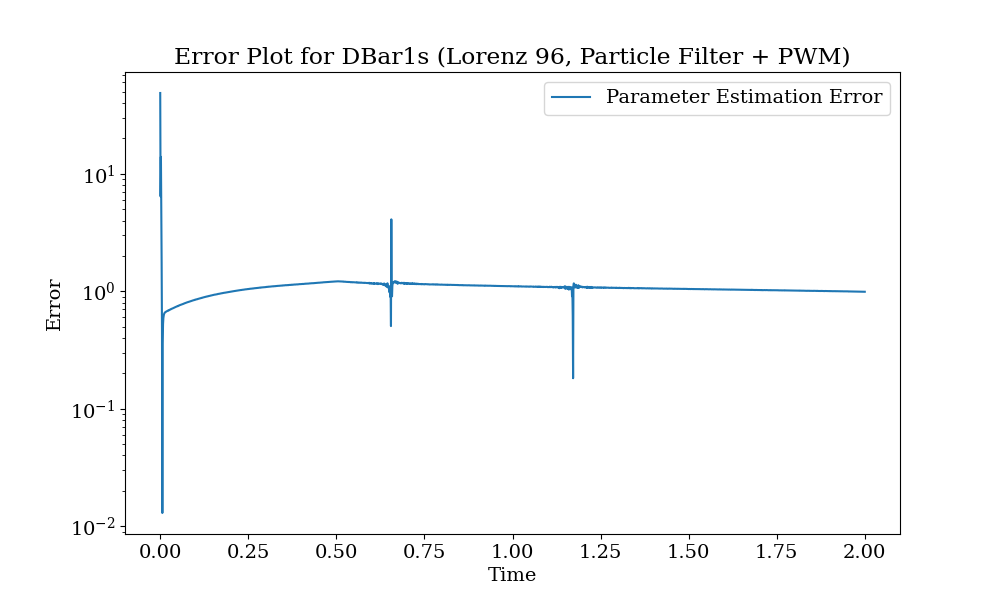} 
        \caption{PF + PWM PR Error (96)}
        \label{fig:subfig4}
    \end{subfigure}
    \caption{Different PR methods paired with stochastic PF DA method at $SD = 10^{-6}$.}
    \label{fig:PF_PR_quad}
\end{figure}

\begin{figure}[h!]
    \centering
    \begin{subfigure}[b]{0.45\textwidth}
        \centering
        \includegraphics[width=1\textwidth]{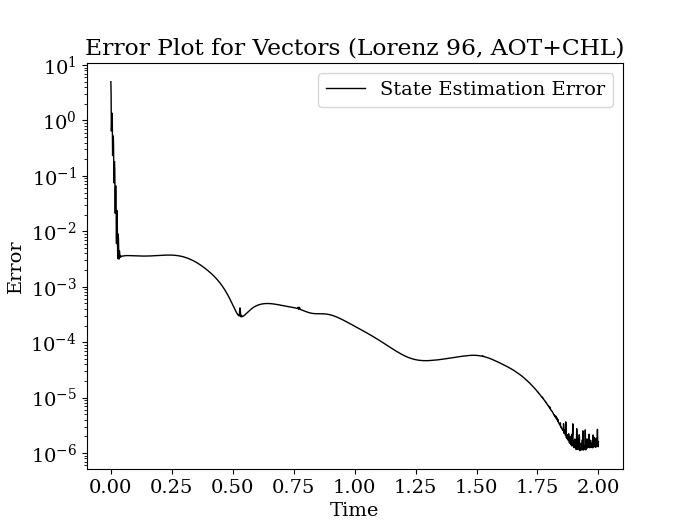} 
        \caption{AOT + CHL DA Error (96)}
        \label{fig:subfig1}
    \end{subfigure}
    \hspace{.1in}
    \begin{subfigure}[b]{0.45\textwidth}
        \centering
        \includegraphics[width=1.1\textwidth]{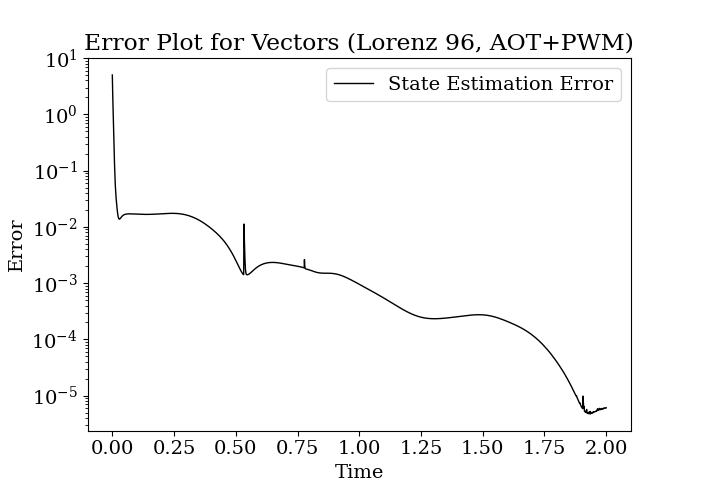} 
        \caption{AOT + PWM DA Error (96)}
        \label{fig:subfig2}
    \end{subfigure}
    \begin{subfigure}[b]{0.45\textwidth}
        \centering
        \includegraphics[width=1\textwidth]{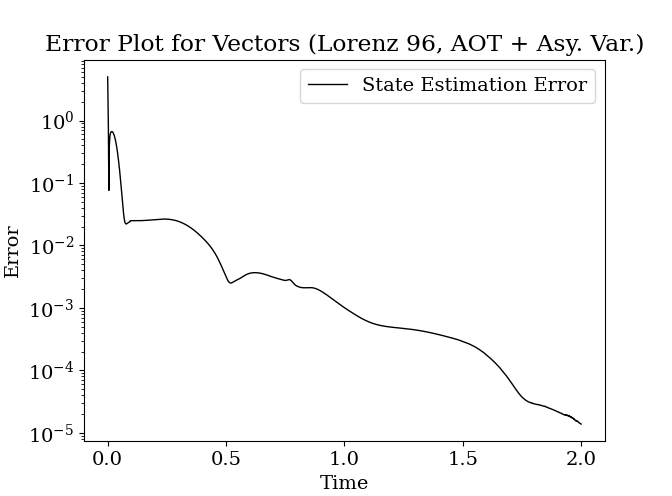} 
        \caption{AOT + AV DA Error (96)}
        \label{fig:subfig3}
    \end{subfigure}
    \hspace{.1in}
    \begin{subfigure}[b]{0.45\textwidth}
        \centering
        \begin{overpic}[width=1.2\textwidth,tics=10]{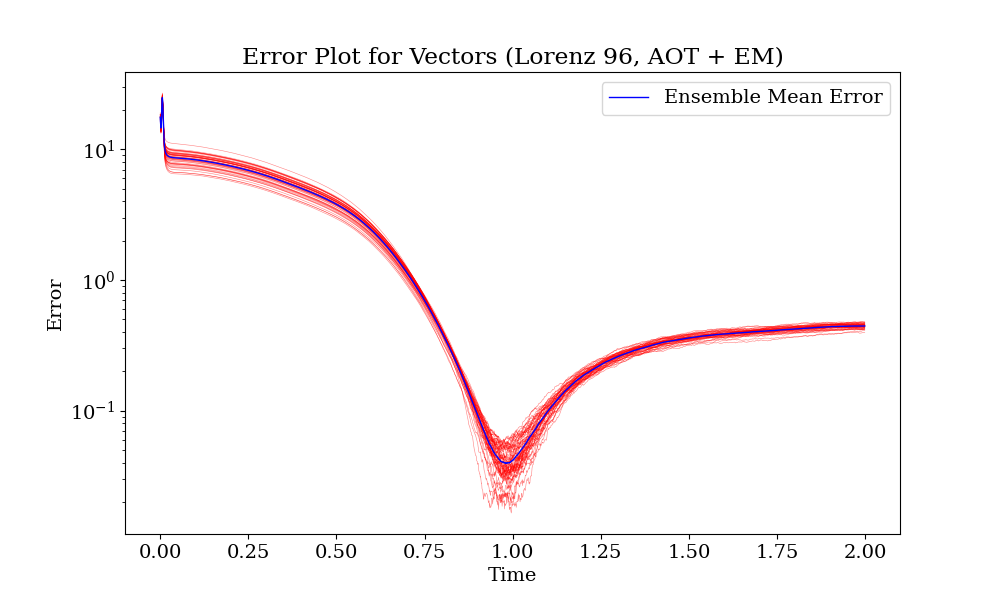}
        \put(58,46){\fontsize{4pt}{7pt}\selectfont\scalebox{0.8}[1.1]{\textcolor{red}{----} Ensemble Individual Errors}}
         \put(66,46){%
    \color{gray!30}%
    \fbox{\phantom{\fontsize{2pt}{4pt}\selectfont\scalebox{0.55}[0.05]{\textcolor{red}{---} Ensemble Individual Errors}}}}
  
        \end{overpic}
        \caption{AOT + EM DA Error (96)}
        \label{fig:subfig4}
    \end{subfigure}
    \caption{State estimation when using deterministic AOT DA method at $SD = 10^{-6}$.}
    \label{fig:AOT_quad_state}
\end{figure}

\begin{figure}[h!]
    \centering
    \begin{subfigure}{0.48\textwidth}
        \centering
        \begin{overpic}[width=1.1\textwidth,tics=10]{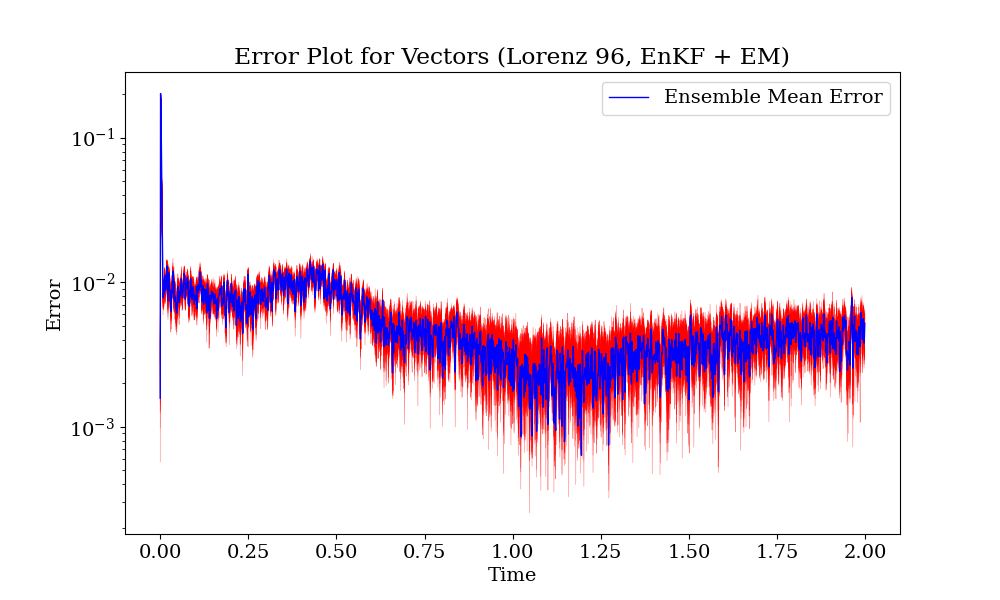}
        \put(57,46){\fontsize{6pt}{8pt}\selectfont\scalebox{0.75}[0.95]{\textcolor{red}{----} Ensemble Individual Errors}}
         \put(56,46){%
    \color{gray!30}%
    \fbox{\phantom{\fontsize{2pt}{4pt}\selectfont\scalebox{0.55}[0.05]{\textcolor{red}{---} Ensemble Individual Errors}}}}
  
        \end{overpic}
        \caption{EnKF + EM DA Error (96)}
        \label{fig:subfig1}
    \end{subfigure}
    \hfill
    \begin{subfigure}{0.48\textwidth}
        \centering
        \begin{overpic}[width=1.1\textwidth,tics=10]{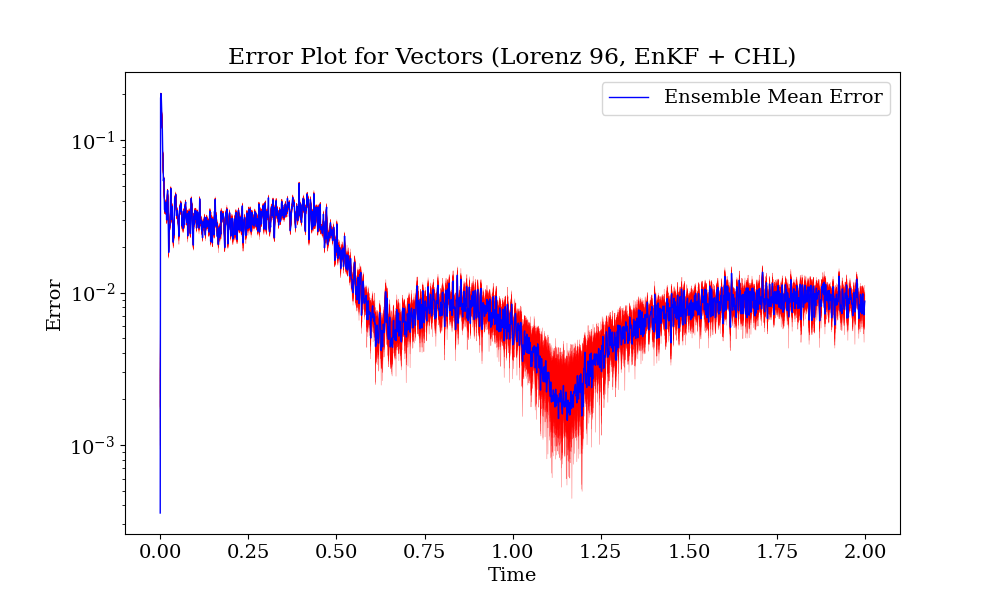}
        \put(57,46){\fontsize{6pt}{8pt}\selectfont\scalebox{0.75}[0.95]{\textcolor{red}{----} Ensemble Individual Errors}}
         \put(54,46){%
    \color{gray!30}%
    \fbox{\phantom{\fontsize{2pt}{4pt}\selectfont\scalebox{0.55}[0.05]{\textcolor{red}{----} Ensemble Individual Errors}}}}
  
        \end{overpic}
        \caption{EnKF + CHL DA Error (96)}
        \label{fig:subfig2}
    \end{subfigure}
    \vskip\baselineskip
    \begin{subfigure}{0.5\textwidth}
        \centering
        \begin{overpic}[width=1.2\textwidth,tics=10]{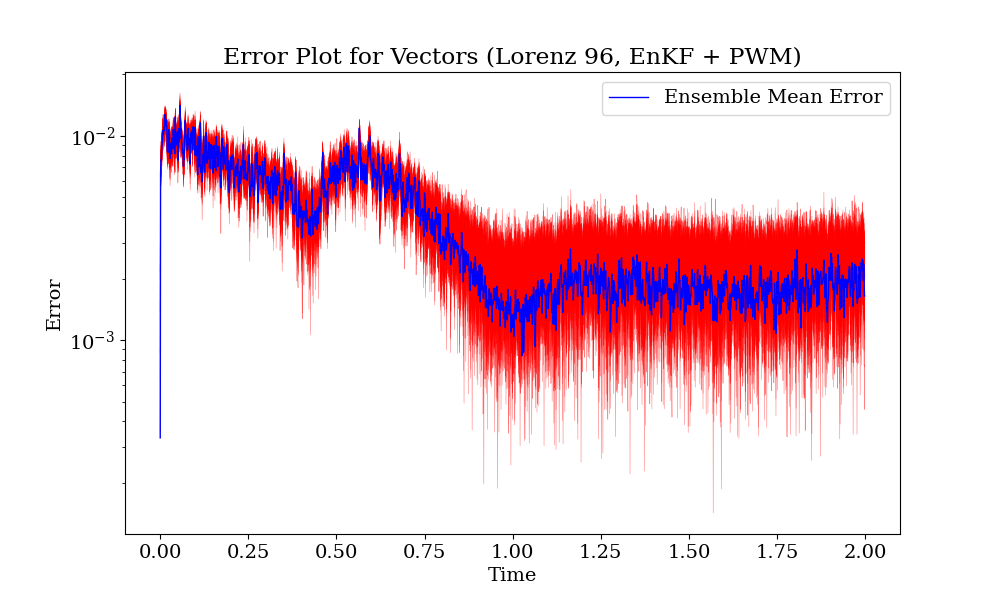}
        \put(57,46){\fontsize{6pt}{8pt}\selectfont\scalebox{0.75}[0.95]{\textcolor{red}{----} Ensemble Individual Errors}}
         \put(56,46){%
    \color{gray!30}%
    \fbox{\phantom{\fontsize{2pt}{4pt}\selectfont\scalebox{0.55}[0.05]{\textcolor{red}{---} Ensemble Individual Errors}}}}
  
        \end{overpic}
        \caption{EnKF + PWM DA Error (96)}
        \label{fig:subfig3}
    \end{subfigure}
    \caption{State estimation when using stochastic EnKF DA method at $SD = 10^{-6}$.}
    \label{fig:EnKF_tri_state}
\end{figure}

\begin{figure}[h!]
    \centering
    \begin{subfigure}{0.45\textwidth}
        \centering
        \begin{overpic}[width=1\textwidth,tics=10]{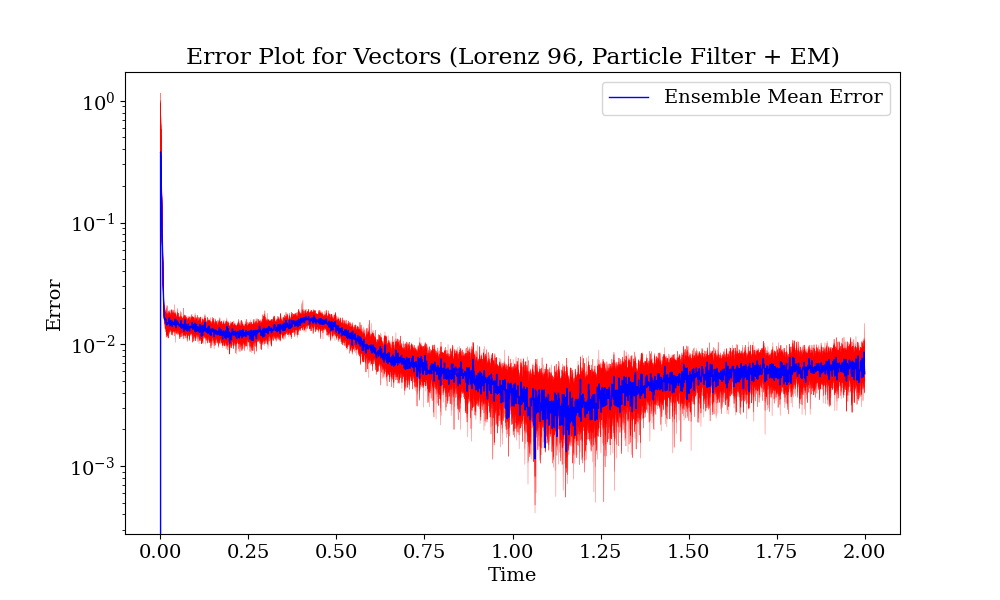}
        \put(59,46){\fontsize{6pt}{8pt}\selectfont\scalebox{0.75}[0.95]{\textcolor{red}{----} Ensemble Individual Errors}}
         \put(58,46){%
    \color{gray!30}%
    \fbox{\phantom{\fontsize{6pt}{8pt}\selectfont\scalebox{0.75}[0.95]{\textcolor{red}{----} Ensemble Individual Errors}}}}
        \end{overpic}
        \caption{PF + EM DA Error (96)}
        \label{fig:subfig1}
    \end{subfigure}
    \begin{subfigure}{0.45\textwidth}
        \centering
        \begin{overpic}[width=1\textwidth,tics=10]{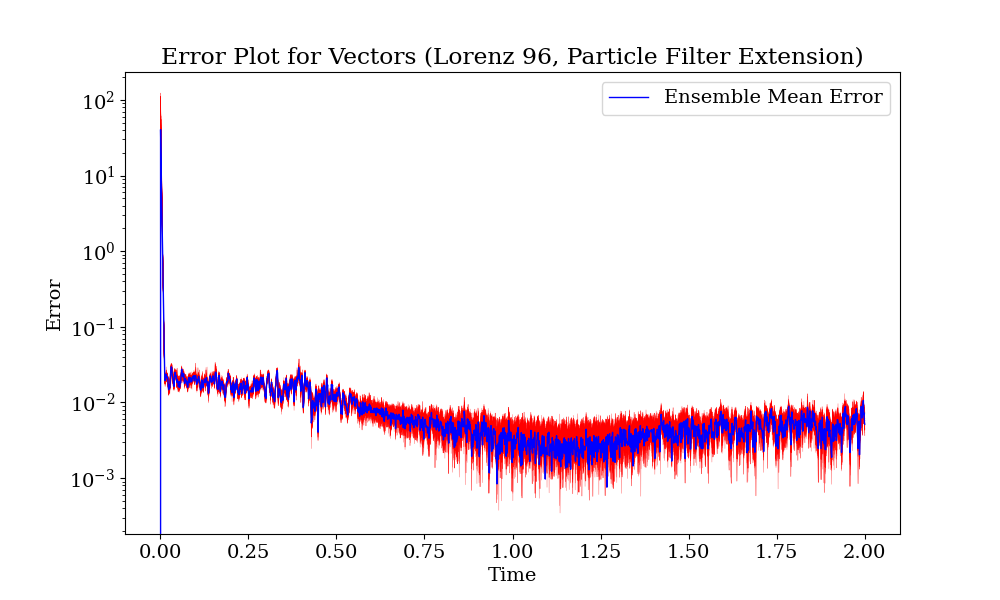}
        \put(59,46){\fontsize{6pt}{8pt}\selectfont\scalebox{0.75}[0.95]{\textcolor{red}{----} Ensemble Individual Errors}}
         \put(58,46){%
    \color{gray!30}%
    \fbox{\phantom{\fontsize{2pt}{4pt}\selectfont\scalebox{0.55}[0.2]{\textcolor{red}{---} Ensemble Individual Errors}}}}
        \end{overpic}
        \caption{PF Extension DA Error (96)}
        \label{fig:subfig2}
    \end{subfigure}
    \vskip\baselineskip
    \begin{subfigure}{0.45\textwidth}
        \centering
        \begin{overpic}[width=1.1\textwidth,tics=10]{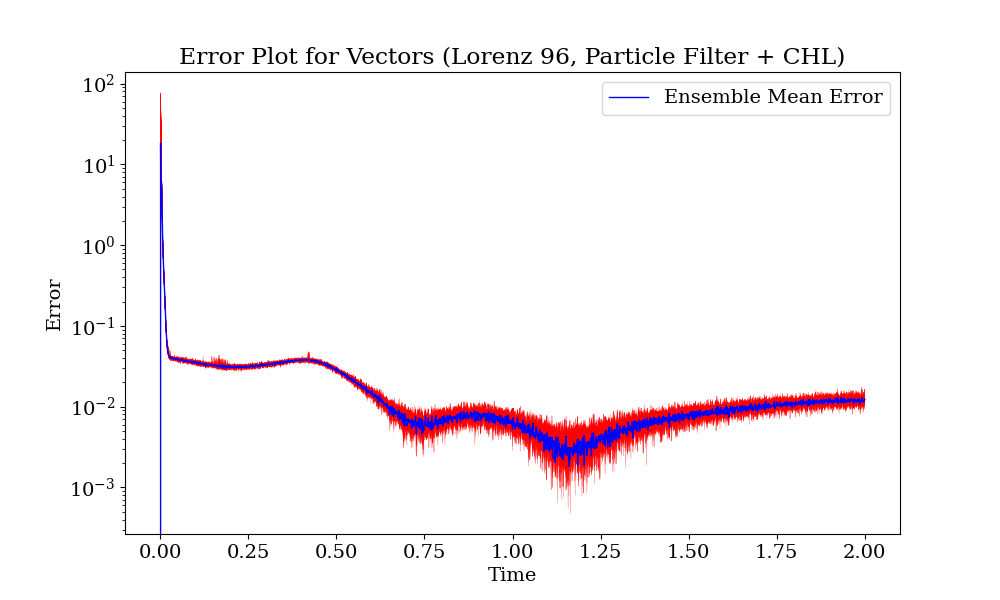}
        \put(59,46){\fontsize{2pt}{4pt}\selectfont\scalebox{0.6}[0.72]{\textcolor{red}{---} Ensemble Individual Errors}}
         \put(58,46){%
    \color{gray!30}%
    \fbox{\phantom{\fontsize{2pt}{4pt}\selectfont\scalebox{0.55}[0.2]{\textcolor{red}{---} Ensemble Individual Errors}}}}
        \end{overpic}
        \caption{PF + CHL DA Error (96)}
        \label{fig:subfig3}
    \end{subfigure}
    \begin{subfigure}{0.45\textwidth}
        \centering
        \begin{overpic}[width=1\textwidth,tics=10]{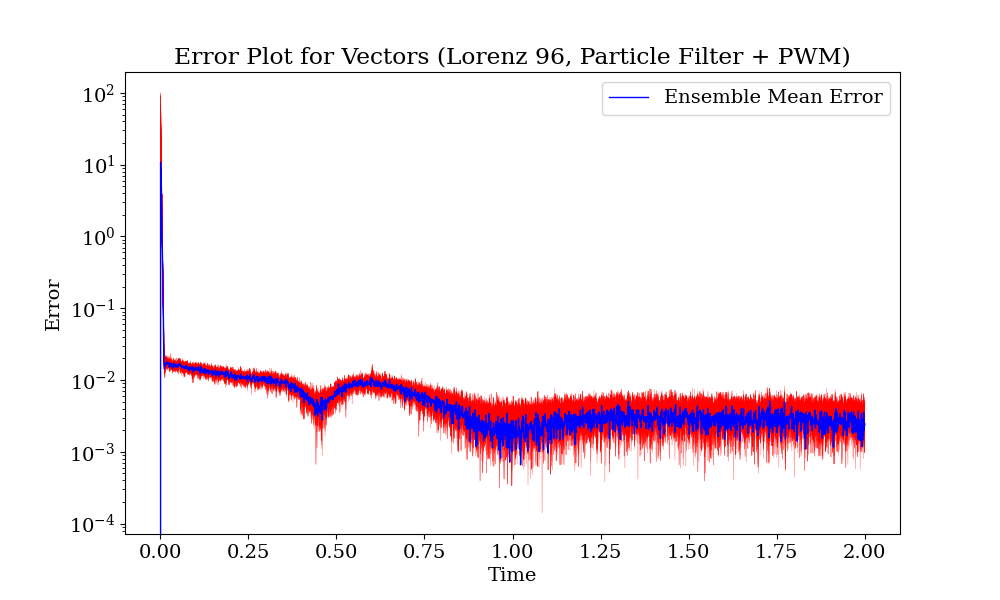}
        \put(59,46){\fontsize{2pt}{4pt}\selectfont\scalebox{0.6}[0.72]{\textcolor{red}{---} Ensemble Individual Errors}}
         \put(58,46){%
    \color{gray!30}%
    \fbox{\phantom{\fontsize{2pt}{4pt}\selectfont\scalebox{0.55}[0.2]{\textcolor{red}{---} Ensemble Individual Errors}}}}
        \end{overpic}
        \caption{PF + PWM DA Error (96)}
        \label{fig:subfig4}
    \end{subfigure}
    \caption{State estimation when using stochastic PF DA method at $SD = 10^{-6}$.}
    \label{fig:PF_quad_state}
\end{figure}

Comparing purely deterministic (AOT + CHL, AOT + PWM, AOT + AV) and purely stochastic methods (EnKF + EM, EnKI, PF + EM, PF Extension, ETKF + EM), we see from Table~\ref{PR_10-6_table} that purely deterministic methods are very computationally efficient and from Tables~\ref{table:96_PR_mean} and \ref{table:96_PR_variance} that they are more accurate and stable at lower noise levels in PR (see also Figures~\ref{fig:purely_deterministic_PR}, \ref{fig:EnKF_w_PR}, \ref{fig:EnKI_PR_96}, and \ref{fig:PF_PR_quad}). As observation noise increases, AOT + PWM and AOT + AV remain accurate in PR while instability starts to arise, which aligns with results in Section~\ref{sec:63_results}. ETKF + EM demonstrates better performance than EnKF + EM and PF + EM overall, which aligns with the fact that ETKF is a state of the art algorithm \cite{soaLETKF}. Moreover, using ETKF with CHL or PWM in the mix-and-match scenarios has helped the PR algorithms return more robust results with higher observation noise levels. State estimation of purely deterministic methods remain both more accurate and stable across all noise levels (see Tables~\ref{table:96_state_mean} and \ref{table:96_state_variance}, and Figures~\ref{fig:AOT_quad_state}, \ref{fig:EnKF_tri_state}, and \ref{fig:PF_quad_state}). Among the the purely deterministic methods, AOT + PWM and AOT + AV significantly outperform AOT + CHL at most times, as predicted in \cite{Newey_Whitehead_Carlson_2025,Pachev_Whitehead_McQuarrie_2021concurrent}.  
Purely stochastic methods have exhibited steady performance across various noise levels even though they are less accurate.  For ease of interpretation, the smallest values in each column of Tables~\ref{table:63_PR_mean}, \ref{table:63_PR_variance}, \ref{table:63_state_mean}, and \ref{table:63_state_variance} are highlighted in blue.

For methods involving CHL PR, results sometimes converge on the whole but encounter occasional large fluctuations in the latter half of the observed time frame, leading to large error mean and variance values. We point out that having large error means and variances is not equivalent to being non-convergent, which also happens and we denote this case with ``NC'' in the tables.
\begin{table}[h!]
\centering
\caption*{\textbf{Runtime for All Methods}}

\begin{tabular}{| l | c | c |}
\hline
\textbf{DA Method} & \textbf{PR Method} & \textbf{Runtime}\\
\hline \hline
\multirow{4}{*}{AOT} 
 & CHL & $\sim$ 10 seconds\\  \cline{2-3}
 & PWM & $\sim$ 10 seconds\\  \cline{2-3}
 & AV & $\sim$ 10 seconds\\ \cline{2-3}
 & EM & {$\sim$ over 10 hours}\\ \hline
\multirow{3}{*}{EnKF} 
 & EM & $\sim$ 8 hours \\  \cline{2-3}
 & CHL & $\sim$ 5 minutes\\ \cline{2-3}
 & PWM & $\sim$ 5 minutes\\ \hline
N/A & EnKI & {$\sim$ 5 minutes}\\ \hline
\multirow{4}{*}{PF} 
 & EM & {$\sim$ over 10 hours}\\ \cline{2-3}
 & Ext. & {$\sim$ 5 minutes}\\ \cline{2-3}
 & CHL & $\sim$ 5 minutes \\ \cline{2-3}
 & PWM & $\sim$ 5 minutes\\ \hline
 \multirow{4}{*}{ETKF}
 & EM & $\sim$ 20 hours \\ \cline{2-3}
 & CHL & $\sim$ 2 minutes\\ \cline{2-3}
 & PWM & $\sim$ 3 minutes\\ \cline{2-3}
 & AV & $\sim$ 2 minutes\\ \hline
\end{tabular}
\caption{Runtime for Different Methods}
\label{PR_10-6_table}
\end{table}

\begin{table}[h!]
\centering
\caption*{\textbf{Parameter Estimation Error Means}}
\adjustbox{margin=0pt,center}{
\begin{tabular}{| l | c | c | c | c | c | c | c |}
\hline
\textbf{DA Method} & \textbf{PR Method} & $SD = 10^{-6}$ & $SD = 10^{-5}$ & $SD = 10^{-4}$ & $SD = 10^{-3}$ & $SD = 10^{-2}$ & $SD = 10^{-1}$ \\ \hline \hline 
\multirow{4}{*}{AOT}
& CHL & \textcolor{blue}{0.00158} & 0.00375 & 0.02805 & 0.25430 & 2.46536 & 11.95500\\ \cline{2-8}
& PWM & 0.00721 & \textcolor{blue}{0.00352} & \textcolor{blue}{0.00755} & \textcolor{blue}{0.00774} & \textcolor{blue}{0.02401} & 0.28916\\ \cline{2-8}
& AV & 0.00384 & 0.00525 & 0.00932 & 0.01725 & 0.28979 & 2.14344\\ \cline{2-8}
& EM & 2.35773 & 2.26555& 2.57697 & 2.27540 & 2.56691 & 0.85120\\ \hline

\multirow{3}{*}{EnKF}
& EM & 0.20414 & 0.29894 & 0.31457 & 0.59823 & 0.52580 & 0.87289\\ \cline{2-8}
& CHL & 2.09508 & 0.66855 & 6.11502 & 5.69854 & NC & NC\\ \cline{2-8}
& PWM & 0.94121 & 1.14847 & 1.09894 & 1.18271 & 1.12574& 3.63515\\ \hline
N/A & EnKI & 0.04416 & 0.03626 & 0.04613 & 0.02869 & 0.03000 & \textcolor{blue}{0.05702}\\ \hline
\multirow{4}{*}{PF}
& EM & 0.12039 & 0.04938 & 0.46484 & NC & NC & NC \\ \cline{2-8}
& Ext. & 0.02662 & 0.02844 & 0.02080 & 0.04331 & 0.05003& 0.00346\\ \cline{2-8}
& CHL & 3.37748 & 1.91984 & 2.05639 & NC & NC & NC \\ \cline{2-8}
& PWM & 1.04544 & 1.03405 & 1.21256 & 1.22203 & 4.30047  & 3.58013\\ \hline
\multirow{4}{*}{ETKF} 
& EM & 0.27481 & 0.03451 & 0.08415 & 0.21504 & 0.04876 & 0.27952 \\ \cline{2-8}
& CHL & 8.42592 & 3.85736 & 7.81056 & 48.85206 & 1884.32046 & NC \\ \cline{2-8}
& PWM & 0.99678 & 0.99145 & 1.01684 & 1.09435 & 1.20638 & 3.82961 \\ \cline{2-8}
& AV & 1.18947 & 1.02410 & 1.05474 & 1.14705 & 1.13688 & 2.42450 \\ \hline
\end{tabular}
}
\caption{PR error means for all methods applied to Lorenz '96.}
\label{table:96_PR_mean}
\end{table}

\begin{table}[h!]
\centering
\caption*{\textbf{Parameter Estimation Error Variances}}
\adjustbox{margin=0pt,center}{
\begin{tabular}{| l | c | c | c | c | c | c | c |}
\hline
\textbf{DA Method} & \textbf{PR Method} & $SD = 10^{-6}$ & $SD = 10^{-5}$ & $SD = 10^{-4}$ & $SD = 10^{-3}$ & $SD = 10^{-2}$ & $SD = 10^{-1}$  \\ \hline \hline 
\multirow{4}{*}{AOT}
& CHL & 0.00005 & 0.00028 & 0.00893 & 1.59218 & 67.74932 & 89.14780 \\ \cline{2-8}
& PWM & 0.00142 & 0.00006 & 0.00273 & 0.00230 & 0.03098 & 1.13799\\ \cline{2-8}
& AV & \textcolor{blue}{0.00001} & \textcolor{blue}{0.00002} & \textcolor{blue}{0.00010} & \textcolor{blue}{0.00020} & 0.06613 & 2.97228\\ \cline{2-8}
& EM & 0.00894 & 0.00857 & 0.01322 & 0.00427 & 0.01283 & 0.00115\\ \hline

\multirow{3}{*}{EnKF}
& EM & 0.00040 & 0.00034 & 0.00037 & 0.00062 & \textcolor{blue}{0.00017} & 0.00193\\ \cline{2-8}
& CHL & 0.20153 & 6.94161& 958.46447 & 199.54083 & NC & NC\\ \cline{2-8}
& PWM & 0.00076 & 0.00233& 0.03779 & 0.08113 & 10.17614 & 682.74123\\ \hline
N/A & EnKI & 0.00008 & 0.00018 & 0.00015 & 0.00012 & 0.00013& 0.00148\\ \hline
\multirow{4}{*}{PF}
& EM & 0.00035 & 0.00049 & 0.00056 & NC & NC & NC\\ \cline{2-8}
& Ext. & 0.33438 & 0.34067 & 0.34984 & 0.33960 & 0.33893& 0.34023\\ \cline{2-8}
& CHL & 2.63853 & 548.23864 & 461.14694 & NC & NC & NC\\ \cline{2-8}
& PWM & 0.00198 & 0.01443 & 4.83808 & 41.29463 & 3219.72773 & 1239.63907\\ \hline
\multirow{4}{*}{ETKF} 
& EM & 0.00038 & 0.00040 & 0.00030 & 0.00039 & 0.00019 & \textcolor{blue}{0.00022} \\ \cline{2-8}
& CHL & 5340.93987 & 3450.25497 & 1140.96265 & 25255.98759 & $\sim 10^{7}$ & NC \\ \cline{2-8}
& PWM & 0.00147 & 0.00760 & 0.01429 & 1.20473 & 2.11753 & 96.58581 \\ \cline{2-8}
& AV & 0.00203 & 0.00287 & 0.00762 & 0.09489 & 0.49732 & 2.87943 \\ \hline
\end{tabular}
}
\caption{PR error variances for all methods applied to Lorenz '96.}
\label{table:96_PR_variance}
\end{table}

\begin{table}[h!]
\centering
\caption*{\textbf{State Estimation Error Means}}
\adjustbox{margin=0pt,center}{
\begin{tabular}{| l | c | c | c | c | c | c | c |}
\hline
\textbf{DA Method} & \textbf{PR Method} & $SD = 10^{-6}$ & $SD = 10^{-5}$ & $SD = 10^{-4}$ & $SD = 10^{-3}$ & $SD = 10^{-2}$ & $SD = 10^{-1}$  \\ \hline \hline 
\multirow{4}{*}{AOT}
& CHL &\textcolor{blue}{0.00005} & \textcolor{blue}{0.00005} & \textcolor{blue}{0.00011} & \textcolor{blue}{0.00073} & 0.00742 & 0.07565\\ \cline{2-8}
& PWM &0.00024& 0.00022 & 0.00026 & 0.00084 & \textcolor{blue}{0.00715} & \textcolor{blue}{0.07390} \\ \cline{2-8}
& AV &0.00030& 0.00048 & 0.00074 & 0.00098 & 0.00767 & 0.07684\\ \cline{2-8}
& EM &0.31189&  0.36731  & 0.32784 & 0.28872 & 0.44669 & 0.34503\\ \hline

\multirow{3}{*}{EnKF}
& EM &0.00337& 0.01223 &0.04844 & 0.16142 & 0.55868 & 1.52472\\ \cline{2-8}
& CHL &0.00687& 0.00877 & 0.03556 & 0.09386 & NC & NC  \\ \cline{2-8}
& PWM &0.00178& 0.00727 & 0.02668 & 0.11445 & 0.31052& 0.75937\\ \hline
N/A & EnKI &N/A& N/A & N/A& N/A&N/A & N/A\\ \hline
\multirow{4}{*}{PF}
& EM &0.00494& 0.05395 & 0.52987 & NC & NC & NC\\ \cline{2-8}
& Ext. &0.00396& 0.00578 & 0.01186 & 0.03467 & 0.10801 & 0.34086\\ \cline{2-8}
& CHL &0.00764&  0.02154 & 0.21863 & NC & NC & NC\\ \cline{2-8}
& PWM &0.00276& 0.02003 & 0.16712 & 1.03525 & 0.85116 &  0.73623\\ \hline
\multirow{4}{*}{ETKF} 
& EM & 0.12045 & 0.15003 & 0.12814 & 0.13104 & 0.13499 & 0.19725 \\ \cline{2-8}
& CHL & 0.00683 & 0.00999 & 0.02226 & 0.05639 & 0.50238 & NC \\ \cline{2-8}
& PWM & 0.00217 & 0.00650 & 0.01882 & 0.02974 & 0.04676 & 0.13521 \\ \cline{2-8}
& AV & 0.00186 & 0.00863 & 0.01710 & 0.03146 & 0.05276 & 0.13762 \\ \hline
\end{tabular}
}
\caption{State estimation error means for all methods applied to Lorenz '96.}
\label{table:96_state_mean}
\end{table}

\begin{table}[h]
\centering
\caption*{\textbf{State Estimation Error Variances}}
\adjustbox{margin=0pt,center}{
\begin{tabular}{| l | c | c | c | c | c | c | c |}
\hline
\textbf{DA Method} & \textbf{PR Method} & $SD = 10^{-6}$ & $SD = 10^{-5}$ & $SD = 10^{-4}$ & $SD = 10^{-3}$ & $SD = 10^{-2}$ & $SD = 10^{-1}$  \\ \hline \hline 
\multirow{4}{*}{AOT}
& CHL & \textcolor{blue}{$1.79 \times 10^{-9}$} & \textcolor{blue}{1.63 $\times 10^{-9}$} & \textcolor{blue}{5.73 $\times 10^{-9}$} & 3.11 $\times 10^{-7}$ & \textcolor{blue}{0.00003} & 0.00326\\ \cline{2-8}
& PWM & $4.30 \times 10^{-8}$ & 2.96 $\times 10^{-8}$ & 3.37 $\times 10^{-8}$ & \textcolor{blue}{2.76 $\times 10^{-7}$} & \textcolor{blue}{0.00003} & 0.00294\\ \cline{2-8}
& AV & $6.00 \times 10^{-8}$ & 1.79 $\times 10^{-7}$ & 3.43 $\times 10^{-7}$ & 3.02 $\times 10^{-7}$ & \textcolor{blue}{0.00003} & 0.00299\\ \cline{2-8}
& EM & 0.01530 & 0.03242 & 0.01844 & 0.02128 & 0.04221 & 0.01454\\ \hline

\multirow{3}{*}{EnKF}
& EM & $1.20 \times 10^{-6}$ & 0.00001 & 0.00014& 0.00147 & 0.02158 & 0.36490\\ \cline{2-8}
& CHL & $7.40 \times 10^{-6}$ & 4.09 $\times 10^{-6}$ & 0.00010 & 0.00017 & NC & NC \\ \cline{2-8}
& PWM & $9.24 \times 10^{-8}$ & 3.32 $\times 10^{-6}$ & 0.00003 & 0.00033 & 0.00089 &  0.02925 \\ \hline
N/A & EnKI &N/A& N/A &N/A &N/A & N/A& N/A\\ \hline
\multirow{4}{*}{PF}
& EM & $1.90 \times 10^{-6}$ & 0.00020 & 0.02435 & NC & NC & NC \\ \cline{2-8}
& Ext. & $2.22 \times 10^{-6}$ & 4.59 $\times 10^{-6}$ & 0.00002 & 0.00016 &  0.00149 & 0.01682\\ \cline{2-8}
& CHL & 0.00001 & 0.00002 & 0.00459 & NC & NC & NC \\ \cline{2-8}
& PWM & $4.40 \times 10^{-7}$ & 0.00002 & 0.00309 & 0.16263 & 0.13162 & 0.08138\\ \hline
\multirow{4}{*}{ETKF} 
& EM & 0.01029 & 0.00457 & 0.00675 & 0.00397 & 0.00981 & 0.00818 \\ \cline{2-8}
& CHL & 0.00001 & 0.00001 & 0.00004 & 0.00054 & 107.08048 & NC \\ \cline{2-8}
& PWM & 4.17 $\times 10^{-7}$ & 2.16 $\times 10^{-6}$ & 0.00001 & 0.00005 & 0.00033 & \textcolor{blue}{0.00259} \\ \cline{2-8}
& AV & $3.88 \times 10^{-7}$ & 0.00001 & 0.00001 & 0.00005 & 0.00029 & 0.00276 \\ \hline
\end{tabular}
}
\caption{State estimation error variances for all methods applied to Lorenz '96.}
\label{table:96_state_variance}
\end{table}

\begin{figure}[h!]
    \centering
    \begin{subfigure}{0.48\textwidth}
        \centering
        \includegraphics[width=1\textwidth]{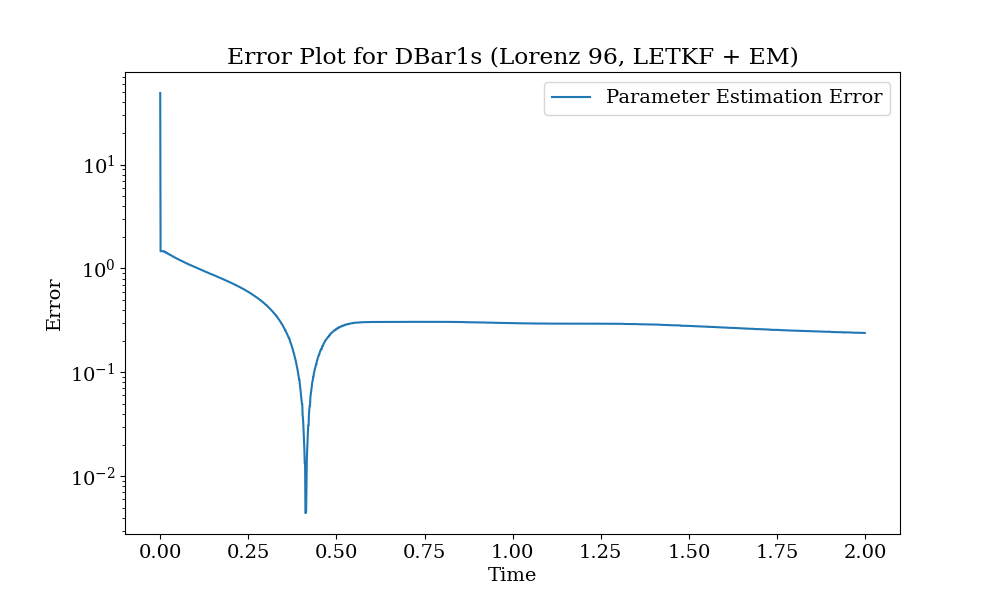} 
        \caption{ETKF + EM PR Error (96)}
        \label{fig:sub1}
    \end{subfigure}
    \begin{subfigure}{0.48\textwidth}
        \centering
        \includegraphics[width=1\textwidth]{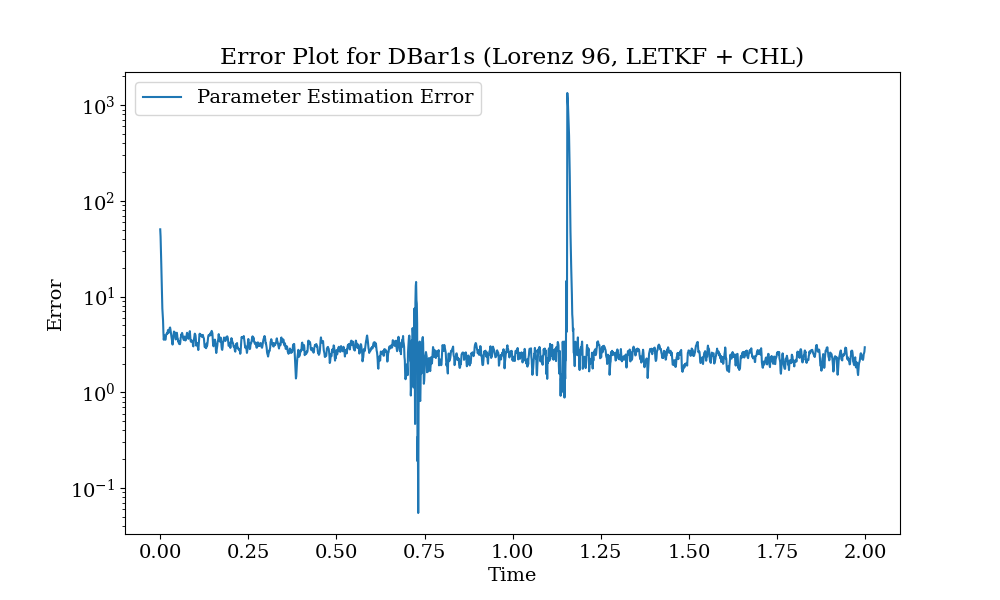} 
        \caption{ETKF + CHL PR Error (96)}
        \label{fig:sub2}
    \end{subfigure}
        \vskip\baselineskip
    \begin{subfigure}{0.48\textwidth}
        \centering
        \includegraphics[width=\textwidth]{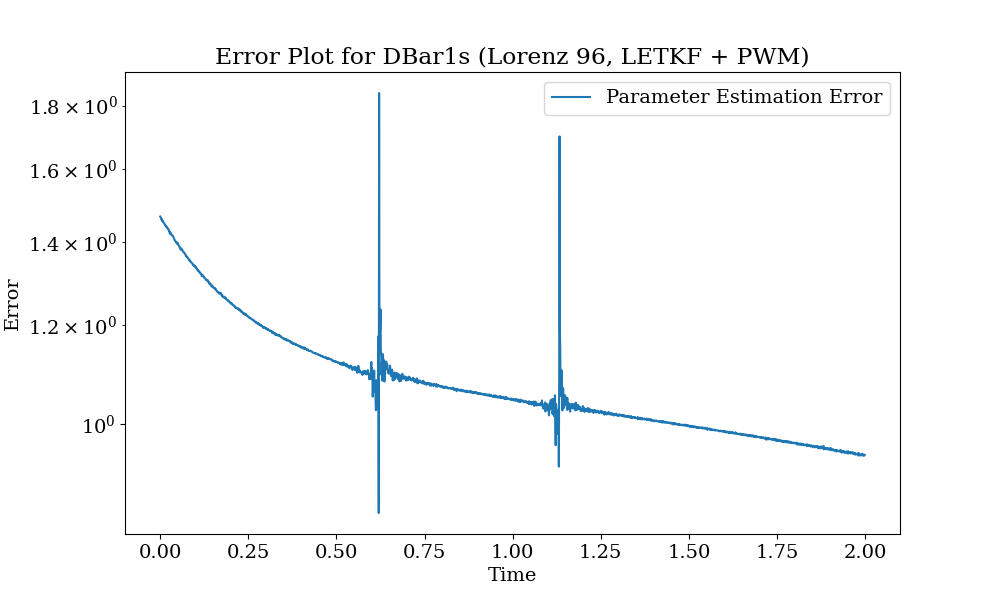} 
        \caption{ETKF + PWM PR Error (96)}
        \label{fig:sub3}
    \end{subfigure}
    \begin{subfigure}{0.48\textwidth}
        \centering
        \includegraphics[width=\textwidth]{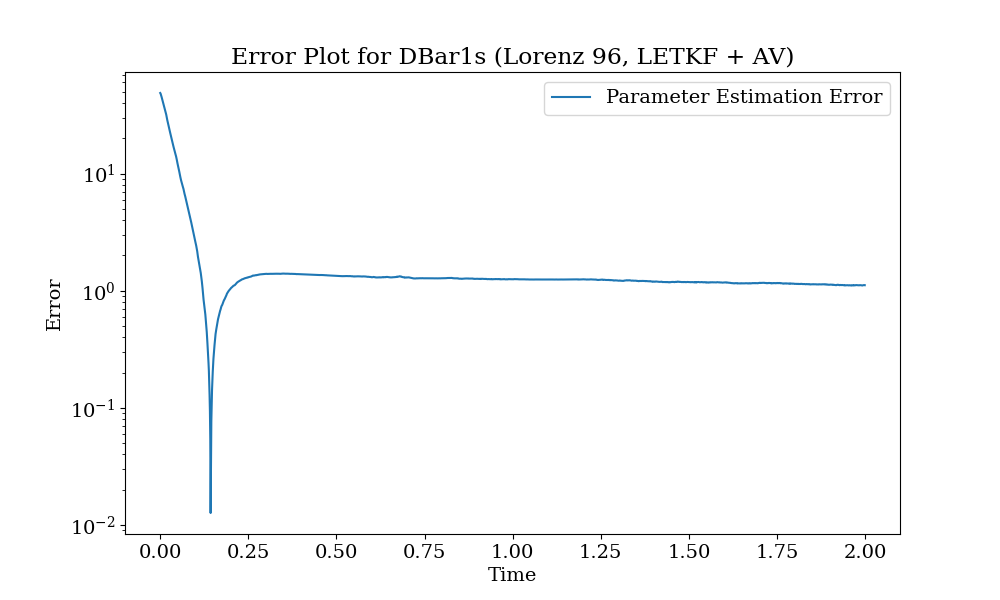} 
        \caption{ETKF + AV PR Error (96)}
        \label{fig:sub4}
    \end{subfigure}
    \caption{Different PR methods paired with stochastic ETKF DA method at $SD = 10^{-6}$.}
    \label{fig:ETKF_w_PR}
\end{figure}





\begin{figure}[h!]
    \centering
    \begin{subfigure}{0.48\textwidth}
        \centering
        \includegraphics[width=1\textwidth]{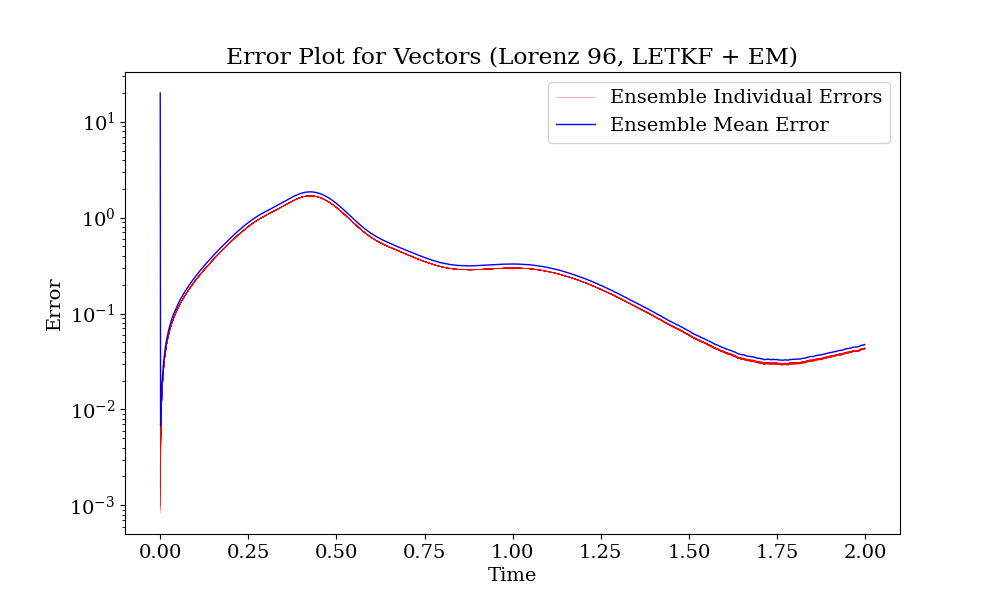}
        \caption{ETKF + EM DA Error (96)}
        \label{fig:s1}
    \end{subfigure}
    \begin{subfigure}{0.48\textwidth}
        \centering
        \includegraphics[width=1\textwidth]{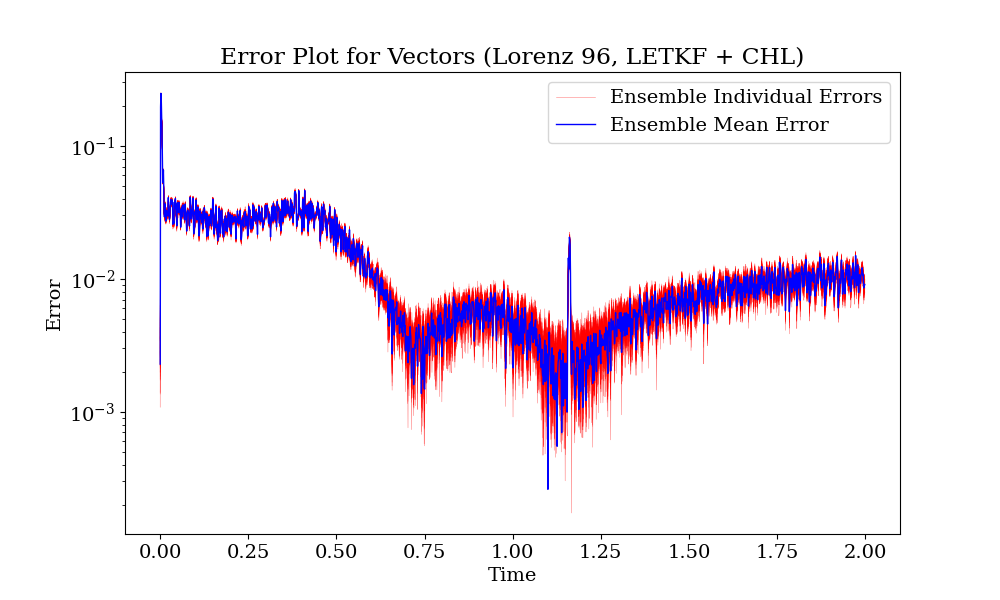} 
        \caption{ETKF + CHL DA Error (96)}
        \label{fig:s2}
    \end{subfigure}
        \vskip\baselineskip
    \begin{subfigure}{0.48\textwidth}
        \centering
        \includegraphics[width=\textwidth]{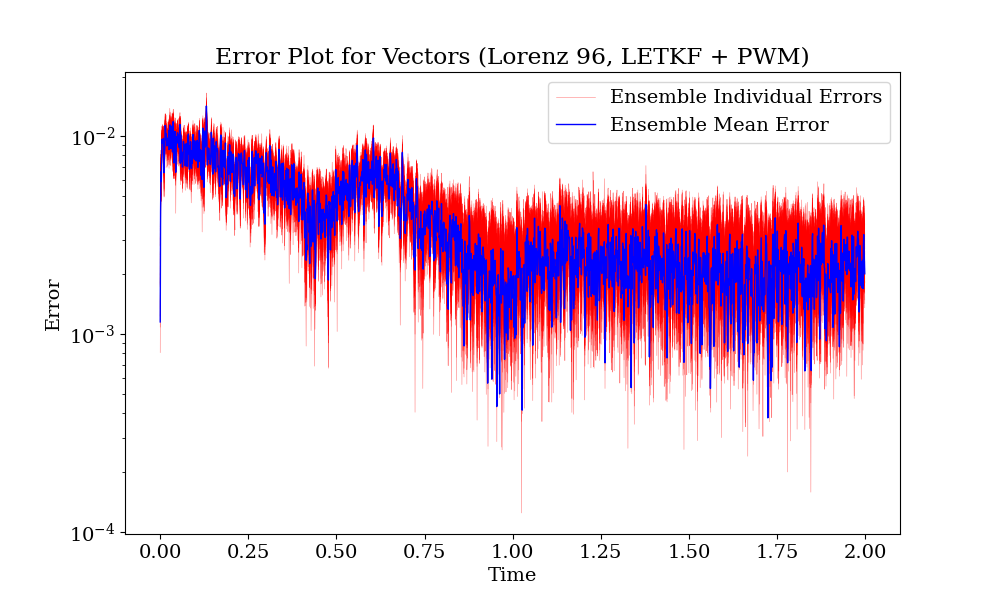} 
        \caption{ETKF + PWM DA Error (96)}
        \label{fig:s3}
    \end{subfigure}
    \begin{subfigure}{0.48\textwidth}
        \centering
        \includegraphics[width=\textwidth]{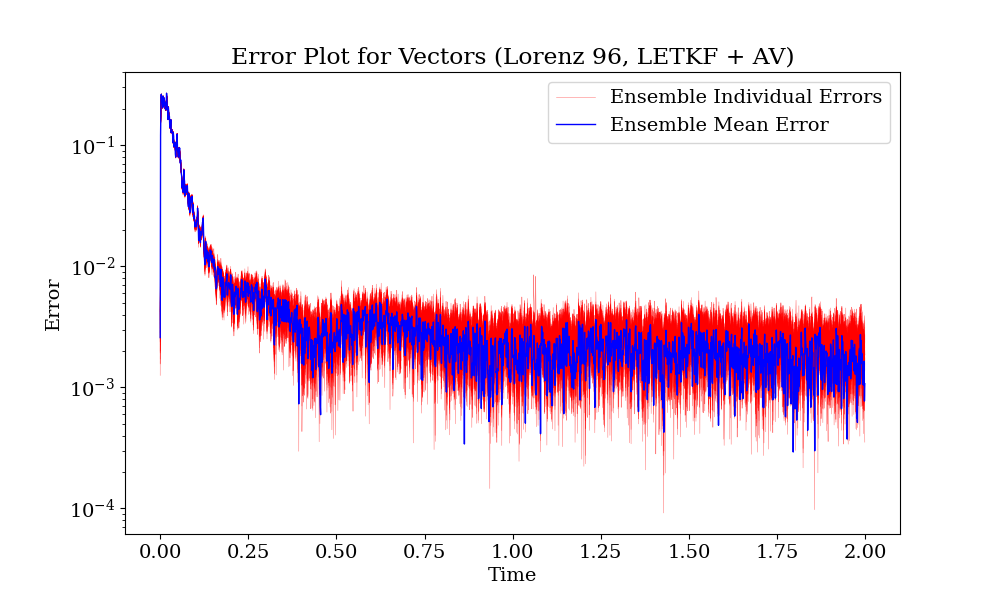} 
        \caption{ETKF + AV DA Error (96)}
        \label{fig:s4}
    \end{subfigure}
    \caption{State estimation when using stochastic ETKF DA method at $SD = 10^{-6}$.}
    \label{fig:ETKF_state}
\end{figure}






\clearpage
\subsection{Comparing Recovery Speed and Threshold}

Table \ref{levels_table} lists how long it approximately takes the PR algorithms to reach the minimum error they can achieve (in our theoretical time domain), and the value for the parameter error mean.  Lorenz '96 is a larger system with greater computational expense, whereas Lorenz '63 converges quickly.  Consequently, we only report the recovery speed and threshold results here for the Lorenz '96 system.

\begin{table}[h!]
\centering
\caption*{\textbf{Rate and Level of Convergence in Parameter Recovery}}

\begin{tabular}{| l | c | c | c |}
\hline
\textbf{DA Method} & \textbf{PR Method} & \textbf{Time (Approximation)} & \textbf{Level of Error}\\
\hline \hline
\multirow{4}{*}{AOT} 
 & CHL & 1.0 & $10^{-3}$\\  \cline{2-4}
 & PWM & 1.0 & $10^{-3}$\\  \cline{2-4}
 & AV & 1.0 & $10^{-3}$ \\ \cline{2-4}
 & EM & 0.5 & $10^{0}$\\ \hline
\multirow{3}{*}{EnKF} 
 & EM & 0.5 & $10^{-1}$\\  \cline{2-4}
 & CHL & $\sim 0$ & $10^0$\\ \cline{2-4}
 & PWM & 0.6 & $10^0$\\ \hline
N/A & EnKI & 0.2 & $10^{-1}$\\ \hline
\multirow{4}{*}{PF} 
 & EM & 0.5 & $10^{-1}$\\ \cline{2-4}
 & Ext. & $\sim$ 0 & $10^{0}$\\ \cline{2-4}
 & CHL & $\sim$ 0 & 5\\ \cline{2-4}
 & PWM & $\sim$ 0 & $10^{0}$\\ \hline
 \multirow{4}{*}{ETKF}
 & EM & 0.5 & $10^{-1}$\\ \cline{2-4}
 & CHL & $\sim 0$ & 5\\ \cline{2-4}
 & PWM & $\sim 0$ & $10^{0}$\\ \cline{2-4}
 & AV & 0.25 & $10^{0}$ \\ \hline
\end{tabular}
\caption{All methods applied to Lorenz '96 with $10^{-6}$ noise level.}
\label{levels_table}
\end{table}

The results in Table~\ref{levels_table} suggest that the quickest algorithms to get to a steady level of error are combinations of stochastic DA methods and deterministic PR methods.  The algorithms with the lowest level of error are the deterministic DA methods with the deterministic PR methods.

\clearpage

\subsection{Stability of CHL with Wider Parameter Recovery Windows}
\label{sec:wide_PR_window}
Let $q$ denote the number of numerical time steps between every parameter update, referred to as the parameter recovery window. The DA numerical time steps are kept the same as before. The sets of experiments below are ordered in descending order by $\mu_{4}$, followed by $\mu_{3}$, $\mu_{2}$, and $\mu_{1}$. We only record results that converge to the true parameter value and either improve or maintain the stability of CHL. Recall that the true parameter value is given in Equation~\ref{d1_value}, which gives us $\bar{d_{1}}\approx 1.216$.

Results indicate that widening the PR window and lowering $\mu$ simultaneously helps with stabilizing CHL. A good example is in the last row of Table~\ref{table:aot_chl_window}, although state estimation accuracy and stability are less than ideal. We observe that state estimation error mean and variance remain steady in general, except in the third-to-last row. In Table~\ref{table:EnKF_CHL_window}, we see that similar manipulations help us achieve convergent results, which is already a huge improvement (since they were non-convergent in Section~\ref{sec:96_results}). The positive effect of similar manipulations is most obvious in the case of PF + CHL (see Table~\ref{table:PF_CHL_window}). 
\begin{table}[h!]
\centering
\caption*{\textbf{AOT + CHL Error Statistics at Noise Level $\mathbf{10^{-1}}$}}
\begin{tabular}{| l | c | c | c | c | c |}
\hline
$\mu$ Values & PR Window & PR Mean & PR Variance & State Mean & State Variance\\
\hline \hline
$\mu_{1} = 400, \mu_{2} = 50, \mu_{3} = 50, \mu_{4} = 0$ & $q = 10$ & 10.71439 & 202.38818 & 0.10030 & 0.00585\\ \hline 
$\mu_{1} = 300, \mu_{2} = 50, \mu_{3} = 50, \mu_{4} = 0$ & $q = 10$ & 22.60590 & 3949.72521 & 0.14400 & 0.03631\\ \hline
$\mu_{1} = 400, \mu_{2} = 20, \mu_{3} = 20, \mu_{4} = 0$ & $q = 20$ & 12.25921 & 190.96479 &  0.17331 & 0.01448\\ \hline 
$\mu_{1} = 300, \mu_{2} = 20, \mu_{3} = 20, \mu_{4} = 0$ & $q = 20$ & 16.00945 & 494.15646 & 1.09362 & 2.15663\\ \hline 
$\mu_{1} = 100, \mu_{2} = 20, \mu_{3} = 20, \mu_{4} = 0$ & $q = 30$ & 9.83016 & 229.08173 & 0.20275 & 0.02405\\ \hline 
$\mu_{1} = 100, \mu_{2} = 10, \mu_{3} = 10, \mu_{4} = 0$ & $q = 30$ & 7.56134 & 77.11326 & 0.21530 & 0.02439\\ \hline 

\end{tabular}
\caption{Various experiments when applying AOT + CHL to Lorenz '96.}
\label{table:aot_chl_window}
\end{table}

\begin{table}[h!]
\centering
\caption*{\textbf{EnKF + CHL Error Statistics at Noise Level $\mathbf{10^{-1}}$}}
\begin{tabular}{| l | c | c | c | c | c |}
\hline
$\mu$ Values & PR Window & PR Mean & PR Variance & State Mean & State Variance\\
\hline \hline
$\mu_{1} = 50$ & $q = 10$ & 3.58960 & 22.99186 & 0.60803 & 0.00631\\ \hline
$\mu_{1} = 30$ & $q = 10$ & 2.88188 & 28.41333 & 0.58241 & 0.03770\\ \hline
$\mu_{1} = 10$ & $q = 10$ & 3.05892 & 15.45746 & 1.09511 & 0.03056\\ \hline 
\end{tabular}
\caption{Various experiments when applying EnKF + CHL to Lorenz '96.}
\label{table:EnKF_CHL_window}
\end{table}

\begin{table}[h!]
\centering
\caption*{\textbf{PF + CHL Error Statistics at Noise Level $\mathbf{10^{-1}}$}}
\begin{tabular}{| l | c | c | c | c | c |}
\hline
$\mu$ Values & PR Window & PR Mean & PR Variance & State Mean & State Variance\\
\hline \hline
$\mu_{1} = 100, \mu_{2} = 10, \mu_{3} = 10, \mu_{4} = 0$ & $q = 30$ & 18.03554 & 1030.29241 & 1.54111 & 0.22214\\ \hline 
$\mu_{1} = 100, \mu_{2} = 10, \mu_{3} = 10, \mu_{4} = 0$ & $q = 50$ & 10.56087 & 278.04718 & 1.35102 & 0.16071 \\ \hline 
$\mu_{1} = 10, \mu_{2} = 10, \mu_{3} = 10, \mu_{4} = 0$ & $q = 100$ & 0.36383 & 1.28153 & 0.85863 & 0.09172 \\ \hline
\end{tabular}
\caption{Various experiments when applying PF + CHL to Lorenz '96.}
\label{table:PF_CHL_window}
\end{table}

\subsection{State Estimation Performance without Parameter Recovery}
\label{sec:vanilla_DA}
In this section, we provide a state estimation performance comparison between AOT, EnKF, PF model (original PF), PF Extension, PF w/ AOT (PF method adapted to deterministic PR), and ETKF with no PR involved, where $\overline{d_{1}}$ is always the true value as given in \eqref{d1_value}. Overall, the runtime of the DA methods is not influenced by the magnitude of noise. AOT is much faster than the stochastic DA methods, and among the stochastic methods, EnKF outperforms the three PF DA methods, as can be seen in Table~\ref{table:vanilla_da_1}. \textbf{By closely examining Tables \ref{table:vanilla_da_2} and \ref{table:vanilla_da_3}, we see that the deterministic AOT DA has both higher accuracy and stability than all stochastic methods at all noise levels. ETKF is the fastest among the stochastic DA methods, and remains a strong option as observation noise becomes large. }
\begin{table}[h!]
\centering
\caption*{\textbf{Runtime of DA Methods (in seconds)}}

\begin{tabular}{| l | c | c | c | c | c | c |}
\hline
\textbf{DA Method} & $SD = 10^{-6}$ & $SD = 10^{-5}$ &  $SD = 10^{-4}$ & $SD = 10^{-3}$ & $SD = 10^{-2}$ & $SD = 10^{-1}$\\
\hline \hline
AOT & 8.5 & 9.8 & 10.6 & 8.3 & 9.7& 9.7\\ \hline
EnKF & 194.6 & 240.2 & 221.2 & 220.1 & 207.1 & 220.3\\ \hline
PF Model & 251.6 & 269.3 & 235.6 & 261.5 & 244.6 &  252.0\\ \hline
PF Extension & 261.6 & 245.9 & 267.9 & 247.7 & 243.1 & 247.9\\ \hline
PF w/ AOT & 268.2 & 271.4 & 248.2 & 272.9 & 257.4 & 250.0\\ \hline
ETKF & 144.1 & 141.8 & 143.4 & 147.6 & 143.5 & 144.6\\ \hline
\end{tabular}
\caption{Runtime of vanilla DA methods at different observation noise levels.}
\label{table:vanilla_da_1}
\end{table}

\begin{table}[h!]
\centering
\caption*{\textbf{State Estimation Error Means for Different DA Methods}}

\begin{tabular}{| l | c | c | c | c | c | c |}
\hline
\textbf{DA Method} & $SD = 10^{-6}$ & $SD = 10^{-5}$ &  $SD = 10^{-4}$ & $SD = 10^{-3}$ & $SD = 10^{-2}$ & $SD = 10^{-1}$\\
\hline \hline
AOT & 0.00014 & 0.00020 & 0.00023 & 0.00075 & 0.00682 & 0.07248\\ \hline
EnKF & 0.00292 & 0.01253 & 0.04272 & 0.15918 & 0.42868 & 0.67619\\ \hline
PF Model & 0.00507 & 0.05023 & 0.45288 & 4.64569 & 7.05398 & 6.51156\\ \hline
PF Extension & 0.00447 & 0.00533 & 0.01171 & 0.03475 & 0.10585 & 0.34425\\ \hline
PF w/ AOT & 0.00303 & 0.02079 & 0.23976 & 0.79650 & 1.20813 & 1.39112\\ \hline
ETKF & 0.10877 & 0.17696 & 0.10488 & 0.11174 & 0.14850 & 0.21743\\ \hline
\end{tabular}
\caption{State estimation error means of vanilla DA methods at different observation noise levels.}
\label{table:vanilla_da_2}
\end{table}

\begin{table}[h!]
\centering
\caption*{\textbf{State Estimation Error Variances for Different DA Methods}}

\begin{tabular}{| l | c | c | c | c | c | c |}
\hline
\textbf{DA Method} & $SD = 10^{-6}$ & $SD = 10^{-5}$ &  $SD = 10^{-4}$ & $SD = 10^{-3}$ & $SD = 10^{-2}$ & $SD = 10^{-1}$\\
\hline \hline
AOT & 1.93 $\times 10^{-8}$ & 2.73 $\times 10^{-8}$ & 3.44 $\times 10^{-8}$ & 2.87 $\times 10^{-7}$ & 2.57 $\times 10^{-5}$ & 0.00276\\ \hline
EnKF & 7.37 $\times 10^{-7}$ & 0.00001 & 0.00006 & 0.00110 & 0.00898 & 0.01231\\ \hline
PF Model & 1.90 $\times 10^{-6}$ & 0.00016 & 0.01587 & 1.28102 & 2.59803 & 5.60876\\ \hline
PF Extension & 1.35 $\times 10^{-6}$ & 2.87 $\times 10^{-6}$ & 0.00002 & 0.00016 & 0.00165 & 0.01552\\ \hline
PF w/ AOT & 4.84 $\times 10^{-7}$ & 0.00002 & 0.00610 & 0.09400 & 0.25083 & 0.29242 \\ \hline
ETKF & 0.00546 & 0.01368 & 0.00571 & 0.00631 & 0.01022 & 0.00936\\ \hline
\end{tabular}
\caption{State estimation error variances of vanilla DA methods at different observation noise levels.}
\label{table:vanilla_da_3}
\end{table}

\FloatBarrier
\section{Conclusions \& Future Work}
We have shown through extensive computational experiments that deterministic PR algorithms applied to nonlinear ordinary differential equations (chaotic systems) exhibit huge potential through their performances. The deterministic PR methods are much more accurate and stable than stochastic methods at lower noise levels, and are generally better or similar to stochastic methods at larger noise levels. Stochastic methods are not necessarily the most accurate or stable, but their performance is less prone to change as noise levels vary.  This calls for an investigation of hybridization of these algorithms, which the authors will be exploring in a future work.  In particular, 
based on the results, the deterministic PR algorithms that appears the most robust to noise are the PWM and AV-LM algorithms, and thus this study motivates future work on analytical investigations of the convergence of mean and variance of parameter estimates from the the PWM and AV-LM algorithms and variants in the presence of noise.
We also note that we have clearly focused on recovering parameters which appear only \textit{linearly} in the systems, and it will be critical in future work to investigate parameters on which the systems are nonlinearly dependent.

\section{Acknowledgments}
A.W. was funded by the Caltech SURF program.  E.C. was supported in part by the Department of Defense Vannevar Bush Faculty Fellowship, under ONR award N00014-22-1-2790. F.H. is
supported by start-up funds at the California Institute of Technology and by NSF CAREER Award 2340762.
ChatGPT was used for (a) transcribing LaTeX representations of derivatives to Python code representation, (b) formatting code, (c) building the Ensemble Kalman Filter code, where code was fixed through error-correcting commands, and (d) assist with calculating derivatives and using Mathematica.  The authors would also like to thank Prof. Andrew Stuart  and Prof. Jared Whitehead for their valuable suggestions and discussions in this work.

\appendix
\section{Analytical Derivatives of the Lorenz '63 System}
\begin{table}[hbt!]
\centering
\begin{tabular}{|l|l|}

\hline
Order  & Expression\\
\hline \hline
1 & $\sigma (x_2 - x_1)$\\  \hline
2 & $\sigma(\rho+\sigma)x_1 - \sigma x_1x_3 - \sigma(\sigma + 1)x_2$\\  \hline
3 & $\sigma\big[x_2(\sigma\rho + \sigma^{2} + \sigma + 1) - x_1(2\sigma\rho + \rho + \sigma^{2}) + x_1x_3(2\sigma + 1 + \beta) - \sigma x_2x_3 - x_1^{2}x_2\big]$\\  \hline
4 & \makecell{$\sigma \big[x_1(\sigma\rho^{2}+3\sigma^{2}\rho +2\sigma\rho + \rho + \sigma^{3}) -x_1x_3(2\sigma\rho +3\sigma^{2} + 2\sigma +1+3\sigma\beta +\beta+\beta^{2})$ \\
$
-x_2(2\sigma\rho +2\sigma^{2}\rho +\sigma^{3}+\sigma^{2}+\sigma+1) + 2x_2x_3\sigma(\sigma + 1 + \beta) + x_1^{2}x_2(4\sigma +\beta+2)$ \\
$+ \sigma x_1x_3^{2} -3\sigma x_1x_2^{2} -\rho x_1^{3} + x_1^{3}x_3 \big]$}\\  \hline
5 & \makecell{$\sigma\big[x_2(\sigma^{2}\rho^{2}+3\sigma^{3}\rho +4\sigma^{2}\rho + 3\sigma\rho +\sigma^{4} + \sigma^{3} + \sigma^{2} + \sigma + 1)$ \\
$-x_1(3\sigma^{2}\rho^{2}+4\sigma^{3}\rho +3\sigma^{2}\rho +2\sigma\rho +\sigma^{4} + 2\sigma\rho^{2} + \rho)$ \\
$-x_2x_3(2\sigma^{2}\rho +3\sigma^{3} +4\sigma^{2}+3\sigma+5\sigma^{2}\beta +5\sigma\beta +3\sigma\beta^{2} )$ \\
$+ x_1x_3(6\sigma^{2}\rho + 4\sigma^{3} +3\sigma^{2}+2\sigma + +6\sigma^{2}\beta +3\sigma\beta +4\sigma\beta^{2}+4\sigma\rho\beta +\beta +\beta^{2}+\beta^{3}+4\sigma\rho +1)$  \\
$-x_1^{2}x_2(11\sigma\rho +11\sigma^{2}+10\sigma +3 +5\sigma\beta +2\beta +\beta^{2})
 -x_1x_3^{2}(3\sigma^{2}+4\sigma\beta+2\sigma)$ \\
$+x_1x_2^{2}(13\sigma^{2}+12\sigma +4\sigma\beta)+ x_1^{3}(7\sigma\rho+\rho\beta + 2\rho) - x_1^{3}x_3(7\sigma+2\beta+2)$ \\
$+\sigma^{2}x_2x_3^{2} + 11\sigma x_1^{2}x_2x_3 - 3\sigma^{2}x_2^{3}+x_1^{4}x_2\big]$
}\\  \hline
\end{tabular}
\caption{First five derivatives of $x_1$.}
\end{table}
\begin{table}[h!]
\centering
\begin{tabular}{|l|l|}
\hline
Order  & Expression\\
\hline \hline
1 & $\rho x_1 - x_1x_3 -x_2$\\  \hline
2 & $x_2(\sigma\rho + 1) - x_1(\sigma\rho +\rho) - \sigma x_2x_3 + x_1x_3(\sigma + \beta + 1)-x_1^{2}x_2$\\  \hline
3 & \makecell{$x_1(\sigma\rho^{2}+\rho+\sigma^{2}\rho + \sigma\rho) - x_1x_3(2\sigma\rho + 1+\sigma^{2}+2\sigma\beta+\sigma+\beta^{2}+\beta) - x_2(2\sigma\rho+1+\sigma^{2}\rho)$ \\ $+ \sigma x_1x_3^{2} + x_2x_3(2\sigma + 2\sigma\beta + \sigma^{2}) - 3\sigma x_1x_2^{2} + x_1^{2}x_2(3\sigma + \beta + 2) - \rho x_1^{3} + x_1^{3}x_3$}\\  \hline
4 & \makecell{$x_2(\sigma^{2}\rho^{2} + 3\sigma\rho + \sigma^{3}\rho + 2\sigma^{2}\rho + 1) -x_1(2\sigma^{2}\rho^{2} + \sigma\rho + \sigma^{3}\rho + \sigma^{2}\rho + 2\sigma\rho^{2} + \rho)$\\
$-x_2x_3(2\sigma^{2}\rho + 3\sigma + \sigma^{3} + 3\sigma^{2}\beta + 2\sigma^{2} + 3\sigma\beta^{2} + 5\sigma\beta)$ \\
$+ x_1x_3(4\sigma^{2}\rho + \sigma + \sigma^{3} + 3\sigma^{2}\beta + \sigma^{2} + 3\sigma\beta^{2} + 2\sigma\beta + 4\sigma \rho\beta + \beta + \beta^{2}+\beta^{3} + 4\sigma\rho + 1)$\\
$-x_1^{2}x_2 (11\sigma\rho + 3 + 7\sigma^{2} + 4\sigma\beta + 8\sigma + \beta^{2} + 2\beta) + \sigma^{2}x_2x_3^{2} -x_1x_3^{2}(2\sigma^{2} + 4\sigma\beta +2\sigma) + 11\sigma x_1^{2}x_2x_3$\\
$+ x_1x_2^{2}(12\sigma + 4\sigma\beta + 10\sigma^{2}) -3\sigma^{2}x_2^{3} + x_1^{3}(6\sigma\rho + \rho\beta +2\rho) -x_1^{3}x_3(6\sigma + 2\beta + 2) + x_1^{4}x_2$}\\  \hline
5 & \makecell{$x_1(\sigma^{2}\rho^{3} + 3\sigma\rho^{2} +3\sigma^{3}\rho^{2} + 4\sigma^{2}\rho^{2} +\rho + \sigma^{2}\rho + \sigma^{4}\rho + \sigma^{3}\rho +\sigma\rho)$\\
$-x_1x_3(3\sigma^{2}\rho^{2} + 6\sigma\rho + 6\sigma^{3}\rho + 8\sigma^{2}\rho + 1 + 11\sigma^{2}\rho\beta + 7\sigma\rho\beta^{2}+9\sigma\rho\beta + \sigma^{4}+\sigma^{3}+\sigma^{2}+\sigma$ \\
$+ 4\sigma^{3}\beta + 6\sigma^{2}\beta^{2} + 3\sigma^{2}\beta + 2\sigma\beta + 3\sigma\beta^{2} + 4\sigma\beta^{3}+\beta^{4}+\beta^{3}+\beta^{2}+\beta)$ \\
$-x_2(3\sigma^{2}\rho^{2} +4\sigma\rho +2\sigma^{3}\rho + 3\sigma^{2}\rho + 1 + 2\sigma^{3}\rho^{2}+\sigma^{4}\rho)$ \\
$+ x_1x_3^{2}(3\sigma^{2}\rho +3\sigma + 3\sigma^{3} + 11\sigma^{2}\beta + 4\sigma^{2} + 11\sigma\beta^{2} + 9\sigma\beta )$ \\
$+ x_2x_3(6\sigma^{2}\rho + 4\sigma + 2\sigma^{3} + 7\sigma^{2}\beta + 3\sigma^{2} + 9\sigma\beta^{2}+9\sigma\beta+4\sigma^{3}\beta + 6\sigma^{2}\beta^{2} + 4\sigma \beta^{3} + 4\sigma^{3}\rho$ \\
$+ \sigma^{4} + 6\sigma^{2}\rho\beta) -x_1x_2^{2}(33\sigma^{2}\rho + 33\sigma +25 \sigma^{3} + 15\sigma^{2}\beta + 50\sigma^{2} + 5\sigma\beta^{2} + 17\sigma\beta)$\\
$+ x_1^{2}x_2(64\sigma^{2}\rho + 15 \sigma + 15\sigma^{3} + 11\sigma^{2} \beta +24\sigma^{2} + 5\sigma\beta^{2} + 10\sigma\beta + 15\sigma\rho\beta + 3\beta + 2\beta^{2} + \beta^{3}$\\
$+ 45\sigma\rho + 4) -x_1^{3}(11\sigma\rho^{2} + 3\rho +25 \sigma^{2}\rho + 7\sigma\rho\beta + 14\sigma\rho +\rho\beta^{2} + 2\rho\beta)$\\
$+ x_1^{3}x_3(22\sigma\rho + 3 + 25\sigma^{2} + 16\sigma\beta +14\sigma +3\beta^{2} + 4\beta) - \sigma^{2}x_1x_3^{3}-x_2x_3^{2}(3\sigma^{2} + 6\sigma^{2}\beta +2\sigma^{3})$\\
$+33\sigma^{2}x_1x_2^{2}x_3 - x_1^{2}x_2x_3(64\sigma^{2} + 33\sigma\beta +45\sigma) -11\sigma x_1^{3}x_3^{2}+15\sigma x_1^{3}x_2^{2}+x_2^{3}(21\sigma^{2}+10\sigma^{3} + 4\sigma^{2}\beta)$\\
$-x_1^{4}x_2(10\sigma +2\beta + 3)+\rho x_1^{5} - x_1^{5}x_3$}\\  \hline
\end{tabular}
\caption{First five derivatives of $x_2$.}
\end{table}
\begin{table}[h!]
\centering
\begin{tabular}{|l|l|}
\hline
Order  & Expression\\
\hline \hline
1 & $x_1x_2 -\beta x_3$\\  \hline
2 & $\sigma x_2^{2} + \rho x_1^{2} - x_1^{2}x_3 + \beta^{2}x_3 - (\sigma + \beta + 1)x_1x_2$\\  \hline
3 & \makecell{$x_1x_2(4\sigma\rho + \beta^{2} + \sigma^{2} + 2\sigma + \sigma\beta + \beta + 1) - 4\sigma x_1x_2x_3 -x_2^{2}(3\sigma + \sigma^{2} + \sigma\beta)-x_1^{2}(3\sigma\rho + \rho\beta + \rho)$\\
$+x_1^{2}x_3(3\sigma + 2\beta + 1) - x_1^{3}x_2 - \beta^{3}x_3 pr  $}\\  \hline
4 & \makecell{$x_2^{2}(4\sigma^{2}\rho + \sigma\beta^{2} + \sigma^{3} + 4\sigma^{2} + \sigma^{2}\beta + 3\sigma\beta + 7\sigma)$\\
$- x_1x_2(12\sigma^{2}\rho + \sigma\beta^{2} + \sigma^{3} + \beta^{3} + 3\sigma^{2} + \sigma^{2}\beta+2\sigma\beta + 3\sigma + 12\sigma\rho + \beta^{2} +\beta + 1+4\sigma\rho\beta)$ \\
$+ x_1^{2}(4\sigma\rho^{2} + \rho\beta^{2} + 7\sigma^{2}\rho + 4\sigma\rho + 3\sigma\rho\beta +\rho\beta + \rho)$ \\
$- x_1^{2}x_3(8\sigma\rho + 3\beta^{2} + 7\sigma^{2} + 4\sigma + 8\sigma\beta+2\beta + 1) - 4\sigma^{2}x_2^{2}x_3 + x_1x_2x_3(12\sigma^{2}+12\sigma +10\sigma\beta)$ \\
$+ 4\sigma x_1^{2}x_3^{2} - 7\sigma x_1^{2}x_2^{2} + x_1^{3}x_2(6\sigma+2\beta +2) - \rho x_1^{4} + x_1^{4}x_3 + \beta^{4}x_3$}\\  \hline
5 & \makecell{$x_1x_2(16\sigma^{2}\rho^{2} + 4\sigma\rho\beta^{2} + 28\sigma^{3}\rho + 40\sigma^{2}\rho + 12\sigma^{2}\rho\beta + 12\sigma\rho\beta + 28\sigma\rho + \sigma^{2}\beta^{2} + \sigma^{4} + 4\sigma^{3} + \sigma\beta^{3}$ \\
$+ \beta^{4} + \beta^{3} + \beta^{2} +\beta + 1 + \sigma^{3}\beta + 3\sigma^{2}\beta + 6\sigma^{2} + 2\sigma\beta^{2} + 3\sigma\beta + 4\sigma)$ \\
$- x_1x_2x_3(32\sigma^{2}\rho + 18\sigma\beta^{2} + 28\sigma^{3} + 40\sigma^{2} + 40 \sigma^{2}\beta + 32\sigma\beta + 28\sigma)$ \\
$- x_2^{2}(20\sigma^{2}\rho + 3\sigma\beta^{2} + 5\sigma^{3} + 11\sigma^{2} + 4\sigma^{2}\beta + 7\sigma\beta + 15\sigma + 12\sigma^{3}\rho + \sigma^{2}\beta^{2} + \sigma^{4} + \sigma\beta^{3} + \sigma^{3}\beta$ \\
$+ 4\sigma^{2}\rho\beta) - x_1^{2}(20\sigma^{2}\rho^{2} + 3\sigma\rho\beta^{2} + 15\sigma^{3}\rho + \rho\beta^{3} + 11\sigma^{2}\rho + 7\sigma^{2}\rho\beta + 4\sigma\rho\beta + 5\sigma\rho$ \\
$+ 12\sigma\rho^{2} + \rho\beta^{2} + \rho\beta + \rho + 4\sigma\rho^{2}\beta) + x_1^{2}x_3(40\sigma^{2}\rho + 15 \sigma\beta^{2} + 15\sigma^{3} + 4\beta^{3} + 11\sigma^{2} + 24\sigma^{2}\beta$ \\
$+ 10\sigma\beta + 5\sigma + 24\sigma\rho + 3\beta^{2} + 2\beta  + 1+ 22\sigma\rho\beta) - x_1^{3}x_2(26\sigma\rho + 3\beta^{2} + 25 \sigma^{2} + 16\sigma + 14\sigma\beta$ \\
$+ 4\beta + 3) + 16\sigma^{2}x_1x_2x_3^{2} + x_2^{2}x_3(20\sigma^{2} + 14 \sigma^{2}\beta + 12\sigma^{3}) - 18\sigma^{2}x_1x_2^{3} - x_1^{2}x_3^{2}(20\sigma^{2} + 12\sigma + 18\sigma\beta)$ \\
$+ x_1^{2}x_2^{2}(44\sigma^{2} + 32\sigma + 16\sigma\beta) + 26\sigma x_1^{3}x_2x_3 + x_1^{4}(10\sigma\rho + 2\rho\beta + 2\rho) - x_1^{4}x_3(10\sigma + 3\beta + 2)$ \\
$+ x_1^{5}x_2 - \beta^{5}x_3$}\\  \hline
\end{tabular}
\caption{First five derivatives of $x_3$.}
\end{table}
\clearpage

\section{Analytical Derivatives of the Lorenz '96 System}
\begin{table}[h!]
\centering
\begin{tabular}{|l|l|}
\hline
Order  & Expression\\
\hline \hline
1 & $u_{i+3} (u_{i+1} - u_{i+2}) - \overline{d_{i}} u_{i} + F + \sum_{j=1}^{5} \gamma_{i} u_{i} \mathbcal{u}_{i,j}$\\  \hline
2 & \makecell{$-\overline{d_{i}} F - \sum_{j=1}^{5} \gamma_{j} \gamma_{i} u_{i}^3 
    - \overline{d_{i}} u_{i+1} u_{i+3} + \overline{d_{i}} u_{i+2} u_{i+3}$ \\
    $- u_{i+3} (-u_{i} u_{i+1} - u_{i} u_{i+2} + u_{i} u_{i+3} + u_{i+1} u_{i+3})$ \\
    $+ \sum_{j=1}^{5} \left( F \gamma_{j} + \gamma_{j} u_{i+1} u_{i+3} - \gamma_{j} u_{i+2} u_{i+3} \right) \mathbcal{u}_{i,j}$ \\
    $+ u_{i} \left( \overline{d_{i}}^2 - \sum_{j=1}^{5} \left( d_{j} \gamma_{j} \mathbcal{u}_{i,j} + 2 \overline{d_{i}} \gamma_{j} \mathbcal{u}_{i,j} + \gamma_{j}^2 \mathbcal{u}_{i,j}^2 \right) \right)$}\\  \hline
3 & \makecell{$u_{i+3} \left( \frac{d^2u_{i+1}}{dt^2} - \frac{d^2u_{i+2}}{dt^2} \right)$ \\
    $+ \sum_{j=1}^{5} \gamma_{i} \left( u_{i} \mathbcal{u}_{i,j}'' + \frac{du_{i}}{dt} \mathbcal{u}_{i,j}' + \frac{d^2u_{i}}{dt^2} \mathbcal{u}_{i,j} \right)$}\\  \hline
\end{tabular}
\caption{First three derivatives of $u_{i}$. Note that $u_{i+k} = u_{i+k-4}$ when $i+k > 4$.}
\end{table}
\section{Derivation of Parameter Recovery for Lorenz' 96}
\subsection{CHL Formula}\label{appendix:CHL}
To derive the CHL version of parameter recovery for the parameter $\overline{d_{1}}$, we start with the following set of equations, which encompasses the original dynamics regulated by $\overline{d_{1}}$ and the corresponding nudged equation.
\begin{equation}
\begin{system}
\dot{u_{1}} = u_{4}(u_{2} - u_{3}) + \sum_{j=1}^{5}\gamma_{1}u_{1}v_{1,j} - \overline{d_{1}}u_{1} + F,\\ 
\tilde{u_{1}} = \tilde{u_{4}}(\tilde{u_{2}} - \tilde{u_{3}}) + \sum_{j=1}^{5}\gamma_{1}\tilde{u_{1}}\tilde{v_{1,j}} - \tilde{\overline{d_{1}}}\tilde{u_{1}} + F - \mu_{1}(\tilde{u_{1}} - u_{1}). 
\label{eqtn:originalAndNudged}
\end{system}
\end{equation}
Set $\Delta d = \tilde{\overline{d_{1}}} - \overline{d_{1}}$, $a = \tilde{u_{1}}-u_{1}$, $b = \tilde{u_{2}}-u_{2}$, $c = \tilde{u_{3}}-u_{3}$, and $d = \tilde{u_{4}} - u_{4}$. Then 
\begin{equation}
\dot{a} = \dot{\tilde{u_{1}}} - \dot{u_{1}}.
\label{eqtn:step1}
\end{equation}
By plugging in the set of equations in \ref{eqtn:originalAndNudged} into Equation~\ref{eqtn:step1}, replacing $u_{1} = \tilde{u_{1}} - a, u_{2} = \tilde{u_{2}} - b, u_{3} = \tilde{u_{3}} - c, u_{4} = \tilde{u_{4}} - d$, and cancelling terms, we get
\begin{equation}
\dot{a} = \sum_{j=1}^{5}\gamma_{1}\tilde{u_{1}}\tilde{v_{1,j}} - \tilde{\overline{d_{1}}}\tilde{u_{1}}-\mu_{1}a+d\tilde{u_{2}}+bu_{4}-bd -d\tilde{u_{3}}-c\tilde{u_{4}}+cd - \sum_{j=1}^{5}\gamma_{1}(\tilde{u_{1}}-a)v_{1,j} + \overline{d_{1}}(\tilde{u_{1}} - a).
\end{equation}
Multiplying this equation by $a$ on both sides, and assuming quadratic terms are negligible except for when multiplied by the forcing constant $\mu_{1}$, we obtain
\begin{equation}
\dfrac{1}{2}\dfrac{d}{dt}a^{2} = a\sum_{j=1}^{5}\gamma_{1}\tilde{u_{1}}\tilde{v_{1,j}} - a\tilde{\overline{d_{1}}}\tilde{u_{1}}-\mu_{1}a^{2} - a\sum_{j=1}^{5}\gamma_{1}\tilde{u_{1}}v_{1,j}+a\overline{d_{1}}\tilde{u_{1}}.
\end{equation}
By \cite{Carlson_Hudson_Larios_Martinez_Ng_Whitehead_2021}, the time derivatives will become negligible after a transient time interval, therefore
\begin{equation}
0 \approx \sum_{j=1}^{5}\gamma_{1}\tilde{u_{1}}\tilde{v_{1,j}} - \tilde{\overline{d_{1}}}\tilde{u_{1}}-\mu_{1}a - \sum_{j=1}^{5}\gamma_{1}\tilde{u_{1}}v_{1,j}+\overline{d_{1}}\tilde{u_{1}}.
\end{equation}
After rearranging, we get Equation~\ref{CHL_96_formula}.

\subsection{PWM Formula}\label{appendix:PWM}
Again, we start with the set of equations in \ref{eqtn:originalAndNudged}. In \cite{Pachev_Whitehead_McQuarrie_2021concurrent}, the authors studied general PDEs in the form
\begin{equation}
\mathbf u_{t} + F(\mathbf u) + \sum_{k=1}^{n}\lambda_{k}G_{k}(\mathbf u) = f.
\end{equation}
In our case, let $F(\mathbf u) = -u_{4}(u_{2} - u_{3})-\sum_{j=1}^{5}\gamma_{1}u_{1}v_{1,j}$. Assuming $n=1$, let $\lambda_{k} = \overline{d_{1}}$ and simply let $G_{k}$ be an identity function for our purposes. $f$, the forcing function, corresponds to the forcing constant $F$. We view the interpolation operator $I_{h}$ as an identity operator in our case since we numerically simulate the system and we already know the numerical values of variables at each discrete time step. The nudged equation can be written as
\begin{equation}
\mathbf v_{t} + F(\mathbf v) + \sum_{k=1}^{n}\hat{\lambda_{k}}G_{k}(\mathbf v) = f + \mu_{1}(\mathbf u - \mathbf v).
\end{equation}
Let $\mathbf w = \mathbf u - \mathbf v$, then
\begin{equation}
\dot{\mathbf w} = \dot{\mathbf u} - \dot{\mathbf v} = \dot{\mathbf u} + F(\mathbf v) + \sum_{k=1}^{n}\hat{\lambda_{k}}G_{k}(\mathbf v) - f - \mu_{1}(\mathbf u - \mathbf v)
\end{equation}
which aligns with the last equation on page 4 in \cite{Pachev_Whitehead_McQuarrie_2021concurrent}. The PWM derivation in \cite{Pachev_Whitehead_McQuarrie_2021concurrent} from Equation (2.2) to Equation (2.4) shows that the orthogonality condition in (2.4) is required for convergence between the true and nudged systems. Moreover, as \cite{Pachev_Whitehead_McQuarrie_2021concurrent} stated, for recovering a single parameter, (2.4) uniquely determines the parameter estimate $\hat{\lambda_{k}} = \hat{\lambda_{1}}$. Therefore, from (2.4), we can write
\begin{equation}
0 = \mathbf w (\dot{\mathbf u} + F(\mathbf v) - f + \hat{\lambda_{1}}G_{1}(\mathbf v))
\label{eqtn:pwmDerivation}
\end{equation}
since $n=1$ in our case and hence rewriting Equation~\ref{eqtn:pwmDerivation} using the notation in our paper
\begin{equation}
0 = \dot{u_{1}} - \tilde{u_{4}}(\tilde{u_{2}}-\tilde{u_{3}}) - \sum_{j=1}^{5}\gamma_{1}\tilde{u_{1}}\tilde{v_{1,j}}-F+\tilde{\overline{d_{1}}}\tilde{u_{1}}
\end{equation}
which can be rearranged to become Equation~\ref{eqtn:PWM96approx}.

\subsection{AV Formula}\label{appendix:AV-LM}
Observe that (36) and (37) in \cite{Newey_Whitehead_Carlson_2025} are the original and nudged Lorenz '96 system examined in our paper. Referring to Equations (27) and (34), we can write out the Levenberg-Marquardt algorithm for parameter recovery 
\begin{equation}
\overline{d_{1}}^{(k+1)} = \overline{d_{1}}^{(k)} - (W^{T}W + \lambda I)^{-1}W^{T}(\tilde{u_{1}} - u_{1}).
\end{equation}
Since we are examining single-parameter recovery, our sensitivity matrix $W$ will be one-dimensional, given by (39) in \cite{Pachev_Whitehead_McQuarrie_2021concurrent} as $w_{k,1}^{OTF}$, which we can write with our notation as
\begin{equation}
W = \dfrac{1}{\mu_{1}}\sum_{j=1}^{5}\tilde{v_{1,j}}\tilde{u_{1}}.
\end{equation}
This leads to the result
\begin{equation}
\tilde{d_{1}}^{(n+1)} = \tilde{d_{1}}^{(n)} - (s^{2}+\lambda)^{-1}s(\tilde{u_{1}} - u_{1})
\end{equation}
where
\begin{equation}
s = W = \dfrac{1}{\mu_{1}}\sum_{j=1}^{5}\tilde{v_{1,j}}\tilde{u_{1}}.
\end{equation}
This is exactly what we have in equations~\ref{AV_formula} and \ref{eqtn:AVFormulaAffiliate}.

\printbibliography

\end{document}